# The MAGNEX spectrometer: results and perspectives

F. Cappuzzello[1,2], C. Agodi[1], D. Carbone[1], M. Cavallaro[1]

[1]*INFN, Laboratori Nazionali del Sud, Via S. Sofia 62, 95125 Catania, Italy*
[2]*Dipartimento di Fisica e Astronomia, Università di Catania, Via S. Sofia 64, 95125 Catania, Italy*

This article discusses the main achievements and future perspectives of the MAGNEX spectrometer at the INFN–LNS laboratory in Catania (Italy). MAGNEX is a large acceptance magnetic spectrometer for the detection of the ions emitted in nuclear collisions below Fermi energy. In the first part of the paper an overview of the MAGNEX features is presented. The successful application to the precise reconstruction of the momentum vector, to the identification of the ion masses and to the determination of the transport efficiency is demonstrated by in-beam tests. In the second part, an overview of the most relevant scientific achievements is given. Results from nuclear elastic and inelastic scattering as well as from transfer and charge exchange reactions in a wide range of masses of the colliding systems and incident energies are shown. The role of MAGNEX in solving old and new puzzles in nuclear structure and direct reaction mechanisms is emphasized. One example is the recently observed signature of the long searched Giant Pairing Vibration. Finally, the new challenging opportunities to use MAGNEX for future experiments are briefly reported. In particular, the use of double charge exchange reactions toward the determination of the nuclear matrix elements entering in the expression of the half-life of neutrinoless double beta decay is discussed. The new NUMEN project of INFN, aiming at these investigations, is introduced. The challenges connected to the major technical upgrade required by the project in order to investigate rare processes under high fluxes of detected heavy ions are outlined.

# Table of contents

# Introduction

The study of the motion of charged particles through a magnetic field is a well-established technique to explore the microscopic structure of the matter constituents. Early in the study of nuclear physics, magnetic spectrographs were used to analyze electron conversion line and alpha particle spectra from naturally occurring radioactivity. Since 1922, when F. W. Aston discovered the existence of isotopes, six Nobel Prizes were awarded in the field of magnetic spectrometry.

Nuclear physics has taken profit by the use of magnets also to select and detect the charged particles produced in a nuclear reaction [1]. In such a context, the magnetic spectrometry offers several advantages compared to other detection techniques. One of these is the strong selection of the reaction products, based on their momentum over charge ratio, which determines a significant reduction of the experimental background. As a consequence, the use of magnetic spectrometers often allows to detect the reaction products at very forward scattering angles. There, due to the proximity of the particle beam, the counting rates are typically beyond the acceptable rate of the common particle detectors. It is worth to notice that the clearest part of the spectroscopic information is normally expected at such small scattering angles. Another important advantage is that the measurements with magnetic spectrometers are usually characterized by a high achievable mass, angular and momentum resolution, thus permitting accurate studies of nuclear spectroscopy.

On the other hand, the magnetic spectrometers have been usually designed with small overall acceptance. In fact, the presence of large optical elements produces unavoidable aberrations determining a sensitive reduction of the above-mentioned properties. The treatment of the high order aberrations is a long standing issue, that has found a solution scheme only with the advent of sophisticated mathematical approaches based on the differential algebra [2] [3] [4] [5]. Nevertheless, the application of such approaches to practical cases has required a consistent supplementary effort in the last few years. This has mainly been driven by the increasing interest that modern experimental nuclear physics is giving to large acceptance detection systems. Such devices are essential when the measurement conditions are characterized by low detection yields as when dealing with low intensity radioactive ion beams or with investigations of suppressed reaction channels.

An example of large acceptance magnetic spectrometer is MAGNEX, which is installed at the INFN - Laboratori Nazionali del Sud (LNS) in Catania (Italy). MAGNEX was designed to investigate several processes, also characterized by very low yields, in different fields of nuclear physics, ranging from nuclear structure to the characterization of reaction mechanisms in a wide interval of energies and masses. The problem of the aberrations is faced by an accurate design of the spectrometer layout, which minimizes their content [6] [7] [8], an efficient ray-reconstruction technique [9], based on the differential algebraic methods of COSY INFINITY [10], and the use of a specialized focal plane detector for the measurement of the optical phase space parameters. The method is powerful enough to treat the problem up to high order ($10^{th}$ order in the case of MAGNEX) and general enough to be extended to other fields of magnetic spectrometry and transport lines.

In this article, an overview of the features of the MAGNEX spectrometer and of the originality of the adopted high order ray-reconstruction technique is presented. The successful application to the precise reconstruction of the momentum vector [9], to the identification of the ion mass [11] and to the determination of the transport efficiency [12] is shown and discussed in terms of the new opportunities thus opened in nuclear research.

Results from first experimental campaigns are discussed in a broad ensemble of research items, showing the versatility of the device and, sometimes, its unicity for facing and solving new and old puzzles of nuclear structure and reaction mechanisms.

Finally, an overview of the challenging research program conceived for the spectrometer in the next years is introduced. This mainly deals with the study of special heavy-ion induced Double Charge Exchange (DCE) reactions in the view of the possible extraction of nuclear matrix elements useful for neutrinoless double beta decay ($0\nu\beta\beta$) studies. In this context an overview of the foreseen major upgrade of the spectrometer for its application at high luminosity is also given.

# 1 Magnetic spectrometers in Nuclear Reaction studies

Nuclear physicists initially conceived magnetic spectrometers for accurate energy measurements. Nonetheless, it was soon demonstrated that such devices can be designed in order to detect reaction products at very forward angles, including zero degree, or to measure accurate reaction cross sections and/or to identify fast heavy ions. Consequently, magnetic spectrometers quickly became essential tools in nuclear physics laboratories. Different layouts have been established, depending on the optimization of one of these functions. In the following, we will concentrate mainly on momentum spectrometers.

## 1.1 The early stage

An excellent review paper describing the evolution of the ideas and technologies in the field of magnetic spectrometry was written by H. A. Enge in 1979 [13]. We refer to that and the references therein for a comprehensive description of the early stages of the field.

The first spectrometer used for nuclear reaction studies was designed by R. J. Van de Graaff and co-workers at MIT in the early forties [14] [15]. It was used to measure the energies of reaction products emitted at 90°, with respect to the beam, from targets bombarded with low energy deuterons or protons from a 2-MeV

Van de Graaff accelerator. The magnet, with an annular geometry, was a copy of a 180° single focusing spectrometer used by Rutherford for alpha-decay studies [16]. Despite the small momentum range of the MIT spectrometer ($p_{max}/p_{min}$ = 1.04) and the small solid angle (0.5 msr) it allowed a precise measurement of the radius of curvature of the ions orbit. The ions momentum could be extracted, in principle, from the measurement of the magnetic field without any calibration. However, this instrument was active before the advent of precise gaussmeters based on Nuclear Magnetic Resonance (NMR). As a consequence, the idea of absolute instruments was not pursued in the successive developments of magnetic spectrometry. Instead, the standard in magnetic spectrometry was to try to remove the aberrations (mainly second orders ones), thus aiming at an optimized focusing of the ions trajectories along the dispersive direction. The impact points of the trajectories on the focus were typically recorded on special photograph plates, which were then analyzed by microscopy. The necessity of calibrating against known reaction channels or by the use of alpha sources to analyze the energy spectra of the reaction products was common to all the instruments.

Another very representative instrument of this age is the double focusing inhomogeneous-field spectrometer at the California Institute of Technology which constitutes the first attempt to increase the solid angle by requiring a focus condition also in the not dispersive direction (transversal focusing) [17].

## 1.2 Playing with magnet boundaries

In the next two decades, the main trend was to improve the focusing properties of the spectrometers by upgrading the profiling of the Effective Field Boundaries (EFB) of the dispersive magnets. Three key instruments were developed in this line: the Browne-Buechner [18] [19], the Elbek [20] and the split-pole [21] [22] spectrometers.

The MIT broad range (known also as Browne - Buechner) spectrometer consisted of a sector magnet with circular profiled entrance and exit EFBs which allowed to get almost second order aberration free image at the focus in a broad range of accepted momenta $p_{max}/p_{min}$ = 1.5. The spectrometer was corrected for the most important second-order aberrations for one point in the spectrum only (90° deflection). It focused only in the medium plane (no normal-plane focusing) and therefore had a rather low solid angle of acceptance (0.4 msr).

B. Elbek and co-workers designed an improved version of the broad range Browne - Buechner spectrometer. The main idea was the inclination by 35° of the entrance dipole EFBs. This generates a transversal focusing which sensibly increases the solid angle (1 msr) and the momentum bite ($p_{max}/p_{min}$ = 2.1).

J. E. Spencer and H. A. Enge carried the basic ideas of the Elbek instrument a bit further in the split-pole spectrometer, where the dispersive magnet is split in two separate parts in order to have four independent EFBs to be properly profiled. The larger amount of surface defining parameters did allow to get an almost complete compensation of the aberration in a momentum range as large as $p_{max}/p_{min}$ = 2.8 and a solid angle of 2 msr. The latter was then increased up to 8 msr when the ions trajectory angle measurement at the focal plane was implemented [23]. The split-pole spectrometers were among the most popular nuclear reaction devices before 1970 and some of them are still used.

All of these instruments were characterized by a rather inclined focal plane surfaces (63.5° in the Brown-Buechner instrument) to reduce the second order chromatic aberration. This was a minor problem for light particles induced reactions (for example (d,p), (p,d), (p,t) and so on), where energy-loss straggling and multiple scattering phenomena are not very important. D. L. Hendrie and co-workers [24] built one of the first instruments especially designed for heavy ions at Berkeley. In this case, the EFB at the dipole exit was profiled with a rather pronounced concave shape in order to remove the necessity of rotating the focal plane, thus reducing the strong effect of multiple scattering in the dead layers of the tilted detector. The consequence of such an extreme profiling was the strong enhancement of the spherical aberrations, which were compensated by the addition of a quadrupole and a correcting sextupole before the dipole.

The above-mentioned progress in the spectrometer technology was also possible thanks to more advanced algorithms to study the transport of the ions through the magnetic elements. In this regard, two milestones were the development of the matrix multiplication computer program TRANSPORT, developed by K. L. Brown and co-workers [25] and the RAYTRACE code by S. N. Kowalsky and H. A. Enge [26].

One should also mention that since the sixties, based on an old idea of Enge and Buechner, the multi-gap spectrometers appeared. These basically consist of a series of vertically located dipoles arranged in a toroidal geometry around the target [27] [28]. The focal planes are also arranged in such a geometry and the entire system was enclosed in a big vacuum chamber. Such a system allows the simultaneous measurement of the reaction products emitted at all the horizontal angles since it covers almost all the horizontal phase space. This largely enhances the horizontal angular acceptance still preserving the momentum resolution and acceptance. In addition, since all the angles are explored at the same time, the normalization problems in constructing angular distributions are largely overcome. On the other hand, such systems cannot easily house quadrupoles or correcting lenses, due to the reduced spaces, thus renouncing to the increase of the solid angle determined by the transversal focusing of such lenses. Moreover, these devices are much more complicated to handle with compared to single gap spectrometers. This perhaps explains the secondary role played by multi-gap devices in the development of magnetic spectrometry.

## 1.3 The use of correcting lenses

Two major technological upgrades were available since the beginning of the seventies. The companies were able to build electromagnetic elements generating high order field multipolarities (multipoles) with good accuracy and affordable prices. In addition, gas filled focal plane detectors, especially multi-wire proportional counters of many kinds, started to be developed [29].

A new frontier was then opened since it became possible to define more complex magnetic layouts, where the adjustable strength of the correcting lenses strongly enhances the possibility to correct aberrations and to compensate for small imperfections in the first order design. In addition, the use of the new focal plane detectors offered, for the first time, the possibility to measure the trajectory angles at the focus, thus adding a powerful tool for the off-line treatment of the aberrations and kinematic broadening. The identification of heavier ions became a less critical issue, reinforcing the role of magnetic spectrometry in heavy-ion physics.

The most representative instrument of this kind is perhaps the Q3D spectrometer [30] [31], first conceived as a collaboration between Heidelberg, Munich and Princeton universities and then exported with slight modifications in many laboratories in the world. The presence of three dipoles with carefully designed boundaries, including field clamps and flexible equipotential bars (snakes), allows to correct most of the second order aberrations. A multipole element installed after the first dipole gives the supplementary correction of the kinematic broadening. The instrument was then able to guarantee a quite high momentum resolving power (about 104), a maximum solid angle of 14.7 msr and $p_{max}/p_{min}$ ranging from 1.1 to 1.28 in the different versions.

The flexibility of such complex spectrometers also well faces the long-standing problem of kinematic broadening which limited the application of magnetic spectrometry to heavy ions physics. Relevant examples of this kind of device were built in the eighties at the RIKEN (RAIDEN [32]) and JAERI [33] laboratories in Japan and at the GSI laboratory in Germany [34]. In these cases, the solid angles can reach values as high as 13 msr with quite relevant resolving power and momentum acceptance.

One should also mention that since the late fifties, the importance of the matching between beam-lines and spectrometers has been recognized [35] [36]. Detailed matching conditions and experimental procedures were extensively discussed in Refs. [37] [38] [39]. Especially when angular and energy straggling in the target are not a major concern, spectacular results can be achieved, even if in a reduced solid angle and momentum acceptance. Examples are the SPEG spectrometer [40] at GANIL (France), the WS – Grand Raiden [41] at RCNP - Osaka (Japan), and the new BigRIPS – SHARAQ [42] at RIKEN (Japan).

At the beginning of eighties, the advent of the cryogenic technology in the manufacturing of magnets gave a supplementary strength to the use of magnetic spectrometers in application where highly energetic particles have to be detected. Magnetic fields strength of several Tesla became achievable at the price of a reduced flexibility in the design.

The tremendous effort to develop magnetic spectrometers for heavy ions generated quite performing devices in terms of resolving power, being the progresses on the side of angular and momentum acceptance not of the same level. The problem was that the large acceptance condition requires much larger lens gaps, which, in turn, means an enhancement of the magnet volumes and electric consumption, which normally go with the cubic power of the gap. In addition, larger gaps require longer elements if one wants to preserve a good magnet cross-section and avoid that the fringe field contributions to the trajectories become important. When this is not the case, the role of high order aberrations is much emphasized, reducing the overall resolution. The situation is often that to add a correcting multipole element with large bore radius introduces more aberrations than it is intended to compensate. This represents the practical limit for this kind of technology.

## 1.4 Ray-tracing spectrometers

The use of focal plane trackers gives the possibility to build databases of trajectories experimentally selected through appropriate multi-holes diaphragms mounted before the spectrometer. Once the database is built, the subsequent determination of a generic trajectory, obtained removing the diaphragm, can be achieved by suitable low-order interpolation algorithms that look for the most similar trajectories in the database and operate some kind of optimization. Anyway the number of trajectories that can be separated and stored in the database is limited when large acceptance devices are considered. The strong aberrations in fact tend to mix the trajectories and enough distance between the holes of the calibration diaphragm is necessary. Thus, only a partial compensation of the aberrations can be achieved in this way. Such an approach is particular indicated when the large acceptance is requested without any particular emphasis on the momentum resolving power. One example of such a spectrometer is the BigBite at the NIKHEF laboratory (Netherlands) for studies of electron scattering [43]. Other cases developed in the nineties are the PRISMA [44] and VAMOS [45] spectrometers for heavy ions installed at the INFN-LNL (Italy) and GANIL (France) laboratories, respectively. These were conceived with special care to particle identification, in order to select medium to heavy nuclei produced in reactions driven by massive projectiles and targets. They are typically coupled with high-resolution clusters of germanium detectors around the target for precise gamma ray spectroscopy.

Since the end of the eighties, a radically different approach to large acceptance devices was also considered. Instead of pushing to the extremes the minimization of high order aberrations, the idea was to solve the equation of motion of the detected ions. To achieve this result, a very detailed measurement of the magnetic field is essential to set the highly non-linear differential equations. In addition, the focal plane detectors must provide the phase space boundary conditions (position and angle at the focus in a three-dimensional space). This revolutionary approach was possible only thanks to the advent of new mathematical instruments as the differential algebra [2], which allows to compute exact expansions of the transport map around the reference trajectory up to very high order. The accuracy of this procedure implemented in the COSY-INFINITY program [10] is only limited by the eighth order Runge-Kutta integrator with automatic step-size control to a level better than $10^{-10}$. The complete description of the trajectories makes the achievement of the focusing condition less relevant. In principle one can reconstruct a clean momentum spectrum at the target position, starting from a blurred position spectrum measured by the focal plane detector. In a certain sense, this idea of an absolute instrument recovers the old idea of the first annular spectrometer of the MIT, driven by the extraordinary development of the computational power. Of course new problems need to be faced regarding the precise measurement of three-dimensional fields with small step-sizes and large volumes; the development of algorithms for the interpolation of the measured fields; the development of suitable focal plane detectors and many others subtle requirements that an absolute instrument has (for example the extreme importance of the precise alignment of the magnets and focal plane detectors).

The MAGNEX spectrometer, described in this article, was designed within this conceptual framework.

The new possibilities offered to magnetic spectrometry, namely the fruitful use in heavy ion physics and the large acceptance, have significantly contributed to the progress in this field. Several new projects are proposed all over the world, many of which were already approved by funding agencies and are presently under construction. With a reference to momentum spectrometers only, one should mention the European project R3B [46], the Japanese SHARAQ [47] and SAMURAI [48] at RIKEN and the Russian MAVR-CORMA at JINR Dubna.

# 2 The MAGNEX spectrometer

## 2.1 MAGNEX layout

MAGNEX is a large acceptance magnetic spectrometer installed at the INFN-LNS laboratory in Catania [6], [7], [8]. It is a high-performance device merging the advantages of the traditional magnetic spectrometry with those of a large angular (50 msr) and momentum (-14%, +10.3%) acceptance detector. The spectrometer is composed of two large aperture magnets, manufactured by Danfysik A/S, namely a quadrupole [49] followed by a 55° dipole [50] and a Focal Plane Detector (FPD), built in collaboration with GANIL, [51], [52], [53], [54] for the detection of the emitted ions. The apparatus is shown in Figure 2.1 and Figure 2.2.

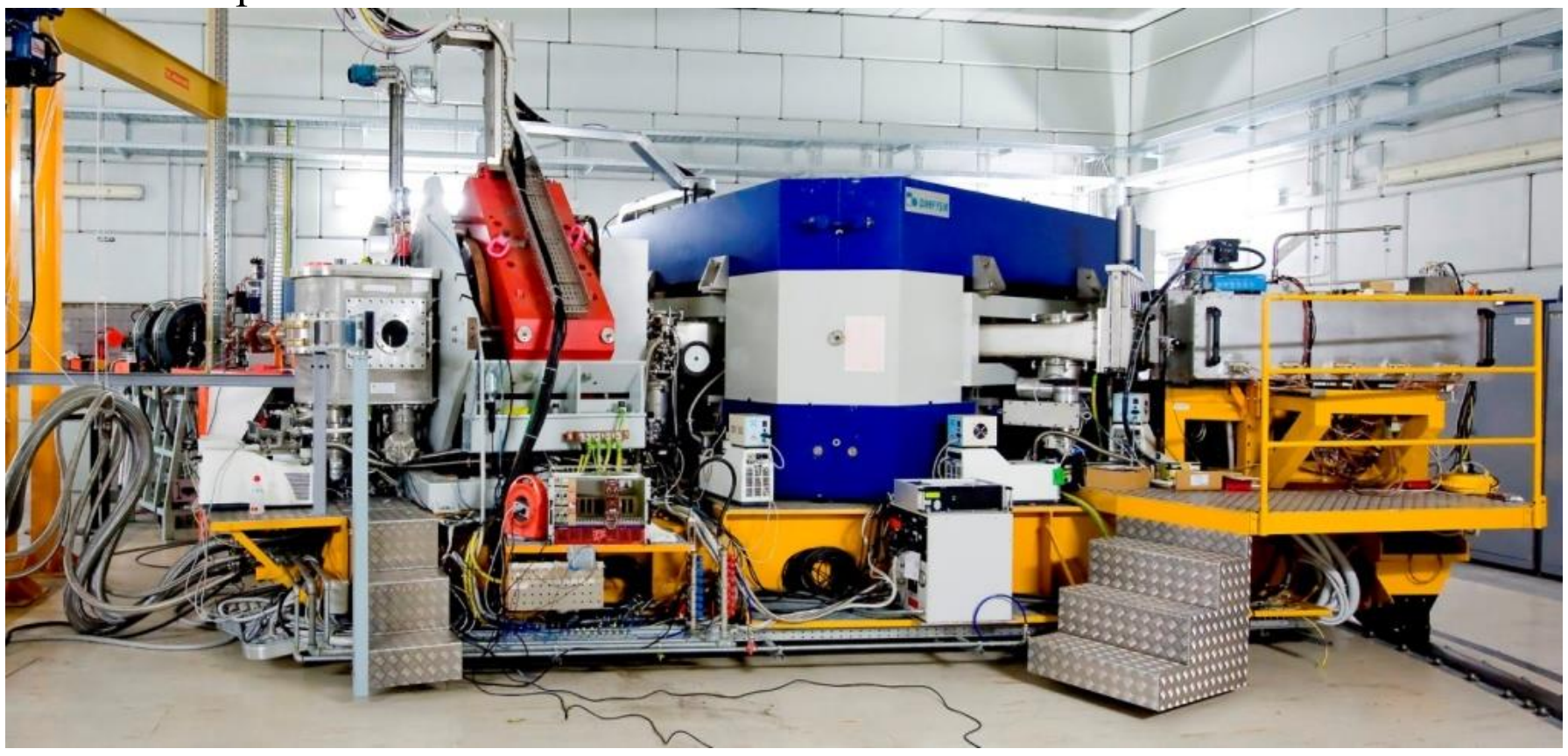

Figure 2.1 MAGNEX at the INFN-LNS. From the left to the right the scattering chamber, the quadrupole (red) and the dipole (blue) magnets, and the FPD chamber are visible.

The quadrupole magnet focuses in the non-dispersive (vertical) direction, while the dipole magnet provides the dispersion and the focusing strength in the dispersive direction (horizontal). The horizontal focus is obtained by the inclination of both the entrance and exit dipole boundaries by an angle of -18°. Two sets of surface coils are inserted between the dipole poles and the vacuum vessel. These coils generate a tunable quadrupole (α coil) and sextupole (β coil) strength. The accepted magnetic rigidities range from $B\rho \sim 0.2$ Tm to $B\rho \sim 1.8$ Tm, corresponding to energies of the detected ions ranging from $E \sim 0.2$ AMeV to $E \sim 40$ AMeV, depending on their mass and charge. The characteristics of the magnets and the main optical parameters are listed in Table 2.1 and Table 2.2, respectively.

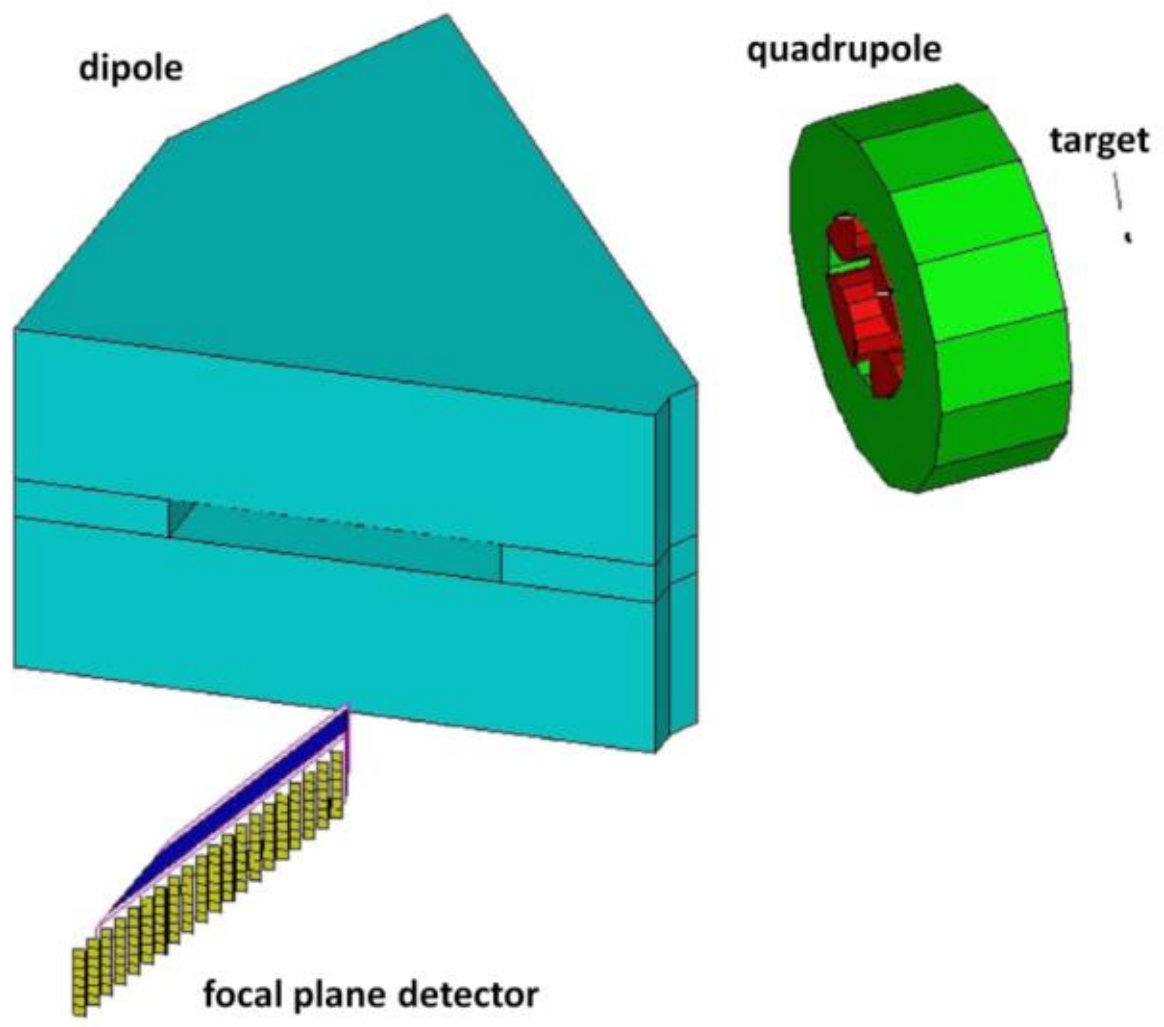

Figure 2.2 Schematic view of MAGNEX.

Table 2.1: Parameters of the dipole and the quadrupole

| **Dipole** | |
|---|---|
| Maximum field | 1.15 T |
| Bending angle | 55° |
| Bending radius $\rho$ | 1.60 m |
| $\rho_{min}$, $\rho_{max}$ | 0.95, 2.35 m |
| Pole gap | 18 cm |
| Entrance and exit pole face rotation | -18° |
| **Surface coils** | |
| Maximum value for $\alpha$ (at 1.15 T) | 0.03 |
| Maximum value for $\beta$ (at 1.15 T) | 0.03 |
| **Quadrupole** | |
| Maximum field strength | 5 T $m^{-1}$ |
| Radius of aperture | 20 cm |
| Effective length | 58 cm |

Table 2.2: Main optical characteristics of MAGNEX spectrometer.

| **Optical characteristics** | **Actual values** |
|---|---|
| Maximum magnetic rigidity (Tm) | 1.8 |
| Solid angle (msr) | 50 |
| Horizontal angular acceptance (mrad) | -90,+110 |
| Vertical angular acceptance (mrad) | ±125 |
| Momentum acceptance (%) | -0.14,+0.1 |
| Central path length (cm) | 596 |
| Momentum dispersion (cm/%) | 3.68 |
| First order momentum resolution | 5400 |
| Focal plane rotation angle (degrees) | 59.2 |
| Focal plane length (cm) | 92 |
| Focal plane height (cm) | 20 |

The simulated horizontal and vertical beam envelope through the spectrometer are shown in Figure 2.3 and Figure 2.4 respectively. A number of trajectories produced by a GEANT simulation [55] with initial conditions distributed through the phase space accepted by the spectrometer are indicated [56].

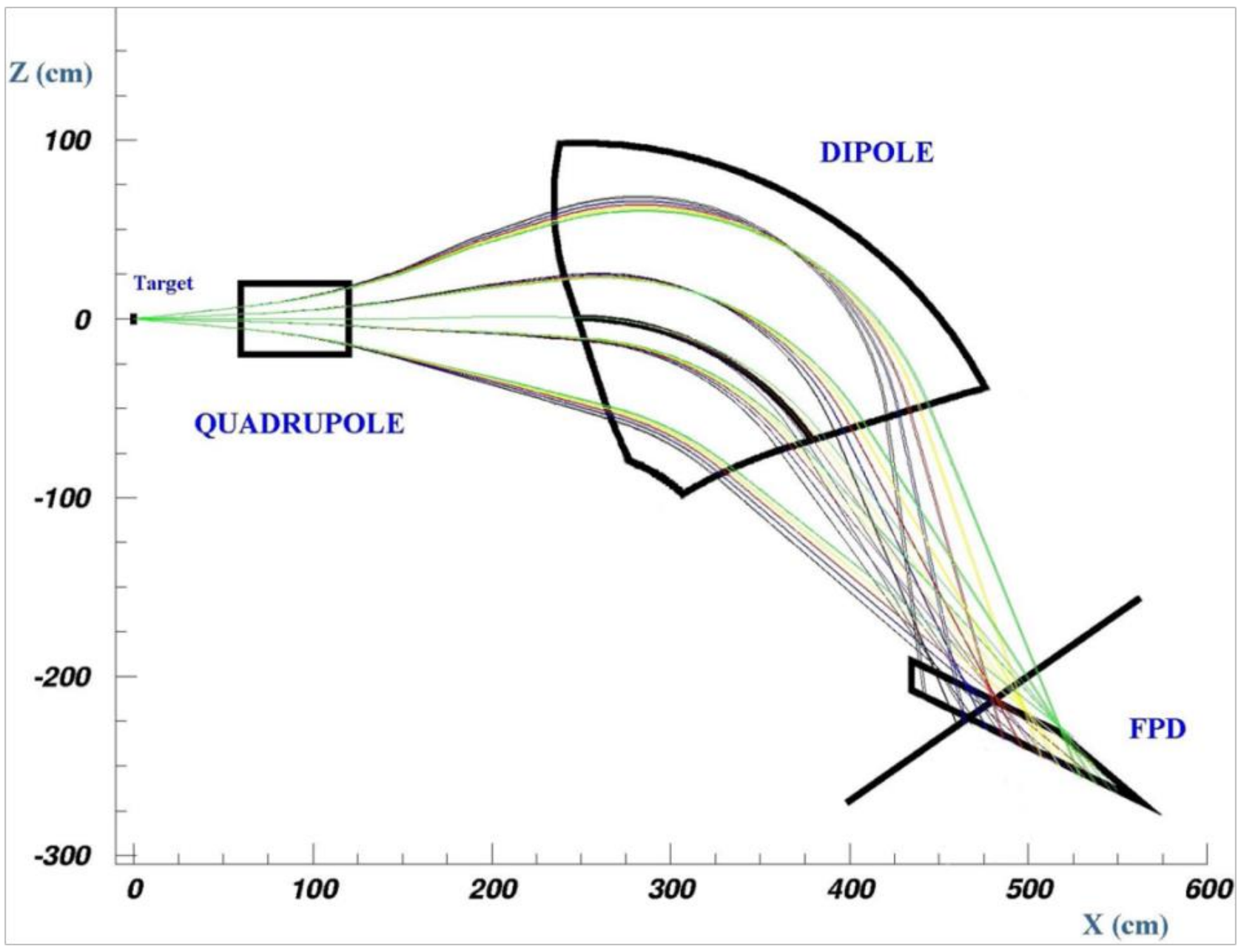

Figure 2.3 Plan view of MAGNEX from a GEANT simulation. The horizontal envelope of 85 particles distributed all over the initial phase space is shown. Rays with different colors have different momentum: $\delta = 0.1$ (green); $\delta = 0.05$ (yellow); $\delta = 0$ (red); $\delta = -0.05$ (blue); $\delta = -0.1$ (black). The contours of the shapes of the dipole and quadrupole fields on the symmetry plane are shown.

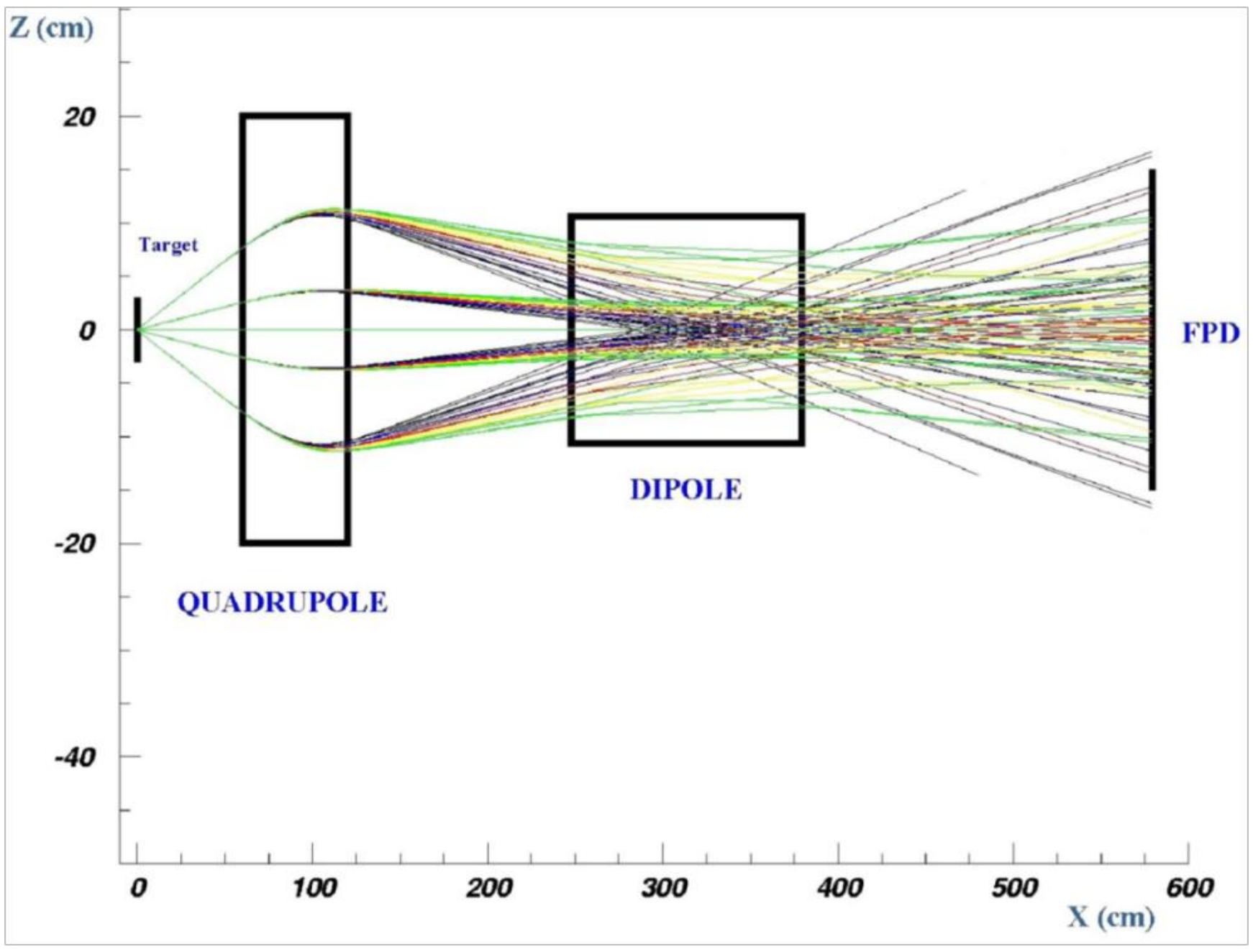

Figure 2.4 Section view of MAGNEX from a GEANT simulation. The vertical envelope of 85 particles distributed all over the initial phase space is shown. Rays with different colours have different momentum: $\delta = 0.1$ (green); $\delta = 0.05$ (yellow); $\delta = 0$ (red); $\delta = -0.05$ (blue); $\delta = -0.1$ (black). The contours of the shapes of the dipole and quadrupole fields on the side plane are shown.

## 2.2 Trajectory reconstruction

The motion of a charged particle beam under the action of a magnetic force can be represented as a general phase space mapping [5] [57]

$$F : P_i \rightarrow P_f \tag{2.1}$$

which connects the final position $P_f \equiv (x_f, \theta_f, y_f, \phi_f, l_f, \delta_f)$ to the initial one $P_i \equiv (x_i, \theta_i, y_i, \phi_i, l_i, \delta_i)$. In eq. (2.1) $x, \theta, y, \phi$ are the horizontal and vertical coordinates and angles at the impact point of the ion trajectory with a plane normal to the central trajectory, $l$ is the trajectory length and $\delta = (p - p_0)/p_0$, is the final fractional momentum, where $p_0$ is the reference momentum and $p$ is the actual one.

Equation (2.1) describes a general non-linear transport relation, where the mapping $F$ depends on the general three-dimensional spatial distribution of the magnetic fields of the actual optical system. More explicitly, it is obtained

$$\begin{aligned} x_f &= F_1(x_i,\theta_i,y_i,\phi_i,l_i,\delta_i) \\ \theta_f &= F_2(x_i,\theta_i,y_i,\phi_i,l_i,\delta_i) \\ y_f &= F_3(x_i,\theta_i,y_i,\phi_i,l_i,\delta_i) \\ \phi_f &= F_4(x_i,\theta_i,y_i,\phi_i,l_i,\delta_i) \\ l_f &= F_5(x_i,\theta_i,y_i,\phi_i,l_i,\delta_i) \\ \delta_f &= \delta_i \end{aligned} \qquad (2.2)$$

The last equation expresses the conservation of the momentum modulus for static magnetic fields in the absence of energy degrading materials. The parameter $l_i$, being essentially constant for thin targets, is not considered in the following. The main mathematical properties of the transport equations for large acceptance spectrometers were discussed in details in ref. [58] [56], where it was also shown that non-linear terms (aberrations) can severely limit the momentum and angular resolution achieved at the focal plane.

In order to get the momentum vector at the target position, it is necessary to invert eqs. (2.2) with respect to the initial set of parameters, which formally gives

$$\boldsymbol{F}^{-1} : \boldsymbol{P}_f \rightarrow \boldsymbol{P}_i \qquad (2.3)$$

It is convenient to formulate the problem in terms of the measured parameters, which can be different for different applications. For example, in the case of MAGNEX the chosen set of measured quantities is $Q_f \equiv (x_f,\ \theta_f,\ y_f,\ \phi_f,\ x_i)$, while the reconstructed vector is $Q_i \equiv (\theta_i,\ y_i,\ \phi_i,\ l_f, \delta)$. The coupled set of inversion equations is

$$\boldsymbol{G}^{-1} : \boldsymbol{Q}_f \rightarrow \boldsymbol{Q}_i \qquad (2.4)$$

An example of the application of such an equation is reported in Section 2.8.

When the size of the $x_i$ interval can be neglected, i.e. assuming a beam horizontally focused on the target, and renouncing to the reconstruction of $l_f$, the eq. (2.4) can be simplified to

$$\boldsymbol{G}'^{-1} : \boldsymbol{Q}'_f \rightarrow \boldsymbol{Q}'_i \qquad (2.5)$$

where $Q'_f \equiv (x_f,\ \theta_f,\ y_f,\ \phi_f)$ is the final reduced phase space vector and $Q'_i \equiv (\theta_i,\ y_i,\ \phi_i, \delta)$ is the reconstructed one.

The achievement of a stable and accurate solution of the eqs. (2.4) or (2.5) strongly depends on the device acceptance. Far from the optical axis and from the reference magnetic rigidity, high order terms in the Taylor expansions of the transport operators are required.

A reliable technique to solve eq. (2.4) was developed at the Michigan State University for the S800 spectrometer [58]. This is based on the differential algebra formalism [2] and allows the calculation of high order transport matrices avoiding lengthy ray-tracing procedures. In particular, the following recurrence formula is used:

$$M_n =_n (A_1^{-1} \circ (I - A_n^* \circ M_{n-1})) \qquad (2.6)$$

where the symbol $=_n$ means that the product is truncated to the $n^{th}$ order, while $A_n$ and $M_n$ are the direct and inverse $n^{th}$ order transport matrices. The COSY INFINITY program [10] contains such an algorithm allowing, at the same time, the inclusion of externally determined magnetic fields. One of the key requirements of this technique is that the magnetic fields must be represented by regular functions of the position coordinates in order to calculate stable high order derivatives. One should also check that the used field model, which could be based on mathematical interpolations, must be compatible with Maxwell equations. These tasks are quite demanding when working with fields obtained by the interpolation of discretely distributed measured data, especially in the region of rapidly changing fields such as the magnet fringes [49] [50] [59] [60].

To summarize, the use of these advanced techniques in a real application does require the detailed description of the magnetic fields crossed by the ions, the exact knowledge of the geometry of the spectrometer and the accurate measurement of the phase space vector at the focal plane $(x_f,\ \theta_f,\ y_f,\ \phi_f)$. The use of refined simulations helps in determining the accuracy level required for all the building blocks, as discussed in refs. [56] for the case of MAGNEX. In particular, the problem of the appropriate description of the magnetic field for the purposes of the algebraic trajectory reconstruction was extensively studied, demonstrating the reliability of this technique even with huge sets of data extracted from discrete three-dimensional field measurements. The high level of accuracy required for the determination of the field is also needed for the position of the beam spot at the target, of the magnets and the FPD. This is achieved by lengthy and accurate (within decimals of millimeter) measurements and alignments of all the elements of the spectrometer, using bubble-levels and theodolites. This is a fundamental point since the measured vectors $(x_f,\ \theta_f,\ y_f,\ \phi_f,\ x_i)$ or $(x_f,\ \theta_f,\ y_f,\ \phi_f)$ used in eq. (4-5) respectively, must be defined in the same reference of the transport matrices ($G$ or $G'$). Indeed, the FPD must guarantee the highest possible resolution in the measurement of the phase space vector in order to preserve a good quality in the reconstructed momentum vector.

Once these procedures are implemented, the practical way to apply the trajectory reconstruction to real data is to compare the measured observables at the focal plane with the simulated ones, which represent a model of the spectrometer response. Small adjustments (~ $10^{-3}$ of the field integral) are allowed in the simulations, in particular in the fringe field regions of the magnetic elements. This accounts for residual discrepancies between the real spectrometer and its model, as the known variation of the magnetic field geometry as a function of its strength, or the effect of slight misalignments of the real elements compared to the simulated ones [50] [49]. The closer is the simulated description of the phase space at the focal plane to the measured one the better is the model of the transport operator and consequently of its inverse.

## 2.3 The Focal Plane Detector

The MAGNEX Focal Plane Detector (FPD) basically consists of a proportional drift chamber divided in five sections, with five proportional counters, four of which are position-sensitive, and a wall of stopping silicon detectors at the back.

The FPD is placed 1.91 m downstream the exit pole-face of the MAGNEX dipole, where the focal plane of the spectrometer is defined. The FPD active volume is confined by a stainless steel vessel with a unique aperture at the front housing a thin Mylar window. No intermediate foils separate the sections, allowing heavy-ions to be detected with energies down to about 0.5 MeV/u.

The FPD vessel is mounted on a movable carriage that translates of ±0.08 m along the spectrometer optical axis in order to adapt the detector position to different focus conditions. The FPD is installed with the entrance surface rotated of $\theta_{tilt} = 59.2°$ with respect to the central trajectory in order to reduce the effect of chromatic aberrations [7].

The gas active volume is 1360 mm wide, 200 mm high and 96 mm deep with a cathode plate below and a Frisch grid above. A schematic drawing of the detector is shown in Figure 2.5.

The Mylar entrance window is 920 mm wide and 220 mm high and has typical thickness ranging from 1.5 μm to 6 μm, depending on the needs. Twenty coated metallic wires 0.5 mm in diameter, arranged horizontally and spaced 10 mm from each other, support it. The gas normally used is N35 isobutane (99.95% pure) at 10 mbar, although various pressures ranging from 5 to 100 mbar can be used and other gases such as $C_3F_8$ or gas mixtures are admitted. The choice of pure isobutane guarantees a reasonable compromise between good localization of the avalanche, stable gain and fast drift velocity. A gas flowing system maintains a stable pressure and preserves the gas purity.

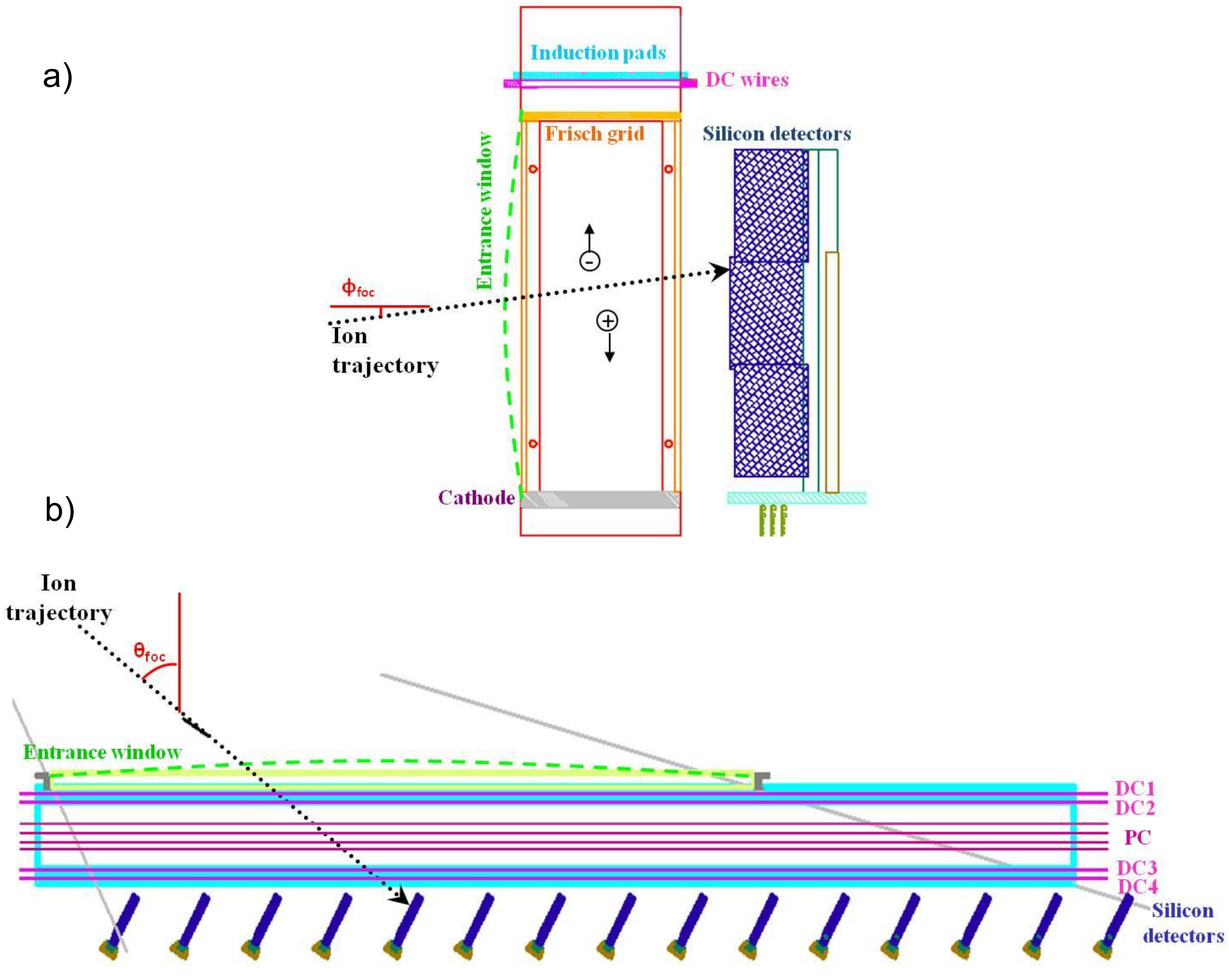

Figure 2.5. Schematic view of the Focal Plane Detector: a) side view; b) top view. From [53].

The Frisch grid is made of 10 gold-plated tungsten wires, 50 μm in diameter, spaced 5 mm between centres. The calculated shielding efficiency of the grid with respect to the anode wires 20 mm above is about 89% [61]. A partition grid guarantees the uniformity of the electric field in the drift region between the cathode (usually at voltage between –900 V and -1500 V) and the Frisch grid (connected to the ground). The equipotential rings of the partition grid partially intercept the beam envelope with an overall efficiency loss of about 1.4% for the detector.

The proportional counter section includes five sets of amplifying wires, four Drift Chambers (DCs) and one Proportional Counter (PC), sequentially defined as $DC_1$, $DC_2$, PC, $DC_3$, $DC_4$ (see Figure 2.5). They are gold-plated tungsten wires located 20 mm above the Frisch grid and spaced 8 mm apart. The DC wires are 20 μm in diameter while the PC ones are 100 μm. Each of the DC counters is made of a unique amplifying wire, while for the PC eight wires are connected in common. The high-voltage to the proportional wires (usually from +600 V to +1300 V) is provided by a unique power supply. A partition grid similar to the one used in the drift region is used to improve the field uniformity also in the proportional section. Due to the low capacitance of the amplifying wires (about 8 pF) and to their proximity, one can expect a certain amount of cross-talk and a high sensitivity to the environmental noises. To reduce such effects, a special care was paid to isolate and shorten the signal transmission lines up to the preamplifiers.

For each of the DC counters, a set of 224 independent induction pads is located 5 mm above, orientated along the spectrometer optical axis, as shown in Figure 2.6. The entire strip patterned electrode is engraved on a six-layered 6 mm thick printed circuit board. Each strip is 8 mm long and 5.9 mm wide and separated by 0.1 mm from its neighbour. This is the result of a compromise between several elements:

i) the need of a thin tracker section to reduce the ballistic effect in the reconstruction of the focal plane impact point [7];

ii) the measurement of the horizontal angle, which requires more than one section;

iii) the problem of differential non-linearities of the strip based detectors, which double the number of position-sensitive sections [62];

iv) the need to maximize the area of the strips to significantly increase the signal to noise ratio.

On the other hand, this slightly enhances the cross-talk between the signals of two adjacent strips. This effect is quantitatively described in Section 2.3.4 and in ref. [54]. To reduce a possible non-linearity in the position measurement, the strips associated with $DC_2$ and $DC_4$ are shifted by half a strip width from the $DC_1$ and $DC_3$ ones, respectively.

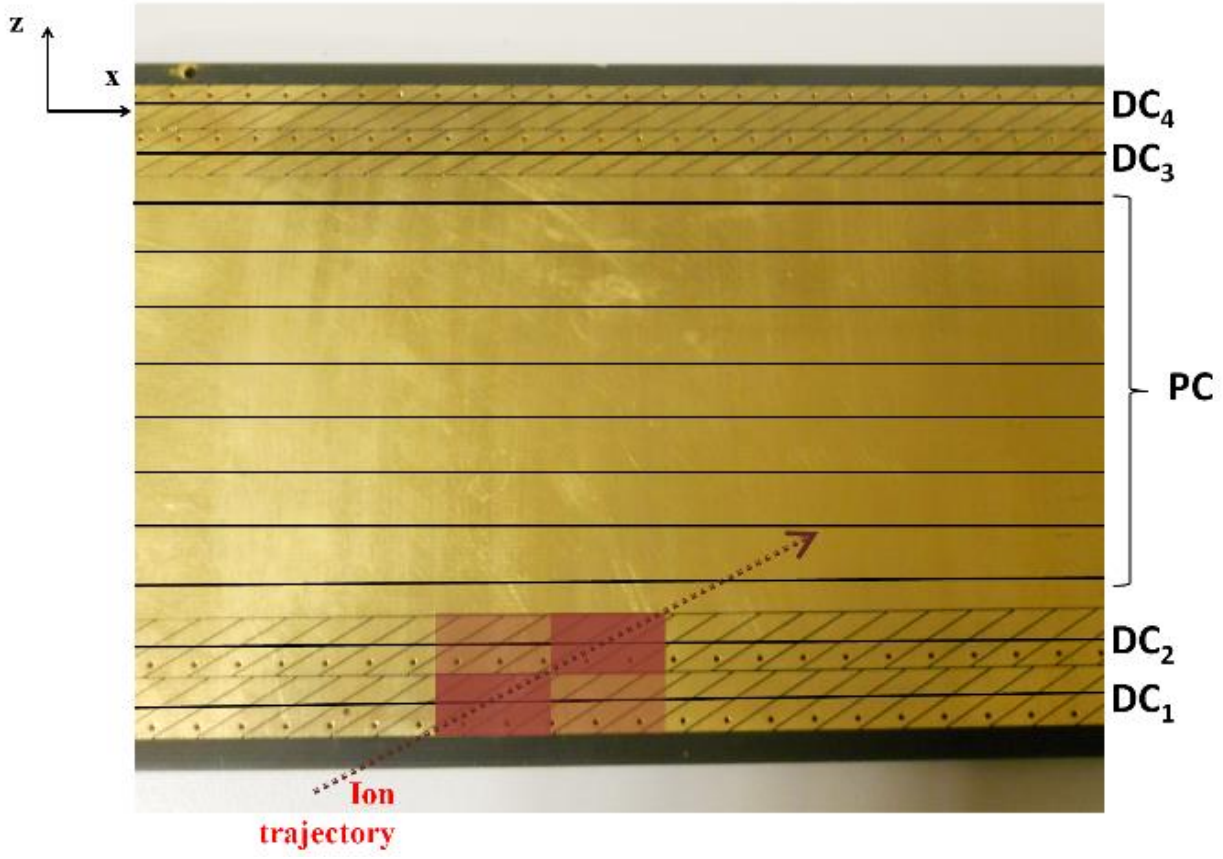

Figure 2.6 Picture of a part of the induction pads. Superimposed, a pictorial representation of the proportional wires $DC_{1\text{-}4}$ and PC and of the induction signal formation is drawn. The arrow represents a typical ion trajectory and the shadowed areas in $DC_1$ and $DC_2$ the regions affected by the charge induction. The dark areas symbolize the main induced signal and the light ones the cross-talk derived signals.

A "wall" of 60 silicon pad detectors, arranged in 20 columns and 3 rows, is located at the back of the gas detector. Each silicon detector has an active area of 70 mm (height) X 50 mm (width) and 500 μm thickness, resulting in a capacitance of about 1 nF. Silicon detectors 1000 μm thick are also available and used if required by the experimental case. The silicon columns are mounted orthogonally to the spectrometer optical axis. The silicon detectors belonging to the same column are 1 mm vertically overlapped in order to minimize the dead spaces between two adjacent detectors. The edges of the detectors are located 15 mm away from $DC_4$. The influence of the electric field generated by such edges on the uniformity of the drift field was studied with finite elements based electrostatic calculations [52]. The results show that the 15 mm distance is a safe working condition. The columns are mechanically supported and electrically connected by a mother board built on a 6 mm thick double sided printed circuit.

### 2.3.1 Principle of operation and read-out electronics

The incident charged particles coming from the dipole cross the FPD window and leave a track of ionized atoms and primary electrons in the gas between the cathode and the Frisch grid, as drawn in Figure 2.5a. Under a uniform electric field of around 50 V/cm, the electrons drift towards the Frisch grid with constant velocities typically in the range of 3-5 cm/μs, depending on the actual voltage and gas pressure [63]. After the grid, the electrons are accelerated in an electric field that becomes much stronger close to the DC and PC wires, where a multiplication by a factor around 100-200 occurs. The avalanche produces a signal proportional to the energy lost by the ions in each section, thus providing five subsequent measurements of the energy loss ($\Delta E_1$, $\Delta E_2$, $\Delta E_{PC}$, $\Delta E_3$, $\Delta E_4$) for each event. The signals, with raise times of about 150 nsec, are shaped and amplified by charge-sensitive pre-amplifiers [64] with sensitivity of about 200 mV/MeV (silicon-equivalent). The amplified signals, proportional to the energy-loss, are used for the particle identification [11]. The logic outputs, extracted only for the DC wires, are used as a STOP for the measurement of the electrons drift time as described below.

The electron avalanche around the DC wires induces a charge on the nearest pads. The signals are then pre-amplified and shaped by an analog multiplexed read-out system based on 16 channels GASSIPLEX chips [65] mounted on the upper side of the board in the gas environment. The multiplexed signals from each of the four DC chains are readout and digitally converted by modules of readout for analog multiplexed signals. The center of gravity of the charge distribution at each DC section is then extracted. By exploiting the regular pattern of the striped anode, it is possible to convert with high accuracy the measured centroid from the variable "pad number" to the more convenient horizontal position $X_1$, $X_2$, $X_3$, $X_4$ in meter units. Thus, four positions are independently determined, allowing the measurement of the horizontal position ($x_f$) and angle ($\theta_f$) of the ion track at the spectrometer focal plane. The observables needed for the ray-reconstruction are the coordinates referred to the optical axis (see Section 2.2). The projection of the optical axis ($x_f = 0$; $\theta_f = 0°$) over the segmented electrode was obtained for each DC section within 0.1 mm accuracy by means of a theodolite.

The charged particles crossing the gas section reach the silicon detector wall. Charge pre-amplifiers similar to the ones used for the wires signals [64], typically with sensitivity selectable from 5 to 90 mV/MeV and working in the gas environment, are used. The outputs are sent to 16-channel shaping amplifiers providing spectroscopic and timing outputs. The former are used to measure the residual energy ($E_r$) of the ions after crossing the gas. The latter are sent to CFDs and give multipurpose timing signals of MAGNEX. For example, they can also be used to measure the Time Of Flight (TOF) of the particles through the spectrometer providing the STOP signal, once a suitable START is available. The logic OR is used as START signal for the electron drift times measurements and also to trigger the data acquisition and to generate the gate for the $\Delta E$ and drift time measurements.

Four subsequent electron drift times in the gas are measured by the interval between the signal generated by the silicon detectors (START) and the DCs (STOP), using four standard TAC + ADC read-out chains. Thanks to the almost constant velocity of the electrons in the gas, these times are used to determine the vertical positions $Y_1$, $Y_2$, $Y_3$, $Y_4$ of the ion tracks at the intercept with the DC wires and consequently the vertical angle. Note that the vertical coordinates need an absolute calibration to be correctly transformed in the optical reference frame. In this way, the vertical position $y_f$ and angle $\phi_f$ of the ion track at the focal plane are determined.

### 2.3.2 Performances of the FPD

The performances of the MAGNEX Focal Plane Detector in terms of achieved resolution, maximum tolerable ion rate, and explored ion mass range were studied in Refs. [53], [66] and are resumed in Table 2.3.

Table 2.3. Achieved performances of the MAGNEX Focal Plane Detector.

| | |
|---|---|
| Horizontal and vertical position resolution (FWHM) | 0.6 mm |
| Horizontal and vertical angular resolution (FWHM) | 0.3° |
| Mass resolution [(a)] | 0.6% |
| Explored ion mass range | from A=1 to A=48 |
| Energy loss resolution [(b)] | 6.3% |
| Maximum incident ion rate (uniform distribution) | 5 kHz |
| Maximum incident ion rate (localized in ~ 1 cm) | 2 kHz |

[(a)] Resolution obtained applying the ray-reconstruction as described in Section 2.5.
[(b)] Resolution measured in the $^{18}O$ region as reported in Ref. [53].

### 2.3.3 Algorithm for the horizontal position calculation

In order to obtain the horizontal position parameters $X_1$, $X_2$, $X_3$, $X_4$ at the focus, it is necessary to perform a relative calibration of the response of the induction pads for each DC detector and then to determine the position of the avalanche of a typical event by extracting the center of gravity of the discrete distribution. A proper centroid-finding algorithm must be adopted to this aim, which accounts for the particular geometrical configuration of the pads with respect to the multiplication wires. In fact, the main consequence of the rotation of the induction pads with respect to the wires is a variable number of excited pads and very different shapes of the charge distribution for different events. With the adopted configuration of the induction pads the typical number of hit pads in each detector is from 5 up to 25, the larger for more inclined trajectories at the focal plane. These two effects are dependent on the horizontal trajectory angle, as it is shown in Figure 2.7, where examples of the induced charge distribution over the DC2 patterned electrode are shown. It was demonstrated in ref. [54] that, in such conditions, any standard centroid-finding algorithm [67], [68] fails in the determination of the centroid.

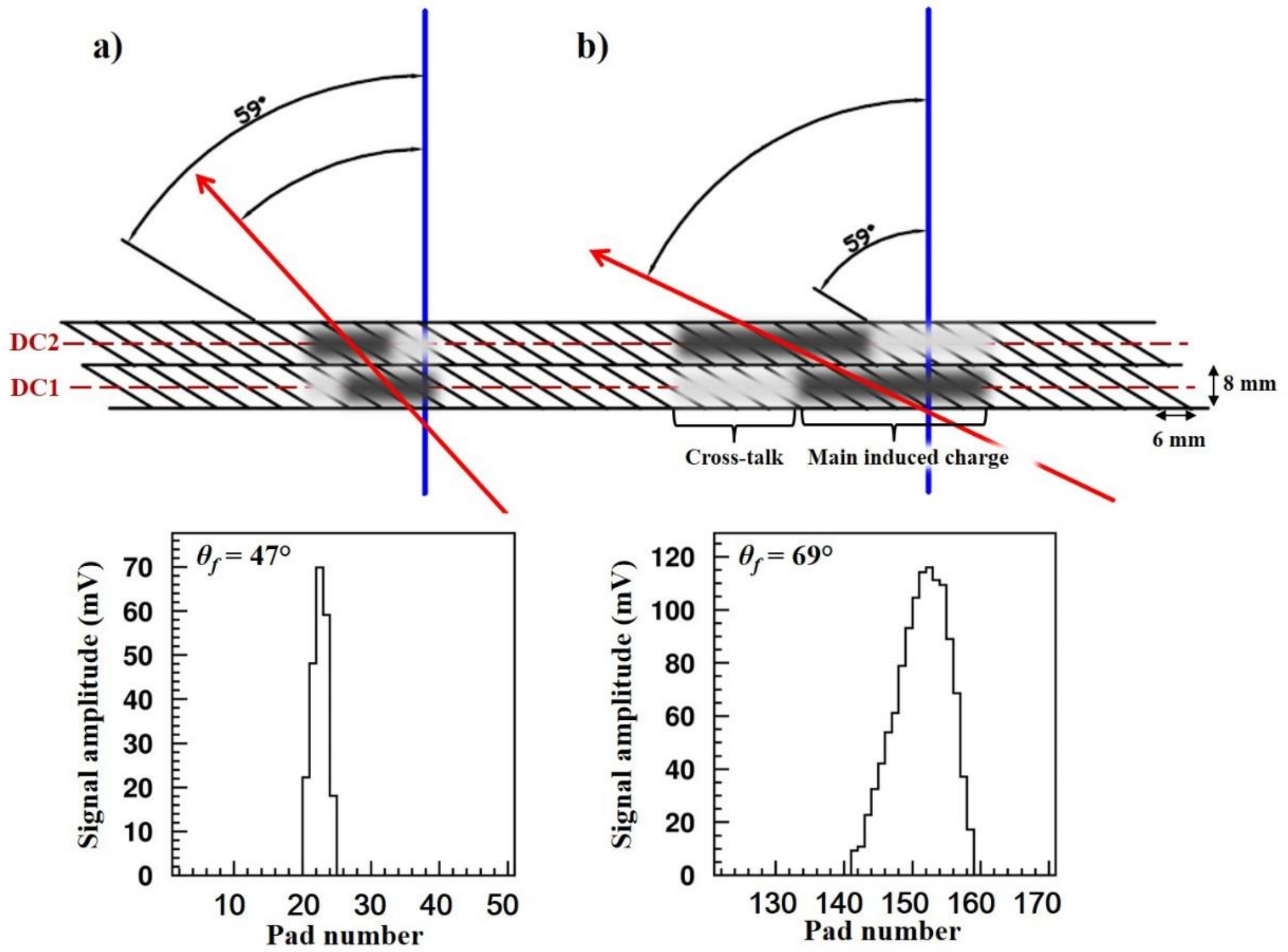

Figure 2.7 Upper panel: induction pads detail. The gray shadow indicates the main induced charge over the electrode after the ion crossing (red arrow). The light gray shadow indicates the cross talk signal induced by the neighbor DC. a) Example of a trajectory with $\theta_{foc} = 47°$; b) example of a trajectory with $\theta_{foc} = 69°$. Lower panel: induced charge distributions over the DC2 patterned electrode corresponding to the trajectories shown in the upper panel.

A new technique for the determination of the horizontal position from the induced charge distribution was presented and applied to the measurement of ejectiles through the MAGNEX FPD in ref. [54]. Such a technique basically consists of the use of the center of gravity algorithm (COG) described in ref. [68] upgraded by the implementation of an iterative procedure to set the proper threshold event by event, even for those with low signal to noise ratio.

In detail, the centroid pad number $\bar{n}$ is calculated by weighting each pad number $n_j$ with the charge $q_j$ induced on that pad:

$$\bar{n} = \frac{\sum(q_j - b)n_j}{\tilde{Q}} \qquad \tilde{Q} = \sum(q_j - b) \qquad (2.7)$$

for $q_j - b > 0$, where a pad is included in the sum if the charge induced on it is above a threshold (bias) $b$. As observed in ref. [68], a careful choice of the threshold improves the quality and the stability of the measurement. The optimal bias level $b$ should be set proportional to the total charge measured $b = kQ = k\sum q_j$, where $k$ can vary between $5\times10^{-3}$ and $2.5\times10^{-2}$.

In order to analyze the charge distribution and test the quality of the chosen threshold and the calculated centroid, the behavior of the standard deviation ($\sigma$) of the distribution is also studied

$$\sigma = \sqrt{\frac{\sum(n_j - \bar{n})^2(q_j - b)}{\tilde{Q}}} \qquad (2.8)$$

for $q_j - b > 0$.

If the threshold is properly set, $\sigma$ should range from ~ 1 (in the case of three bins exceeding the bias level in the distribution) to ~ 5 (for 20 bins). The improving of the standard COG algorithm consists in taking $\sigma$ as the control parameter of an iterative procedure. In particular, when the $\sigma$ parameter exceeds the accepted values ($1 < \sigma < 5$), the bias level is increased of a small quantity in successive iterations, controlled by the $i$ index:

$$b_i = (k + 0.002i)Q \qquad (2.9)$$

The iterations are repeated until the $\sigma$ value becomes smaller than 5. Typically a number of iterations smaller than 40 is enough to obtain the correct bias level. Using this algorithm a precise determination of the centroid of the charge distribution with a very good efficiency (~ 97%) is obtained [54].

### 2.3.4 Cross-talk correction

Another phenomenon that only the optimized COG algorithm can treat is the cross-talk in the induced signals between two neighboring detectors. In the geometry of the MAGNEX FPD the capacitive coupling between a proportional wire and the pads above the neighboring wire is relevant (~ 30%) with respect to that of the same wire with its pads. Consequently, each time an avalanche is produced near a wire a cross-talk signal is induced also in the pads above the neighbouring wire. When dealing with detectors in which the pads are perpendicular with respect to the multiplication wire this phenomenon is not accounted for, since the cross-talk modify only the amplitude of the induced signal, but not the shape of the distribution. On the contrary, with the inclined geometry adopted in the MAGNEX FPD the presence of a cross-talk signal modifies also the shape of the induced charge distribution, making it asymmetric. In fact, the cross-talk induced signal is located only on one side of the main induced charge distribution and becomes larger as the horizontal angles increases, as shown in the upper panel of Figure 2.7.

The presence of a cross-talk affects the measure of the angle $\theta_f$ of the ions at the FPD. The effect is visible in the ($Z$, $X$) correlation plot shown in Figure 2.8. After the application of a least-squared algorithm, the coefficients of a linear fit are extracted, which represent the position ($x_f$) and the angle ($\theta_f$) of the impinging trajectory at the focal plane. As shown in Figure 2.8. A systematic misplacement of $X_1$ and $X_4$ compared to $X_2$ and $X_3$ can be observed. In particular, $X_1$ is pushed above and $X_4$ below the ideal line connecting $X_2$ and $X_3$.

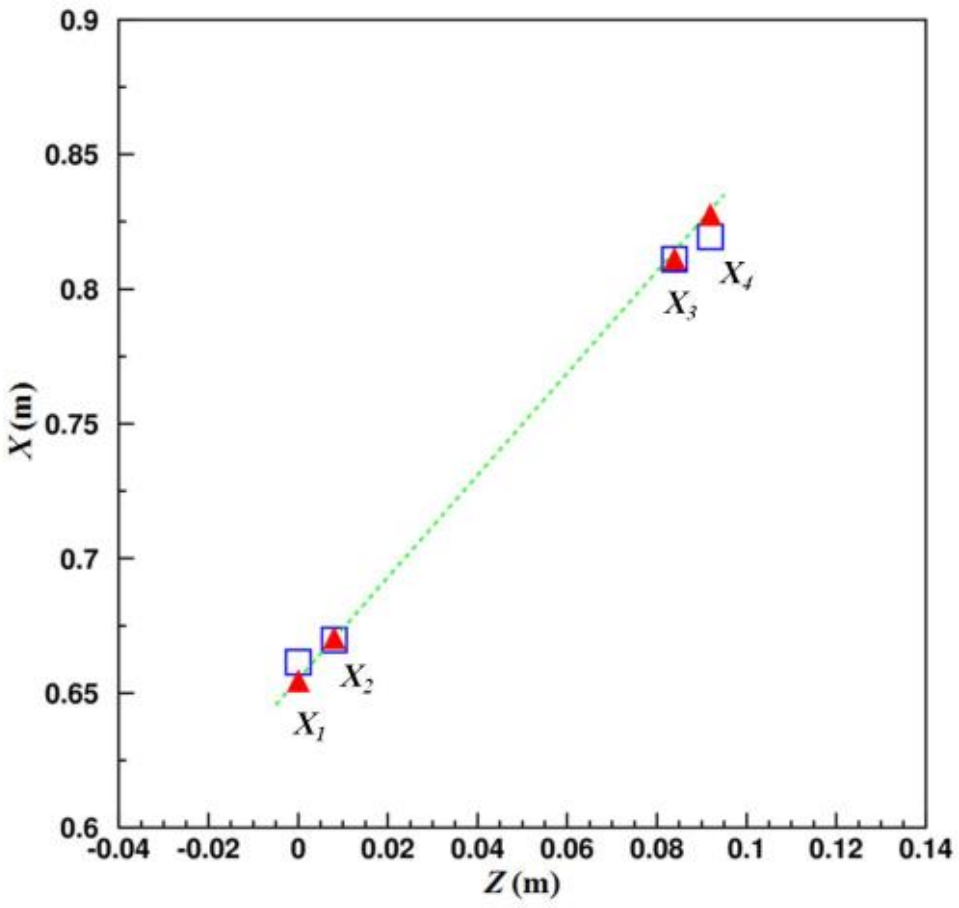

Figure 2.8 Correlation between the centroids of the charge distribution for one event and the $Z_i$ distances between the 4 DCs. $Z = 0$ is assumed for DC1. The blue squares are obtained using the main charge distributions, the red triangles when the cross-talk is subtracted. The green dashed line, drawn to guide the eye, corresponds to the $\theta_f^{simul}$ passing for ($Z_2$, $X_2$) (see text). From [54].

From an inspection of the phenomenon, an overall effect of about 30% of cross-talk between two neighboring DCs (DC1 and DC2, DC3 and DC4) was observed, being negligible for all the other pairs (e.g. DC1 with DC3) [54]. The main distributions for one physical event are shown in Figure 2.9. The charge distributions show a tail on one side and are not symmetric around their center of gravity. Moreover, the integrated charge for DC2 and DC3 is about twice of that observed in DC1 and DC4. The four distributions were scaled for the 0.3 cross-talk factor and superimposed to the main ones of the corresponding cross-talk partner, as shown in Figure 2.9. The factor two difference in the integrated distributions results in a dramatic influence of the cross-talk for the DC1 and DC4 distributions, which is negligible for DC2 and DC3. The DC1 and DC3 distributions are stretched towards larger pad number and the DC2 and DC4 ones in opposite direction. The observed tails in the DC1 and DC4 distributions are well reproduced by the cross-talk generated one, while those of DC2 and DC3 are on both sides, due to the effect of DC1 and PC for DC2 and of DC4 and PC for DC3. However, this effect is very small, rather below the threshold set for the event (green dashed line in Figure 2.9).

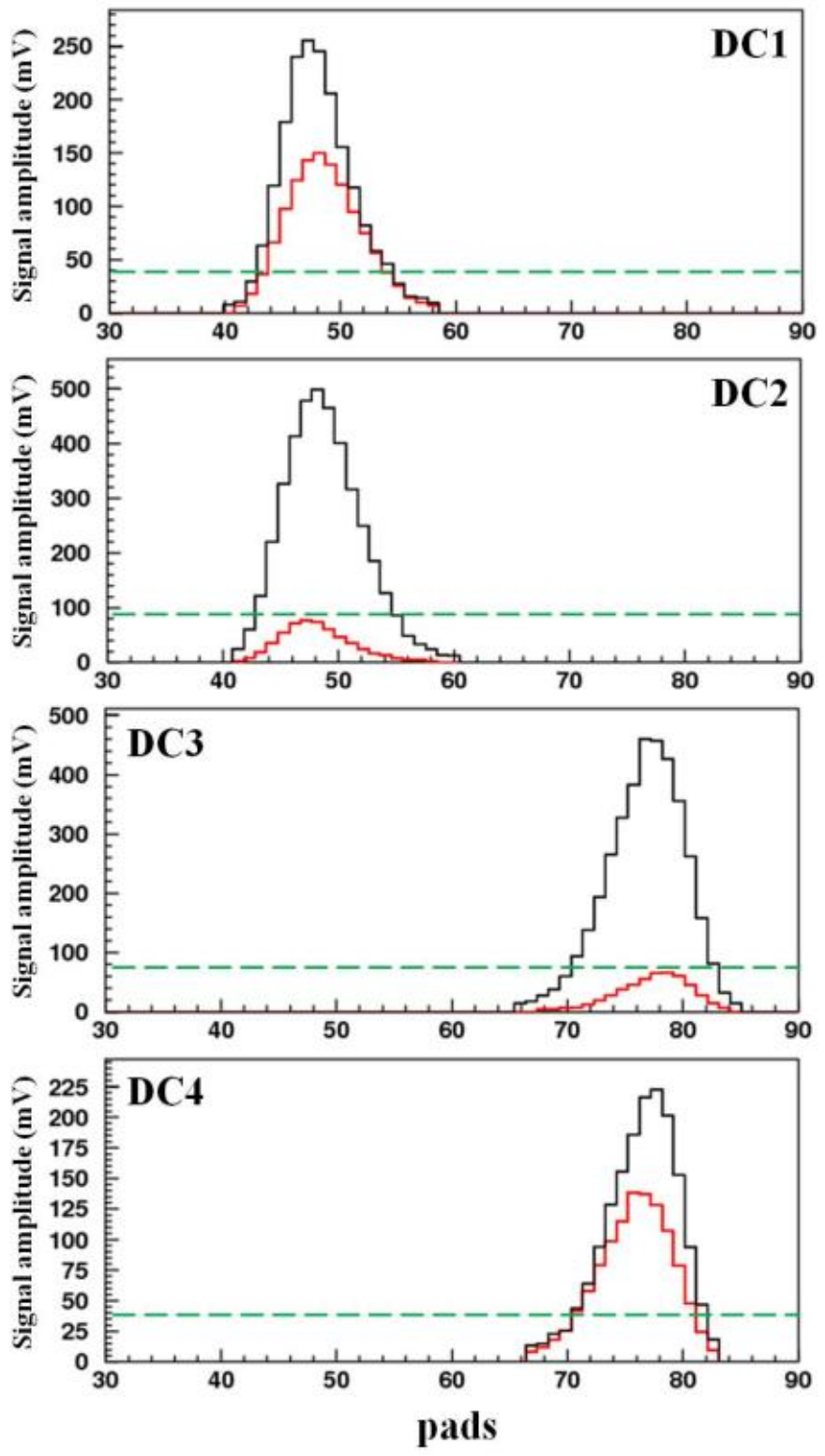

Figure 2.9. Main and cross-talk charge distributions on each DC electrode (black and red histograms respectively) for one event. The threshold set for each DC is shown as a green dashed line. From [54].

Under such conditions, the influence of cross-talk on the $X_2$ and $X_3$ measurements is rather low, whereas in the $X_1$ and $X_4$ cases could be considerably high. As a result, even the computation of the average of the $X_1$ and $X_2$, $X_3$ and $X_4$ quantities does not remove the systematic error in the horizontal position and angle measurements due to these effects. The horizontal angle measurement could result more accurate using only the DC2 and DC3 signals.

From the comparison between experimental data and accurate simulations, developed as described in Sections 2.2 and 2.5, it was found that the most accurate measurement of $\theta_f$ is obtained when only the DC2 and DC3 signals are used ($\theta_{23}$), a worse result when the average of the four signals is taken ($\theta_{1234}$) and the worst using DC1 and DC4 only ($\theta_{14}$). The expected strong effect of the cross talk on DC1 and DC4 signals was thus confirmed. However, a measurable discrepancy $\Delta\theta$ up to about 0.3° between the simulated value and $\theta_{23}$ was still present.

In order to improve the horizontal coordinate and angle measurements, the cross-talk generated distributions was subtracted from the main one for each DC. In the resulting $\theta_f^{corr}$ an accuracy of about 0.1° was achieved, determining an improvement even up to 1.5° compared to the uncorrected $\theta_{1234}$ model and a measurable one with the $\theta_{23}$ result (up to 0.3°). This demonstrates, within the sensitivity of the adopted analysis, the complete removal of the strong cross-talk effects. The effect of the correction is evident in Figure 2.8, where the centroids of the charge distributions are shown with and without the cross-talk subtraction.

## 2.4 Particle identification

A standard way to identify the charged products of a nuclear reaction by a magnetic spectrometer is to measure the energy loss ($\Delta E$) and the residual energy ($E_r$) by the focal plane detector and to additionally determine the Time of Flight (TOF) along the instrument. This requires the generation of a START signal for the TOF, that can be obtained either exploiting the discrete beam time structure or using an appropriate start detector close to the target. However, the use of such a start detector causes some limitations on the use of the spectrometer itself, for example, it makes not possible to measure at very forward angles (including 0°) under beams more intense than $10^5$ - $10^6$ pps. In addition, it introduces a degrading effect on the overall achieved energy and angular resolution, which could overcome the positive effects of the position measurement. Finally, the limited efficiency of the detector, that is particularly important when dealing with low count-rate experiments, and its geometrical encumbrance could limit the use of ancillary detectors. For all of these reasons, for the MAGNEX PID a new technique was developed, which avoid the use of the TOF measurement [11].

In a magnetic spectrometer with dispersive elements the relation

$$\alpha = \frac{\sqrt{m}}{q} = \frac{B\rho}{\sqrt{2T}} \qquad (2.10)$$

in part determines the particles trajectory. In eq. (2.10) $m$ and $q$ are the mass and electric charge of the ion, $T$ is the kinetic energy, while $\rho$ is the radius of curvature of the ion trajectory inside a dispersive element (bending magnet) with field $B$. Such a relation is particularly important whenever $B$ is uniform, since in such cases $\alpha$ depends only on $\rho$ and $T$, which can be accessed experimentally or indirectly estimated. The situation does not change even in presence of real bending magnets [11]. The parameter $\rho$ is not directly measured in the experiments, being replaced for small angular acceptance spectrometers by the measurement of the position $x_f$ of the particles at a plane after the last bending magnet [7]. When high order aberrations become relevant, as for MAGNEX, the correspondence between $\rho$ and $x_f$ is no more unique, even in a rotated focal plane [7]. The measurement of the only $x_f$ parameter is not enough and the complete reconstruction of the ion motion by the trajectory reconstruction technique, described in Section 2.2, is mandatory.

Eq. (2.10) can be transformed as

$$\alpha = \frac{\sqrt{m}}{q} = \frac{B\rho}{\sqrt{2T}} = \frac{p}{q\sqrt{2T}} = \frac{p_0(\delta + 1)}{q\sqrt{2T}} \qquad (2.11)$$

where $p_0$ indicates the reference momentum and $\delta$ the relative deviation of the ion momentum from the reference one. The kinetic energy $T$ in Eq. (2.11) is normally approximated by $T = E_r + E_l + E_d$, i.e. the sum of the residual energy $E_r$, measured at the end of the track, the energy $E_l$ lost by the ions while passing through different active layers of the detector and a remaining part of the energy $E_d$, lost by the ions trough the crossed dead layers. The latter is not measured but can be estimated by calculations.

In order to prove the particle identification capability, an in-beam test was performed. A beam of $^{18}O$, delivered by the INFN-LNS Tandem Van de Graaff accelerator, bombarded a 50 μg/cm$^2$ self-supporting $^{13}C$ target at 84 MeV incident energy. MAGNEX was located at 12° central laboratory angle (actual spanned interval 7° - 19°). The magnetic field was set in order to transport the $^{16}O$ ejectiles, corresponding to the population of $^{15}C$ states in the $^{13}C(^{18}O,^{16}O^{8+})^{15}C$ reaction.

As shown in Figure 2.10 many different ejectiles are produced in the experimental conditions mentioned above. In Figure 2.10a $E_l$ represents the measured energy loss by the ions along the FPD. A poor resolution is noticed, due to the broad distribution of the tracks length of the ions inside the detector, consequence of the wide range of the horizontal angle at the focus (45° < $\theta_f$ < 72°) [7]. The effect of the vertical angle is much smaller since this is always within ± 2° around 0°. A ratio of about 2.5 between the longest and the shortest track is obtained without the measurement of the horizontal angle.

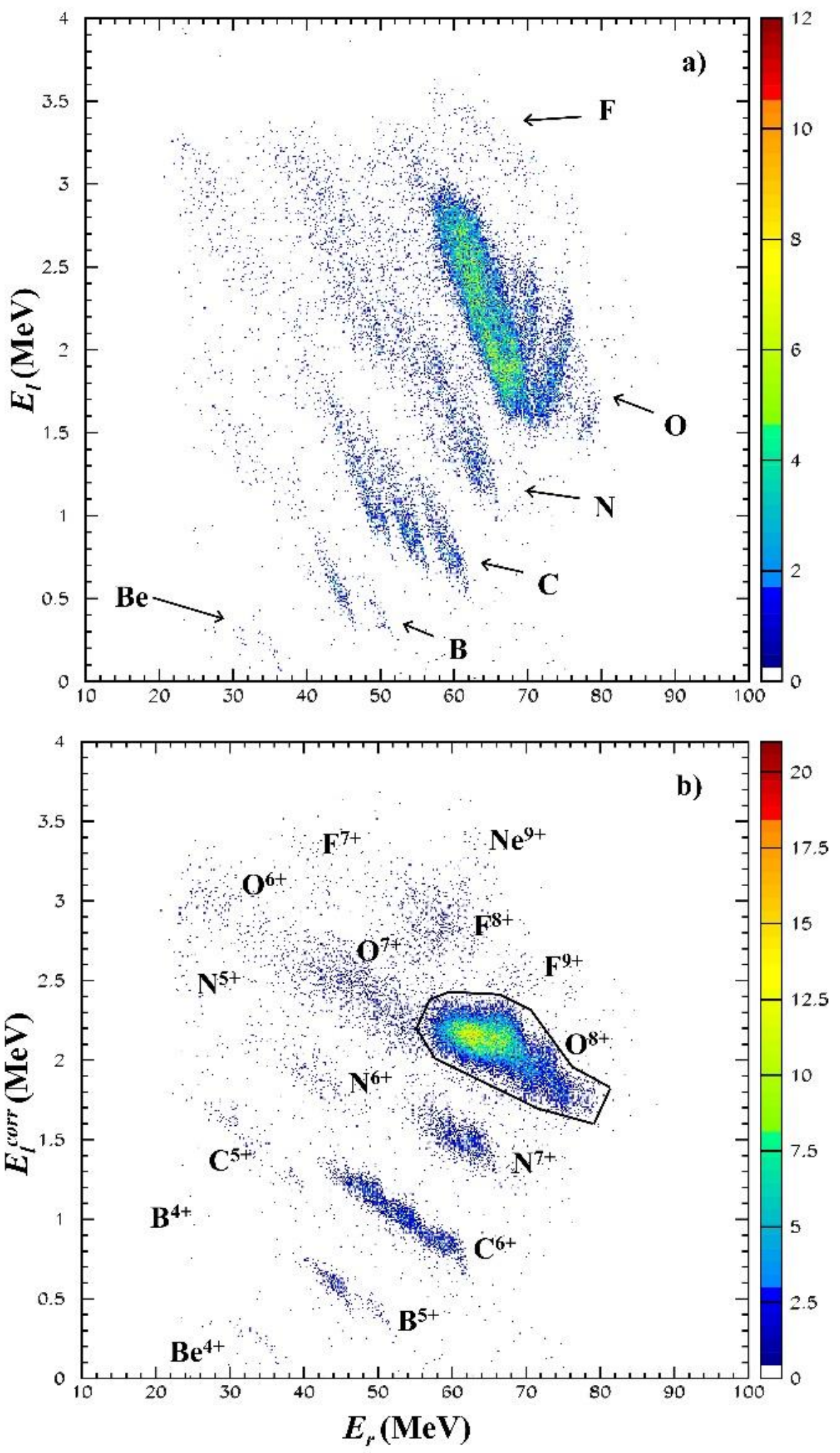

Figure 2.10 Two-dimensional spectra measured in the $^{18}O + ^{13}C$ reaction at 84 MeV and scattering angle of $\theta_{sc} = 7°$-$19°$ for unselected events. Panel a) shows the $E_l$ - $E_r$ spectrum without correction for the trajectory length. Panel b) shows the effects of the correction. The contour line represents the graphical selection on the Oxygen $8^+$ ions. From [11].

The obtained angular resolution of about 0.6° gives a contribution to the accuracy on the energy loss corrected for the track length $E_l^{corr}$ of about 1.3% at 50° incident angle and 2.9% at 70°. In Figure 2.10b $E_l^{corr}$ is plotted versus the $E_r$, showing a drastic improvement in the quality of the data [69]. The observed resolution tends to be worse for heavier ions due to a not linear behavior of the detector for larger signals in this experiment. That is likely due to space charge phenomena around the multiplication wires emphasized by the observed pollution of the isobutane active gas. The resolution in the $E_l^{corr}$ parameter is lower than 4% up to the Nitrogen ions, mainly due to the electronic noise. This gives a contribution of about 0.1% to the overall reconstruction of the kinetic energy and about 2% in the $Z$ parameter ($Z/\Delta Z = 48$), thus allowing the identification up to the cadmium isotopes.

An example of the PID in terms of the $x_f$ - $T$ two-dimensional spectrum gated on the oxygen isotopes in the $E_l^{corr}$ - $E_r$ one, is shown in Figure 2.11a. The isotopes are separated but the loci present a sizeable width. A careful inspection of the $^{17}O$ and $^{16}O$ shows the presence of narrow lines, recognized as known states of the residual $^{14}C$ and $^{15}C$ nuclei. These lines are quite instructive to demonstrate the strong effect of the chromatic aberrations. The different slopes observed for the different states make hard any attempt to upgrade the resolution by simple heuristic transformation of the data.

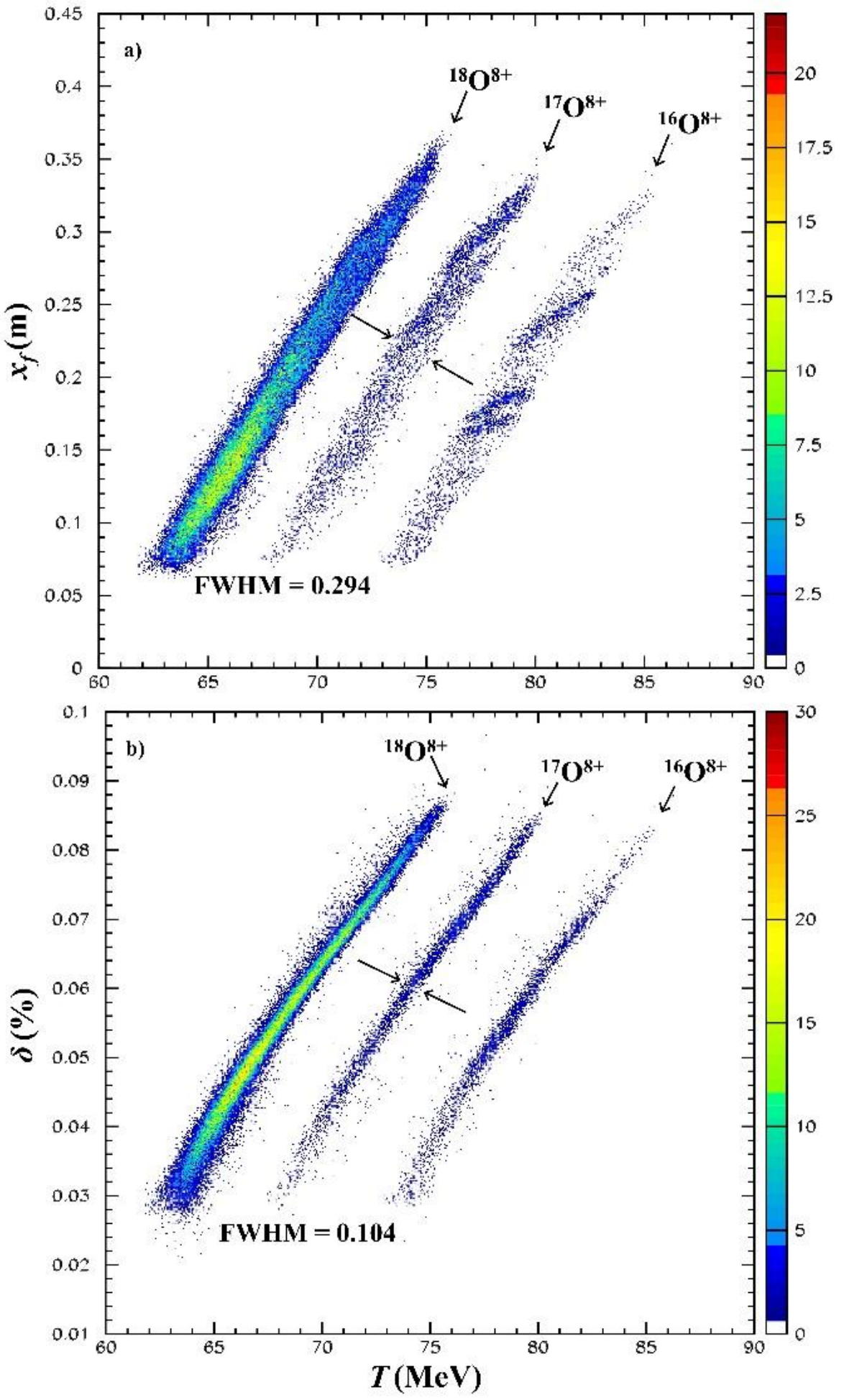

Figure 2.11 Panel a) the $x_f$ - $T$ spectrum for the selected Oxygen $8^+$ ions. Panel b) the $\delta$ - $T$ spectrum for the selected Oxygen $8^+$ ions. The indicated FWHMs refer to the mass spectrum (see text).

In order to estimate the achieved resolution, the selected data were projected into a one-dimensional spectrum of $T^*$ obtained by a transformation which minimizes the width of the projected peaks of the Oxygen loci over the new horizontal axis. A subsequent calibration of the $T^*$ parameter does allow to get an atomic number or mass spectrum in which the FWHM was deduced for the $^{17}O^{8+}$ peak. The value is indicated in Figure 2.11. A corresponding resolution of about 1/72 in the mass of the Oxygen isotopes is measured in this way.

An example of the PID is shown in terms of the $\delta$ - $T$ two-dimensional spectrum gated on the $E_l^{corr}$ - $E_r$ one is shown in Figure 2.11b. The performance of the technique in discriminating the $8^+$ Oxygen isotopes is evident. Similar results are obtained for all the other detected ions. The described $T^*$ transformation procedure was applied again to the spectrum to get an estimate of the achieved resolution. A resolution of 1/160 in the mass of the Oxygen isotopes is measured in this case. The same result is obtained for all the detected ions and was also confirmed in other experiments.

## 2.5 Momentum vector reconstruction

The trajectory reconstruction technique, discussed in the Section 2.2, was applied to the analysis of several experimental data collected using both Tandem Van der Graaff and Cyclotron beams at INFN-LNS. In the following two examples are presented.

In a first experiment a beam of $^{16}O^{7+}$ at 100 MeV total incident energy was focused on a 181 μg/cm$^2$ thick $^{197}$Au target placed in the scattering chamber of MAGNEX, whose optical axis was centered at $\theta_{opt}$ = 18°. A system of collimators was used in order to limit the beam spot size and the angular divergence to 1.2 mm × 0.8 mrad in the horizontal direction and 2.3 mm × 1.8 mrad in the vertical one. A pepper-pot shaped diaphragm, shown in Figure 2.12, was positioned between the target and the spectrometer, thus only ions having a few well-defined trajectories were transmitted. The magnetic field of the quadrupole and dipole elements were set in order to transmit the elastically scattered $^{16}O^{8+}$ ions along the optical axis. The $^{16}O^{8+}$ where identified according to the technique described in Section 2.4 and final phase space parameters at the focal plane were measured both in the horizontal and vertical directions ($x_f$, $\theta_f$, $y_f$, $\phi_f$).

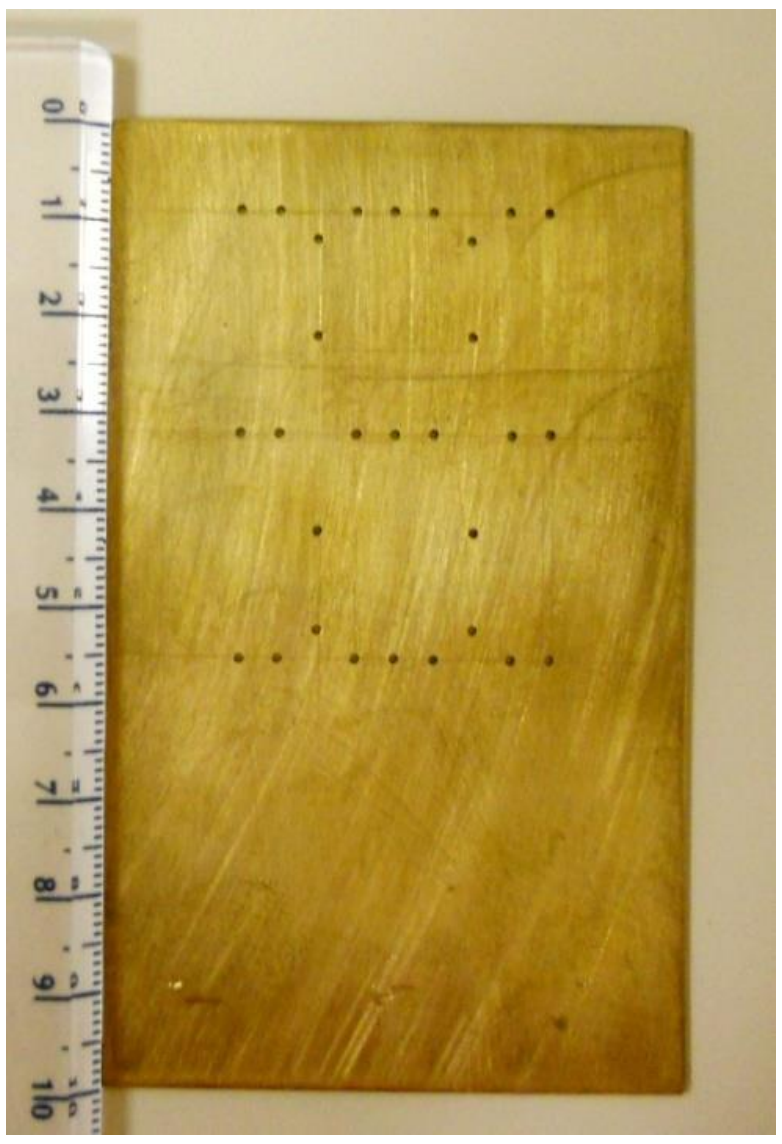

Figure 2.12 Picture of the pepper-pot diaphragm.

In a second experiment a beam of $^{18}O^{6+}$ at 84 MeV bombarding energy collided on a 152 μg/cm$^2$ thick $^{208}$Pb target. MAGNEX was located at $\theta_{opt}$ = 12° and the magnetic fields of the optical elements were tuned, in different runs, in order to transmit the $^{18}O^{8+}$ elastically scattered at different positions along the focal plane.

Two different sets of simulations were developed for representing the above experimental conditions using eq. (2.1). The transport matrices were calculated to the 10$^{th}$ order. The energy loss of the emitted ions across the target and the different dead and active layers of the FPD was calculated by the SRIM program [70]. The geometric structure of the simulated magnetic fields was extracted from the measured ones according to the three-dimensional interpolation algorithm discussed in refs. [59], [71]. In the simulation, the dipole field intensity was that measured by a NMR probe, permanently inserted inside the central region of the magnet. The quadrupole field strength was extracted as the average value of a set of four independent Hall probes mounted 18 cm away from the symmetry axis. As discussed in Section 2.2, the positions of all the elements of the spectrometer (target, magnets, FPD) were accurately measured by optical devices.

An example of the measured bi-dimensional ($x_f$, $\theta_f$) plot is compared to the relative simulation in Figure 2.13a. The effect of the pepper-pot diaphragm is the presence of separated circular loci. All of these events correspond to the elastically scattered ions since this process dominates over other inelastic ones. In a momentum dispersive spectrometer as MAGNEX the horizontal position, measured at the focal plane, depends only on the momentum modulus, if the horizontal focus is achieved and the optical aberrations are negligible. Thus a monochromatic ensemble of particles distributed over the full angular acceptance would define a vertical line in the ($x_f$, $\theta_f$) scatter-plot. Both the experimental and simulated data indicate the not ideal horizontal focus and the strong effect of the aberrations. In fact the observed loci deviate from the expectation in a non-linear way, especially at the borders. Despite the highly non-linear aberrations it is important to notice the rather faithful representation of the experimental data given by the simulations. In particular, the average deviation between the measured position of the spots in the ($x_f$, $\theta_f$) plane and the simulated ones is estimated about -0.5 ± 0.9 mm and 3 ± 5 mrad. This indicates the rather accurate description of the momentum dispersion and horizontal angular magnification of the spectrometer. The different horizontal angles selected by the pepper-pot diaphragm are well separated and from a one-dimensional projection of the data on the $\theta_f$ axis, a FWHM ranging from about 12 mrad at the larger $\theta_f$ to about 15 mrad at the smaller $\theta_f$ is extracted.

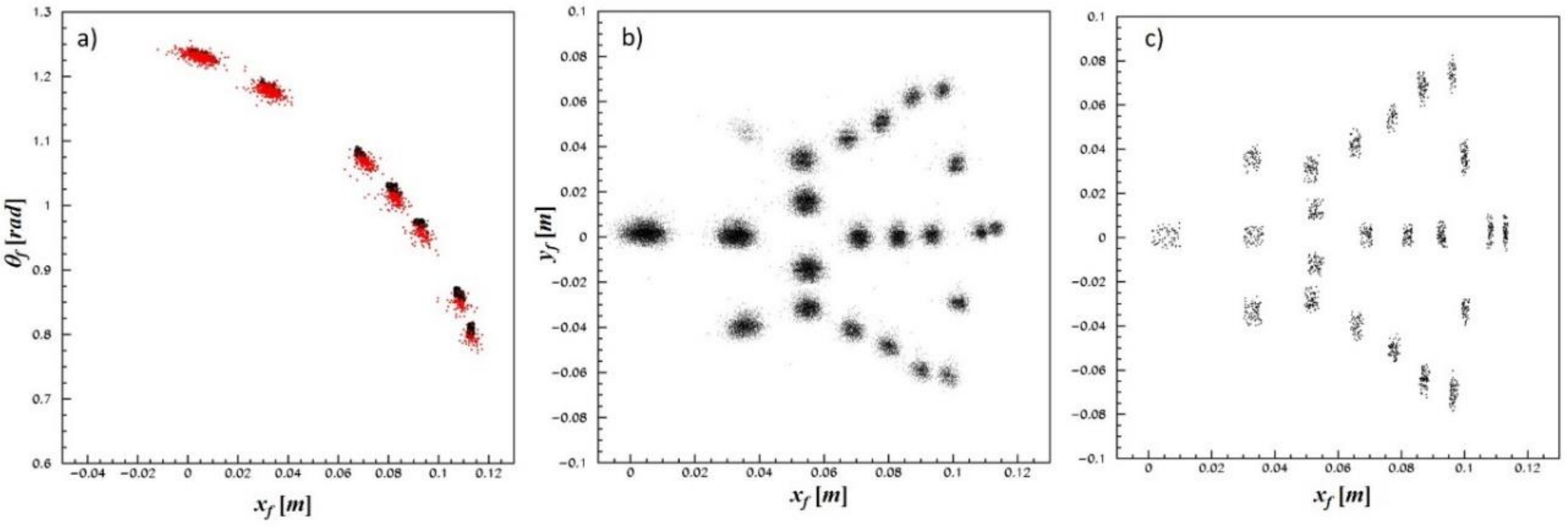

Figure 2.13 a) Comparison of the measured (red dots) and simulated (black dots) two-dimensional plot of the horizontal phase space at the MAGNEX focal plane for the elastic scattering of $^{16}O$ + $^{197}Au$ at 100 MeV and 13° < $\theta_{lab}$ < 24° with the insertion of the pepper-pot diaphragm. Experimental (panel b) and simulated (panel c) ($x_f$, $y_f$) image of the pepper-pot diaphragm at the focal plane.

In Figure 2.13b-c a similar comparison between the experimental data and the simulation is shown for the ($x_f$, $y_f$) phase space scatter plot. In this representation the way the aberrations distribute the data along the $x_f$ is dramatically clear. The loci cannot be represented by any kind of function, since multiple values of $x_f$ are observed for a fixed $y_f$. Consequently, any attempt to linearize the observed data distributions with standard procedures, that could be valid for low acceptance devices [69], becomes very hard. A similar distribution is observed in the ($x_f$, $\phi_f$) phase space. It is also interesting to notice that in Figure 2.13-b the different trajectories selected by the diaphragm define well-separated loci and that 6 of the 29 expected spots are missing. This is a consequence of the not ideal transport efficiency of the spectrometer at the borders of the geometrically accepted solid angle, as will be discussed in Section 2.6.

### 2.5.1 Angle reconstruction

The ($\theta_i$, $\phi_i$) parameters were reconstructed according to the technique described in Section 2.2, using in particular eq. (2.5). An example is shown in Figure 2.14 and compared with that generated by the pepper-pot diaphragm, obtained by simulating a uniformly distributed ensemble of ion trajectories spanning the full solid angle and imposing the geometrical conditions of the pepper-pot hole positions.

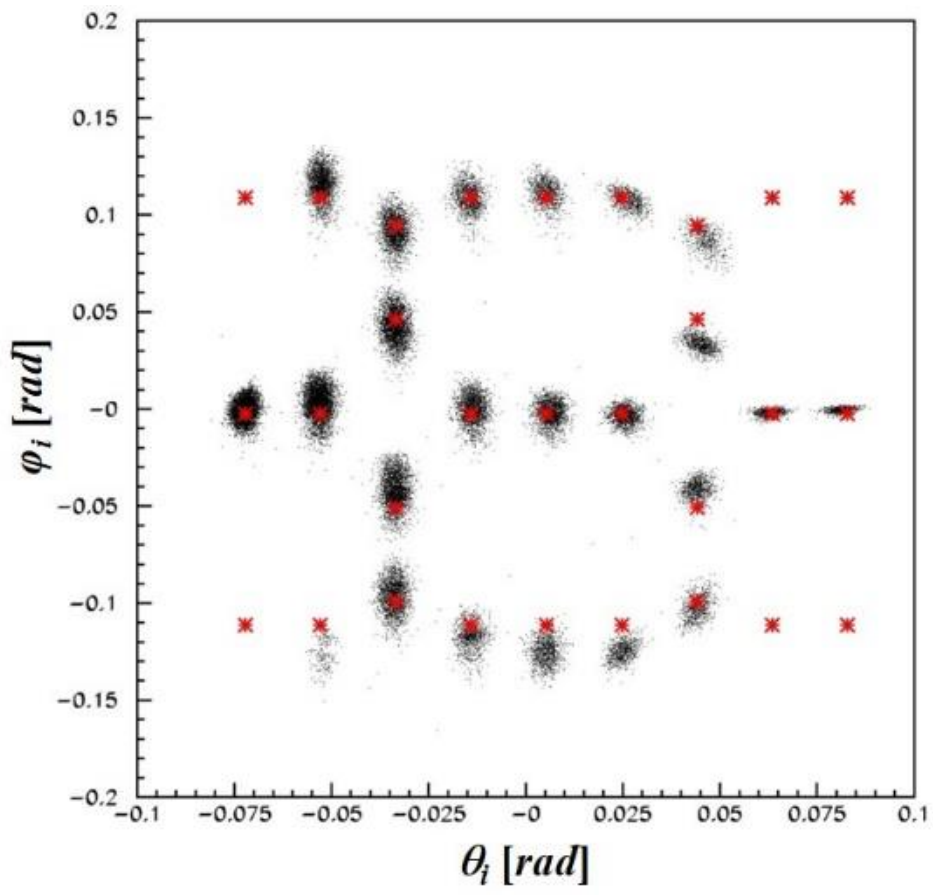

Figure 2.14 Reconstructed bi-dimensional ($\theta_i$, $\phi_i$) angular spectrum for the $^{16}O$ + $^{197}Au$ elastic scattering at 100 MeV, with the insertion of the pepper-pot diaphragm. The red asterisks indicate the expected positions of the loci as deduced by the geometry. From [9].

A clear correspondence between the reconstructed loci and the expected ones is observed. An estimation of the overall geometrical deviation gives an average value of about -0.2 ± 0.7 mrad (-0.01° ± 0.04°) in horizontal and -0.8 ± 5 mrad (-0.05°± 0.30°) in vertical. Thus the present analysis turns out to be an effective tool to evaluate, in a local way, the systematic errors in the determination of the scattering angle and therefore to correct them.

From a projection of the same data over the $\theta_i$ axis, a FWHM ranging between 4.2 mrad and 5.2 mrad (0.24° and 0.30°) is measured for the horizontal angle, including the intrinsic angular opening of about 0.18° defined by the 1 mm diameter holes. An angular resolution between 0.16° and 0.24° is extracted for the $\theta_i$ parameter, which is a quite remarkable result, not achieved even by several low acceptance spectrometers.

A slightly worse result is obtained for the reconstruction of the initial vertical angle $\phi_i$, for which a FWHM ranging from 3 mrad (0.2°) to 20 mrad (1.2°) is measured. In fact the vertical motion is squeezed by the quadrupole lens to an almost point to parallel condition, characterized by a first order angular magnification $R_{44}$ = -0.27 for the central trajectory and globally ranging in the interval [-0.21;-0.29] for the data discussed here. As a consequence, the measured $\phi_f$ are always within the ±2° interval. This emphasizes the effects of the finite resolution in the measurement of $\phi_f$ at the focal plane as one can deduce from the first order formula $\Delta\phi_i \sim \Delta\phi_f / R_{44}$. This behavior is different to that observed in the horizontal motion, where the value of the horizontal angular magnification is larger than 1 and the opposite effect is found.

In the reconstruction of the scattering angle in the laboratory frame ($\theta_{lab}$), both the horizontal and the vertical angles contribute according to basic geometrical relations

$$\theta_{lab} = arccos\frac{\cos(\theta_{opt}) - sin(\theta_{opt})tan(\theta_i)}{\sqrt{1 + tan^2(\theta_i) + tan^2(\phi_i)}} \qquad (2.12)$$

The reconstruction of the vertical angle is of minor importance unless one measures at very forward angles. As an example the overall error induced by an uncertainty of $\Delta\phi_i$ = 1° on the scattering angle is less than $\Delta\theta_{lab}$ = 0.08° at $\theta_{lab}$ = 40° and $\Delta\theta_{lab}$ = 0.8° at $\theta_{lab}$ = 5°.

### 2.5.2 The momentum modulus reconstruction

The momentum modulus $p$, reconstructed to the 10$^{th}$ order using eq. (2.5) is shown in Figure 2.15e correlated with the reconstructed $\theta_{lab}$.

The comparison with the expected behaviour, shown in Figure 2.15e as a red line, demonstrates a striking correspondence with a maximum discrepancy of about 1/2000. A slight deviation is systematically observed at the most forward angles, as result of a less precise reconstruction due to the not ideal focusing obtained there in this experiment (see Figure 2.13-a for the largest $\theta_f$). For the same reason, also the observed widths of the loci become larger at forward angles. At backward angles, a projection of the upper part ($\theta_{lab}$ > 22°) over $p$ axis generates a peaked structure characterized by a FWHM of 1/1850. Slightly worse resolutions are obtained when projecting larger angular ranges.

In order to study the systematic errors in the reconstruction of the momentum modulus the elastic scattering of the $^{18}O$ ions on the $^{208}Pb$ target was explored. Different magnetic settings and a fixed spectrometer position and aperture were set. In such conditions the reference momentum $p_0$ varies and, equivalently, the deviation $\delta$. In this way the linearity of the adopted reconstruction technique was explored in the full accepted phase space.

The variation of $p_0$ was monitored by measuring the magnetic field inside the dipole by a high precision NMR probe ($10^{-5}$ stability), permanently inserted at a fixed position between the magnet poles. The ratio of the dipole fields measured in two different runs is directly connected to the ratio of the reference momenta and determines a precise change $\Delta\delta_B$ in the $\delta$ parameter. Ideally, one should obtain that $\Delta\delta_B = \Delta\delta$, where $\Delta\delta$ represents the experimental value, besides possible discrepancies due to the uncertainties of the adopted technique. In a particular experimental condition both the $^{18}O^{8+}$ and the $^{18}O^{7+}$ were transmitted to the FPD, thus providing a supplementary check. In fact these ions are characterized by the same kinematical behaviour, practically the same mass (neglecting the mass of the electron) and differing only for the charge. As a consequence, the reconstructed $\delta_{8+}$ and $\delta_{7+}$ should differ of the fixed amount $\Delta\delta$ = (8/7 – 1) = 0.143 for these ions.

A bi-dimensional histogram of the reconstructed ($\delta$, $\theta_{lab}$) parameters is shown in Figure 2.16 for the different settings. A highly linear relation is observed, at each $\theta_{lab}$, between the expected and the reconstructed $\Delta\delta$. More precisely, a scaling factor of 1.03 ± 0.04 is extracted, which represents an estimation of the accuracy in the reconstruction of the $\delta$ parameter. The 3% deviation in the reconstructed $\delta$ scale, corresponding to a 3/10000 in the momentum, can be removed by a linear calibration of the data. On the other hand the ±0.04 uncertainty cannot be removed in such a way and would require a non-linear calibration of the momentum scale. A possible way to get this is based on recognized states populated in a specific reaction. Nonetheless, after a sole linear scaling an overall accuracy of about 1/1200 is obtained in the measurement of the momentum modulus.

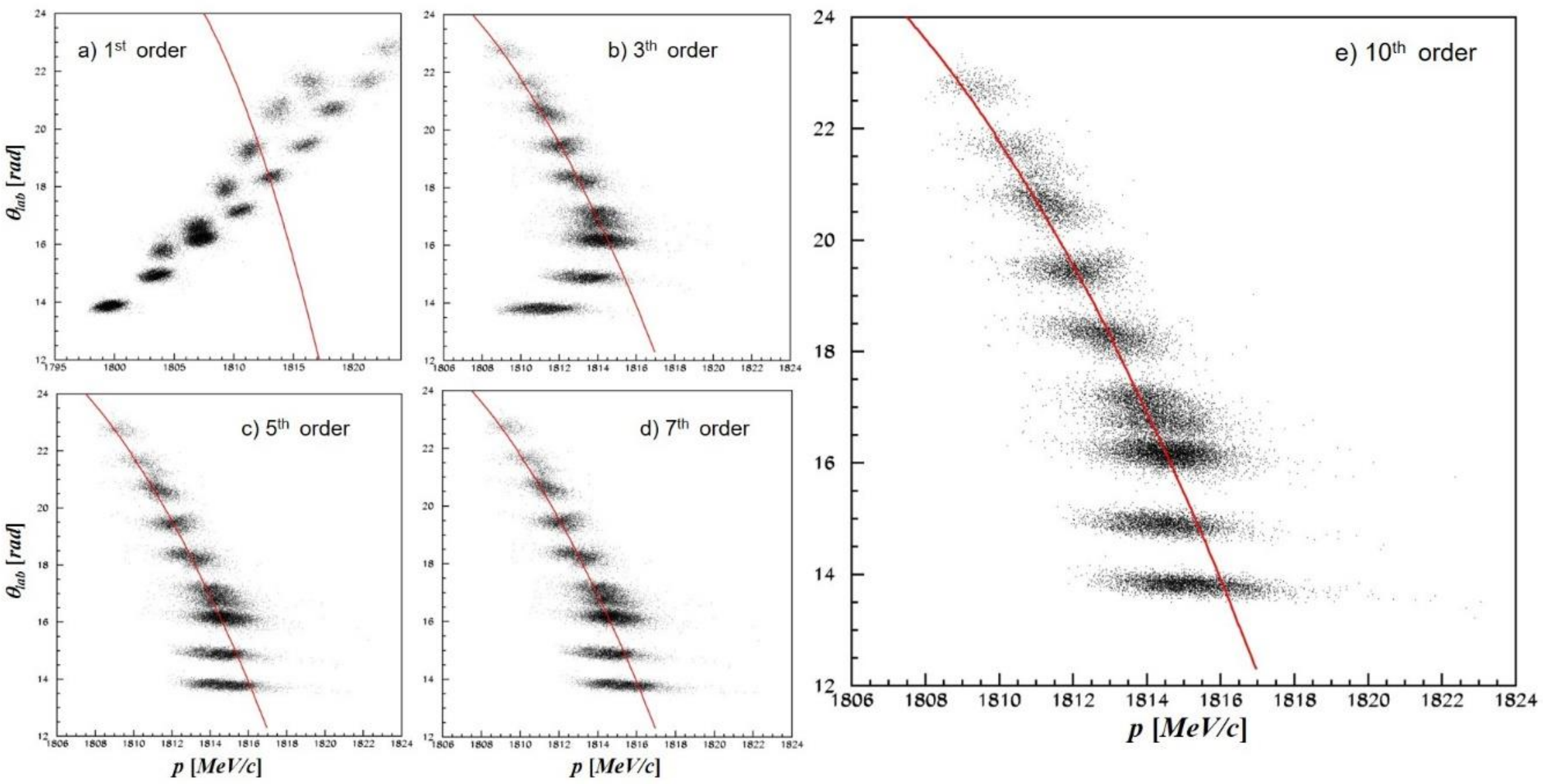

Figure 2.15 Reconstructed ($p$, $\theta_{lab}$) scatter plots in the same experimental condition of Figure 2.13. In the different panels, the order of the reconstruction algorithm was varied. The red line represents the expectation of the kinematics in such conditions. The different loci refer to the different holes in the diaphragm, which select different values of the momentum according to the kinematics.

### 2.5.3 The order of the reconstruction

The effect of the finite order of the algorithm (see eq. (2.6)) on the obtained results was studied for the reconstruction of both the direction and the modulus of the momentum vector. As a general result, the predicted higher sensitivity of the modulus to the non-linear terms was confirmed [56], [71], since the reconstruction of the angles is quite good already at the 1st order.

A specific calculation of the effect of the non-linear terms was done, in order to analyse the reconstruction of the momentum direction. In particular the inverse transport matrices was built removing the first order terms, thus computing the pure contribution of the aberrations to the trajectories shape.

The effect of the aberrations in the reconstruction of the horizontal angle does not exceed 2 mrad for all the cases. Larger values are obtained for the vertical angle, where one can get a contribution of about ± 10 mrad for the extreme rows of the diaphragm. These effects can be almost fully compensated by the inclusion of 3rd order terms in the transport matrices.

For the reconstruction of the momentum modulus, it was not necessary to remove the linear terms by the transport equations. A set of calculations at different orders was performed for the same data discussed above and the results were compared with the expected kinematics. The results are shown in Figure 2.15 and demonstrate that large deviations are obtained even by a 3rd order calculation. The convergence is obtained starting from the 7th order, thus proving the essential role of the reconstruction algorithm for the achievement of the overall performances of the spectrometer.

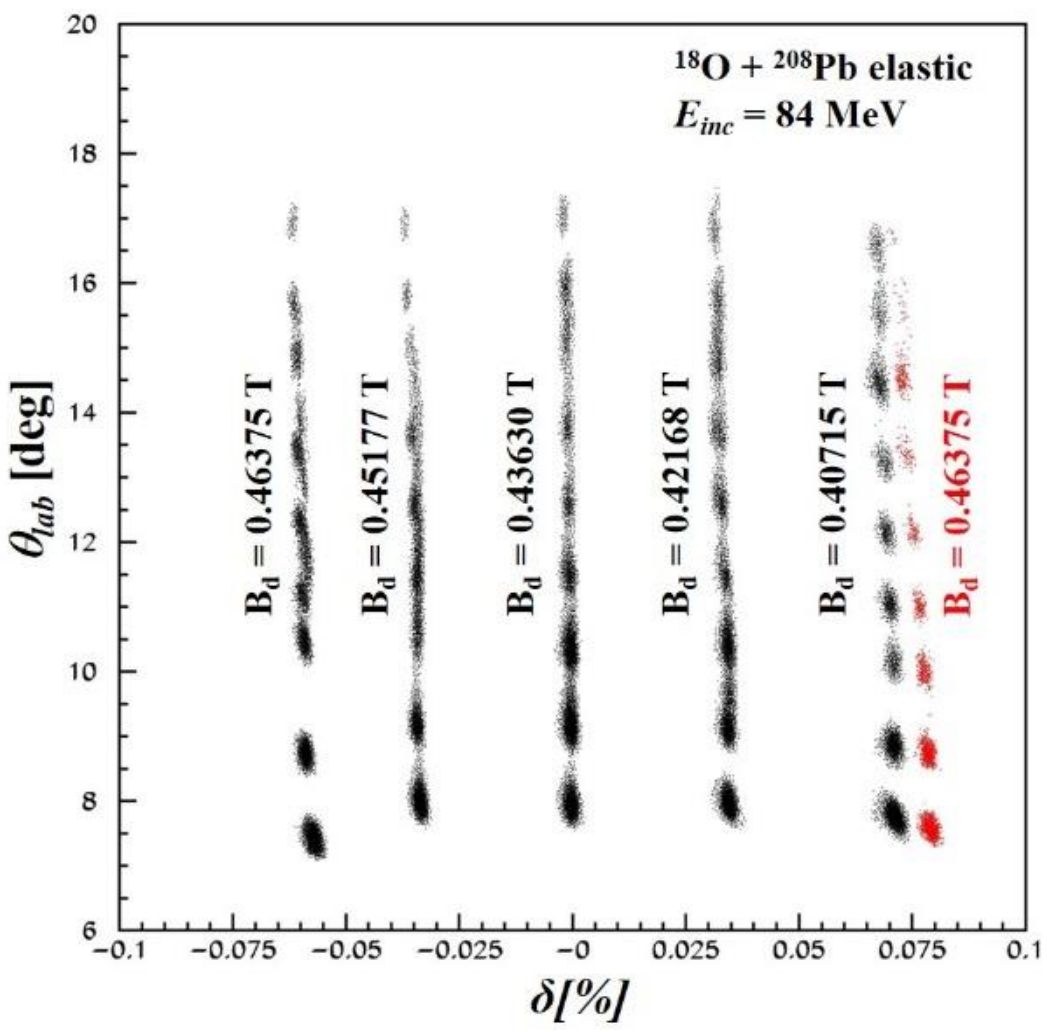

Figure 2.16 ($\delta$, $\theta_{lab}$) scatter plot of the $^{18}O$ ions elastically scattered from a $^{208}Pb$ target at 84 MeV incident energy. The loci showing a vertical correlation are obtained at the indicated value of the field. The black points refer to the $^{18}O^{8+}$ ions, the red to the $^{18}O^{7+}$ ones. From [9].

### 2.5.4 The excitation energy

The corresponding $Q$-values, or equivalently the excitation energy $E_x$, are finally obtained by a missing mass calculation from the reconstructed kinetic energy of the ejectiles. Relativistic kinematics is applied, supposing a binary reaction

$$E_x = Q_0 - Q = Q_0 - K\left(1 + \frac{M_e}{M_r}\right) + E_{beam}\left(1 - \frac{M_b}{M_r}\right) + 2\frac{\sqrt{M_b M_e}}{M_r}\sqrt{E_{beam}K}\cos\theta_{lab} \qquad (2.13)$$

where $M_e$, $M_r$, $M_b$ are the ejectile, residual and beam masses, respectively, $K$ is the kinetic energy of the ejectile and $Q_0$ is the ground state to ground state $Q$-value.

An application of this technique to reconstruct the laboratory scattering angle and the excitation energy of the residual nucleus is shown in Figure 2.17. The data refer to the elastic scattering of a beam of $^{16}O^{7+}$ at 100 MeV incident energy on a 100 μg/cm$^2$ $^{27}Al$ target. The bi-dimensional histogram of the measured $\theta_f$ versus $x_f$ is shown in Figure 2.17-upper panel, in which the well-correlated loci correspond to the population of the ground (at about $x_f = 0$) and of the excited states of $^{27}Al$. The curvature of the loci is due both to the kinematic effect and to the aberrations in the horizontal phase space. It is also distinguishable, with different curvatures, the population of the states of $^{12}C$ and $^{16}O$ due to target impurities.

The reconstructed parameters for the $^{16}O^{8+}$ reaction ejectiles are shown in Figure 2.17-lower panel. The $^{27}Al$ ground and several excited states are well visible as vertical and straight loci, as expected. The tilted loci represent the events relative to the presence of $^{16}O$ and $^{12}C$ contaminants in the target that induce the population of the $^{16}O$ and $^{12}C$ ground and $^{12}C$ excited state at 4.439 MeV. The distinct curvatures are due to the different kinematics of $^{16}O$ scattering on the $^{16}O$ and $^{12}C$ target nuclei. This helps the identification and discrimination of such events, which is one of the qualifying features of the ray-reconstruction technique implemented in MAGNEX.

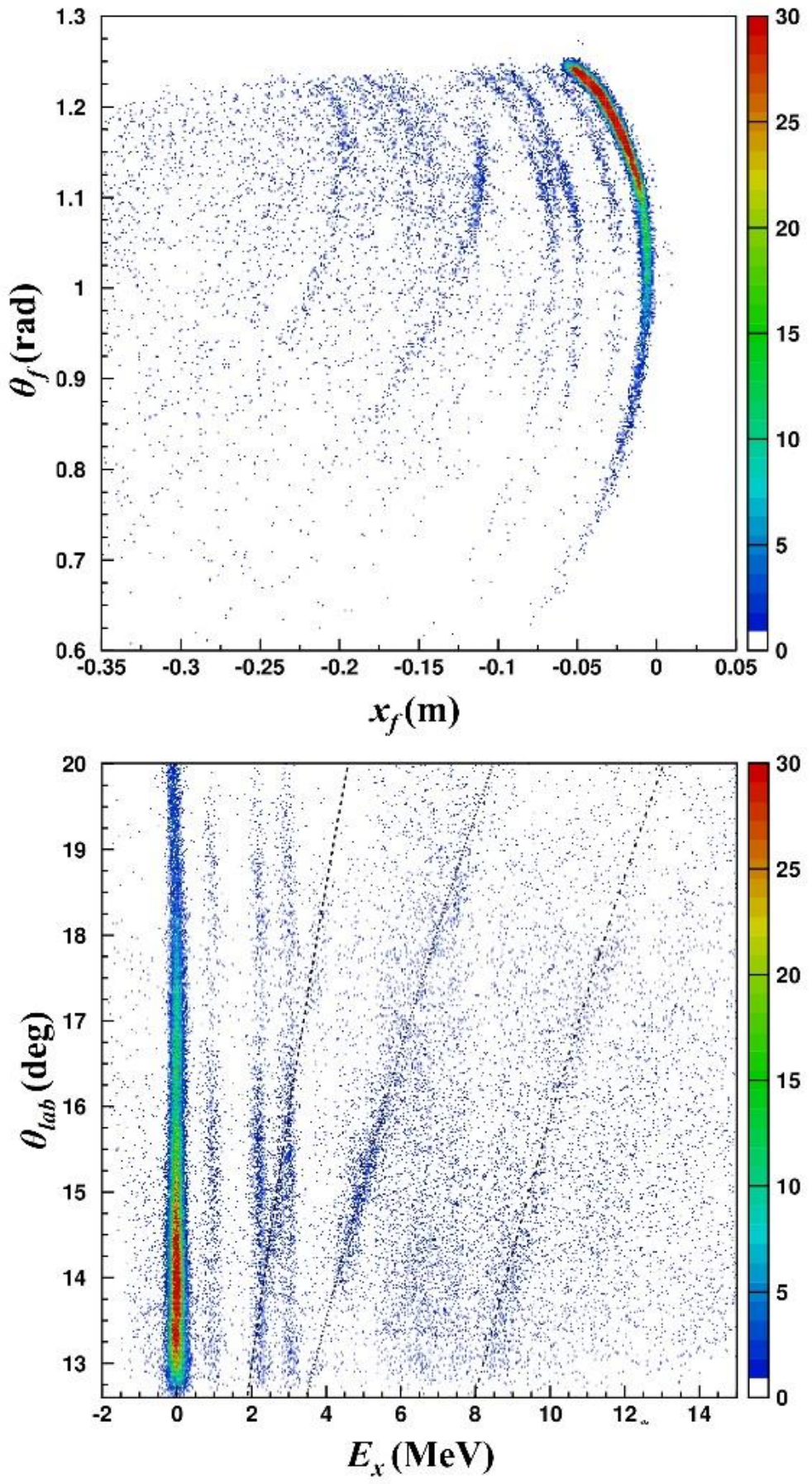

Figure 2.17 The $^{27}Al(^{16}O,^{16}O)^{27}Al$ reaction at 100 MeV and 13° < $\theta_{lab}$ < 21°. Upper panel: plot of the horizontal angle against the horizontal position measured at the focal plane for the $^{16}O^{8+}$ ejectiles. Bottom panel: two-dimensional plot of the reconstructed $\theta_{lab}$ against the $^{27}Al$ excitation energy $E_x$. The black lines represent the calculated kinematics for the population of, from the left, the $^{16}O$ ground state, the $^{12}C$ ground state and the $^{12}C$ excited state at 4.439 MeV.

## 2.6 Measurement of absolute cross sections

When dealing with a large acceptance spectrometer as MAGNEX, a precise estimation of the overall efficiency is mandatory, especially if measurements of absolute cross section are of interest. In particular, the accurate and high resolution measurement of angular distributions requires a precise description of the differential efficiency and solid angle at each angular bin. To achieve this goal a detailed study of the efficiency losses over the accepted phase space was performed [12].

According to the ion optics calculations [7], [56], three points are expected to be sources of efficiency loss:

a) The limited acceptance of the FPD [53];
b) The dipole gate valve, between the dipole and the FPD vessels (see Figure 2.18) ;
c) The vacuum vessels in the region between the quadrupole and the dipole, where the horizontal divergence of the beam envelope is quite large (see Figure 2.18).

In order to determine the efficiency losses in the mentioned critical points, a set of simulations were performed spanning the whole accepted phase space. Practically, the efficiency $\varepsilon = N^*/N$ was deduced as the ratio of the $N^*$ trajectories that reach the active region of the FPD over the $N$ initially generated.

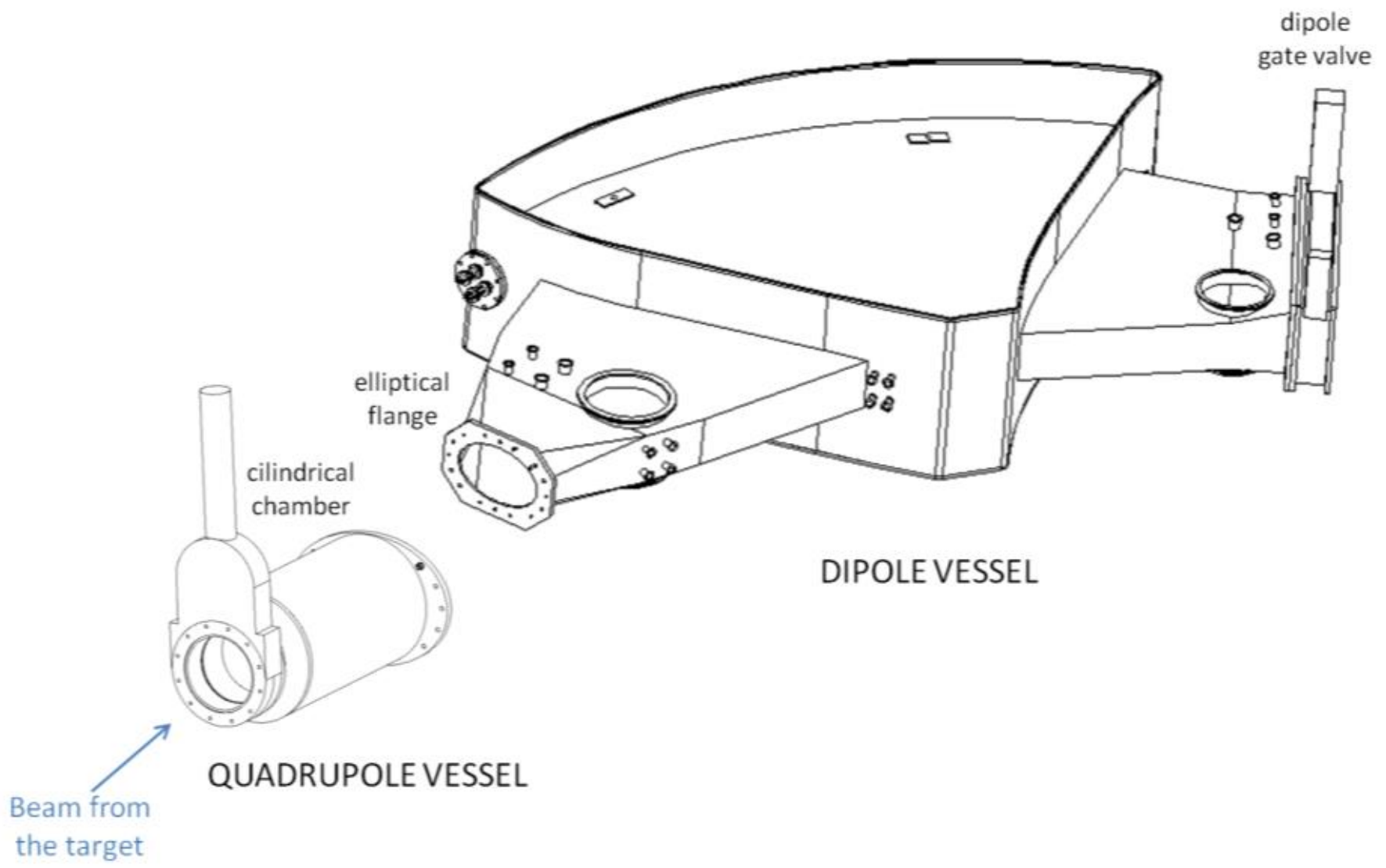

Figure 2.18 Schematic drawing of the quadrupole and dipole vacuum vessel. From [12].

The reliability and the precision of the technique to estimate the differential solid angle of MAGNEX was investigated through the measurement of the absolute cross section of the Rutherford scattering in the full solid angle accepted by the spectrometer.

In the experiment, the elastic scattering of a $^{16}O^{8+}$ beam at 100 MeV incident energy on a $^{197}Au$ target (thickness 181 ± 5 μg/cm$^2$) was explored. The spectrometer axis was positioned at $\theta_{opt}$ = 18° and the aperture diaphragms were set within -90 mrad < $\theta_i$ < 110 mrad and -100 mrad < $\varphi_i$ < 100 mrad in the spectrometer reference frame. This corresponds to accept the ejectile trajectories between 12.8° and 24.9° in the laboratory. A graphite Faraday cup (4 cm deep) having a circular aperture (8 mm diameter) and an electron suppression ring was used to stop the beam. A low noise circuit including a digital integrator was used to determine the collected charge with an intrinsic accuracy better than 0.5%. The overall number of recorded events was corrected for the data acquisition dead time. A factor of 0.79 was measured as the ratio of the number of acquired events over the total event triggers. Finally, the detection efficiency of the FPD was assumed to be 95% uniformly distributed along the phase space [53].

In order to obtain the effective differential solid angle, simulations using the transport technique were performed. A set of trajectories uniformly distributed within the spectrometer acceptance was taken, thus ignoring the effect of the steeply decreasing cross section of the elastic scattering as a function of the laboratory angle. The kinematics of the reaction and the spectrometer tuning parameters were set as the experimental ones. In addition, all the constraints on the FPD acceptance and on the vacuum vessels geometry investigated in ref. [12] were considered.

The resulting ($\theta_i^{lab}$, $\varphi_i^{lab}$) scatter-plot is shown in Figure 2.19(a). The experimental plot obtained by the application of the 10$^{th}$ order transport matrix to the measured data is shown in Figure 2.19(b). The same graphical contour of the transmitted phase space is drawn in the two plots to help the comparison. A remarkable overall agreement is found indicating that the simulation describes also the details of the transmission losses. The small deviations observed in the region at small $\theta_i^{lab}$ and large $\varphi_i^{lab}$ is mainly due to the non-linearity in the ions trajectories passing near the borders of the magnets apertures. The drawn contour was used to deduce the differential solid angle. A step size of $\Delta\theta_{lab}$ = 0.625° was chosen in order to maintain the statistical uncertainties in the number of counts in each bin below 6%. The differential cross section $d\sigma/d\Omega(\theta_{lab})$ was consequently derived at each angular bin.

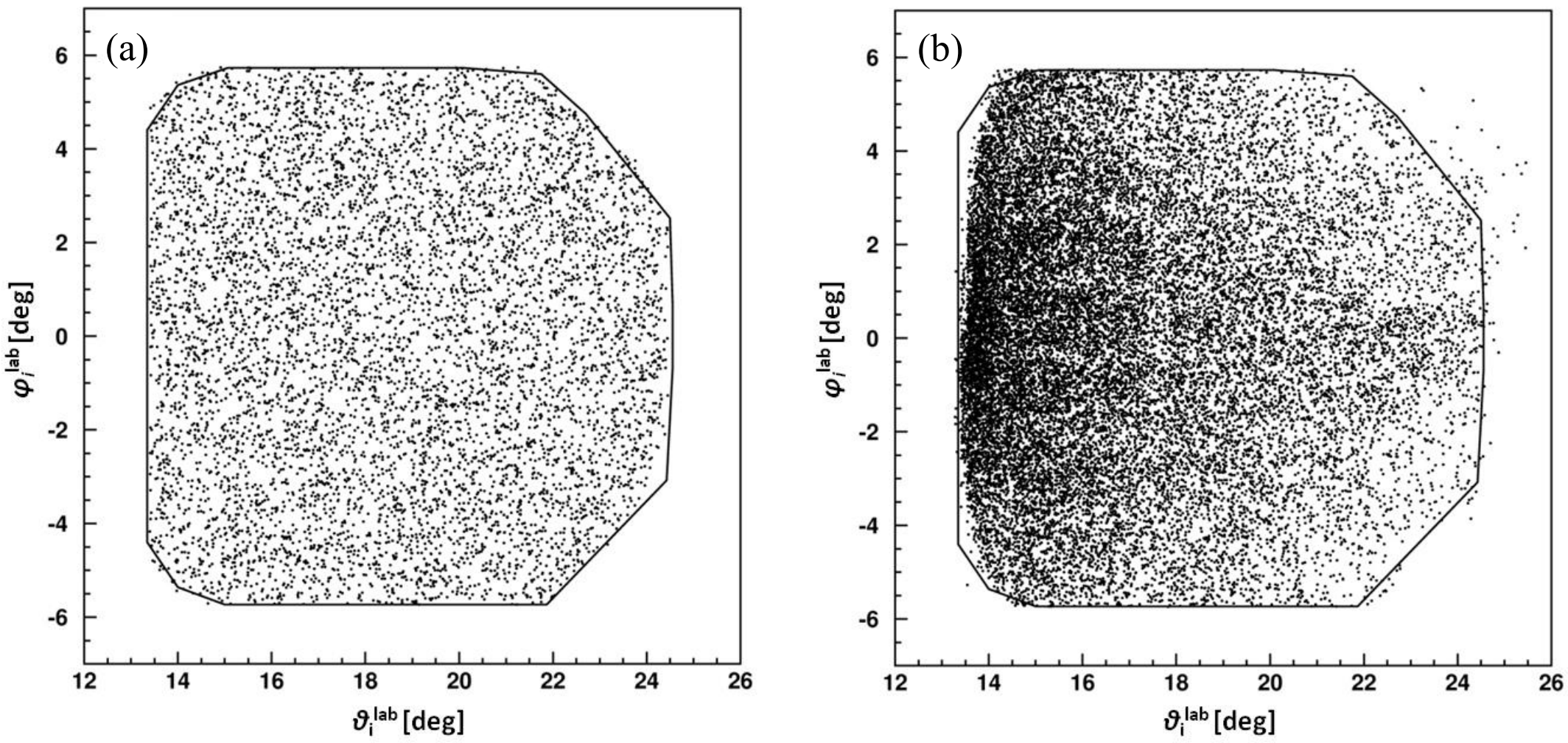

Figure 2.19 (a) ($\theta_i^{lab}$, $\varphi_i^{lab}$) scatter-plot obtained by the simulations for the $^{197}Au(^{16}O,^{16}O)^{197}Au$ reaction under the geometrical constraints described in the text. (b) ($\theta_i^{lab}$, $\varphi_i^{lab}$) scatter-plot obtained by the experimental data using the ray-reconstruction procedure. The same contour line is drawn in the two plots. From [12].

Theoretical calculations in the DWBA formalism were also performed for this reaction using the microscopic Sao Paolo Potential (SPP) [72]. This is known to be rather successful in describing the elastic and inelastic scattering of $^{16}O$ on many heavy targets in a broad range of incident energies. The DWBA calculations indicate that in the explored angular range the Rutherford mechanism is safely dominant, thus simplifying the comparison with the experimental data. This is shown in Figure 2.20, in which the calculated cross section over the Rutherford one ($\sigma_{calc}/\sigma_{Ruth}$) is plotted as a solid blue line.

The experimental $\sigma_{exp}/\sigma_{Ruth}$ corrected for the efficiency losses is superimposed in the plot in Figure 2.20 and more clearly visible in the insert. The error includes both a statistical part (about ± 3.8%) and a component due to the solid angle determination (about ± 2.5% in the central bins). The latter becomes important mainly in the first and last bins where the systematic error on the measurement of the horizontal and vertical angles and the uncertainties on the spectrometer geometry give a major contribution.

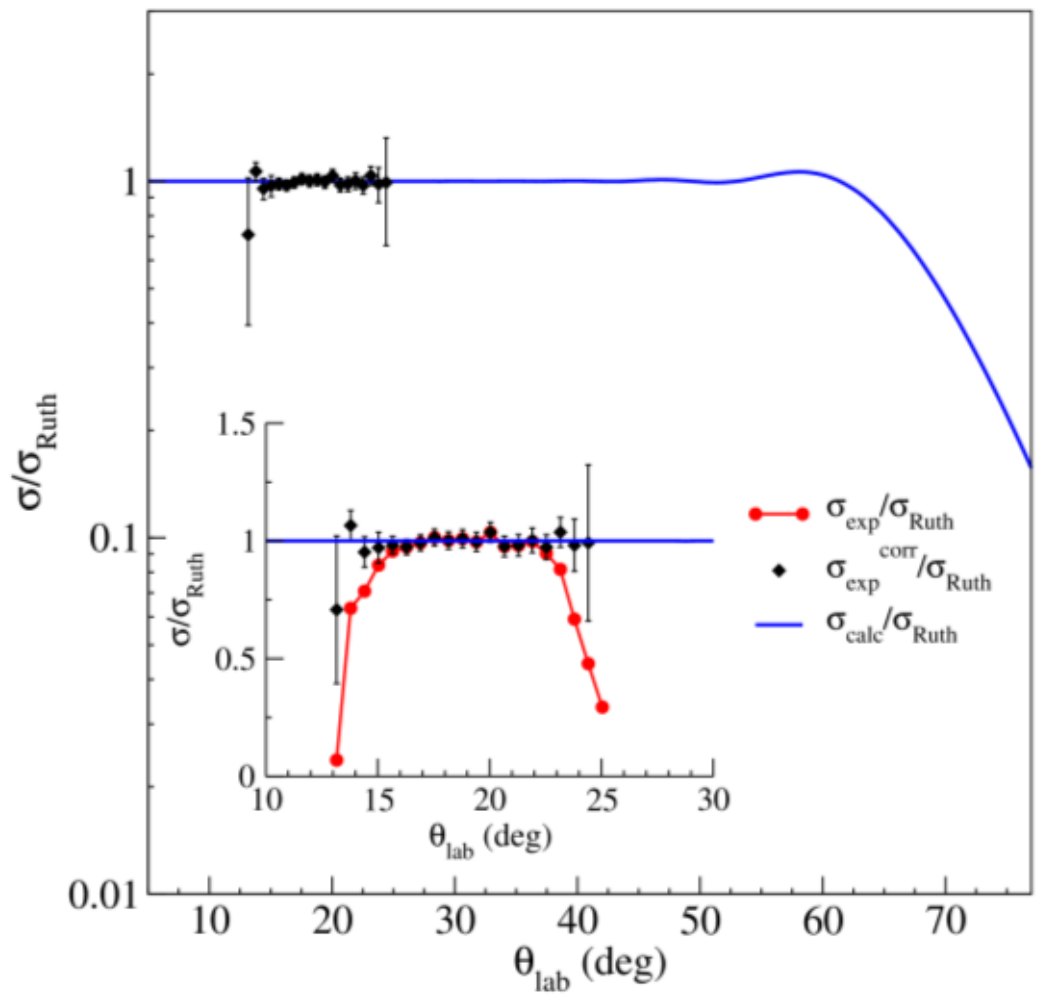

Figure 2.20 Ratio of the cross section over the Rutherford one $\sigma/\sigma_{Ruth}$ obtained for the DWBA calculations described in the text (solid blue line) and for the measured data (black diamonds). In the insert, the ratio of the cross section over the Rutherford one $\sigma/\sigma_{Ruth}$ is shown for the DWBA calculations (solid blue line), for the measured data without the solid angle correction (red circles) and for the measured data with the solid angle correction (black diamonds).

A very good agreement is found all over the angular acceptance. The average value of the measured ratio, weighted according to the error in each experimental bin, results $(\overline{\sigma_{exp}/\sigma_{Ruth}})$ = 0.996 ± 0.015. An average uncertainty of about ± 5% is found in each angular bin. A ± 3% uncertainty in the cross section scale, mainly due to the contribution of the target thickness, is estimated.

A comparison between the measured $\sigma_{exp}/\sigma_{Ruth}$ with and without the correction for the efficiency is shown in the insert of Figure 2.20. The accurate simulation of the spectrometer response and the consequent estimation of the effective differential solid angle allows to sensibly correct the experimental angular distribution at the borders of the accepted phase space. In this way the possibility to use the full acceptance of the spectrometer for accurate measurements of reaction cross sections is demonstrated. The comparison of the measured ratio $\sigma_{exp}/\sigma_{Ruth}$ with the $\sigma_{calc}/\sigma_{Ruth}$ derived by DWBA calculations allows to conclude that the reduction of the transmission efficiency observed at the borders of the accepted solid angle can be properly compensated for. This technique is crucial to allow accurate measurements of cross sections in large acceptance spectrometers as MAGNEX.

## 2.7 Zero-degree measurements

Several scientific cases require the possibility to detect the ejectiles of a nuclear reaction at very forward angles, including zero degree. This opportunity is one of the main advantages of the use of a magnetic spectrometer, which exploits the property of the magnetic field to separate the ejectiles of interest from the beam particles, thanks to the difference in their magnetic rigidity. However, the zero-degree measurements represent a difficult task and no general prescription exists. The beam envelope changes in fact from case to case and the position where the beam is expected to stop downstream of the magnets should be known with sufficient accuracy to avoid hitting the detectors or generating intolerable background. Moreover, the use of start detectors for the time-of-flight measurements makes unfeasible the zero-degree measurement.

In the last few years, the zero-degree mode has been successful applied to MAGNEX in different experiments. In these measurements the optical axis of the spectrometer is usually centred at $\theta_{opt} = 4°$. Thanks to the large angular acceptance of MAGNEX, this setting corresponds to a covered scattering angle range that includes zero-degree, namely $-1° < \theta_{lab} < +10°$. The magnetic fields are set in order to transport the beam along the spectrometer and stop it downstream of the dipole in a "safe" place aside the focal plane detector, where a dedicated Faraday cup is located.

The beam envelope is studied with high accuracy by a $10^{th}$ order transport map, as described in Section 2.2. A set of events, simulating the beam particles, is randomly generated and it is transported through the spectrometer by the application of the direct map. An example of the beam envelope studied to this purpose is shown in Figure 2.21.

Since the beam spot at the focal plane is ~ 30x30 mm$^2$, a large Faraday cup 80 mm wide, 55 mm high, 100 mm deep was built (see Figure 2.21). The use of a Faraday cup also allows the measurement of the beam charge and consequently of the absolute cross-section for the zero-degree measurement.

Many experiments were performed in the last years in the zero-degree mode, using different beams accelerated by both the Tandem and the Superconducting Cyclotron (CS). They are listed in Table 2.4. Examples of the extracted data are shown and discussed in Section 3.6.

Table 2.4. List of experiments performed with MAGNEX at zero degree.

| Beam | Energy (MeV) | Accelerator | Reaction |
|---|---|---|---|
| $^{18}$O | 270 | CS | $^{40}$Ca($^{18}$O,$^{18}$Ne)$^{40}$Ar |
| $^{18}$O | 270 | CS | $^{40}$Ca($^{18}$O,$^{20}$Ne)$^{38}$Ar |
| $^{18}$O | 270 | CS | $^{40}$Ca($^{18}$O,$^{16}$O)$^{42}$Ca |
| $^{18}$O | 270 | CS | $^{11}$B($^{18}$O, $^{18}$Ne)$^{11}$Li |
| $^{18}$O | 270, 450 | CS | $^{116}$Sn($^{18}$O, $^{18}$Ne)$^{116}$Cd |
| $^{18}$O | 270 | CS | $^{116}$Sn($^{18}$O, $^{18}$F)$^{116}$In |
| $^{18}$O | 270 | CS | $^{1}$H($^{18}$O, $^{18}$F)n |
| $^{18}$O | 450 | CS | $^{14}$N($^{18}$O, $^{18}$F)$^{14}$C |
| $^{6}$Li | 16,20,25,29 | Tandem | $^{1}$H($^{6}$Li, $^{4}$He) |
| $^{7}$Li | 38 | Tandem | $^{1}$H($^{7}$Li, $^{4}$He) |
| $^{6}$Li | 25 | Tandem | $^{12}$C($^{6}$Li, $^{2}$H)$^{16}$O |
| $^{16}$O | 512 | CS | $^{9}$Be($^{16}$O, $^{15}$O) |
| $^{4}$He | 53 | CS | $^{4}$He($^{4}$He,$^{4}$He)$^{4}$He$^{*}$ |

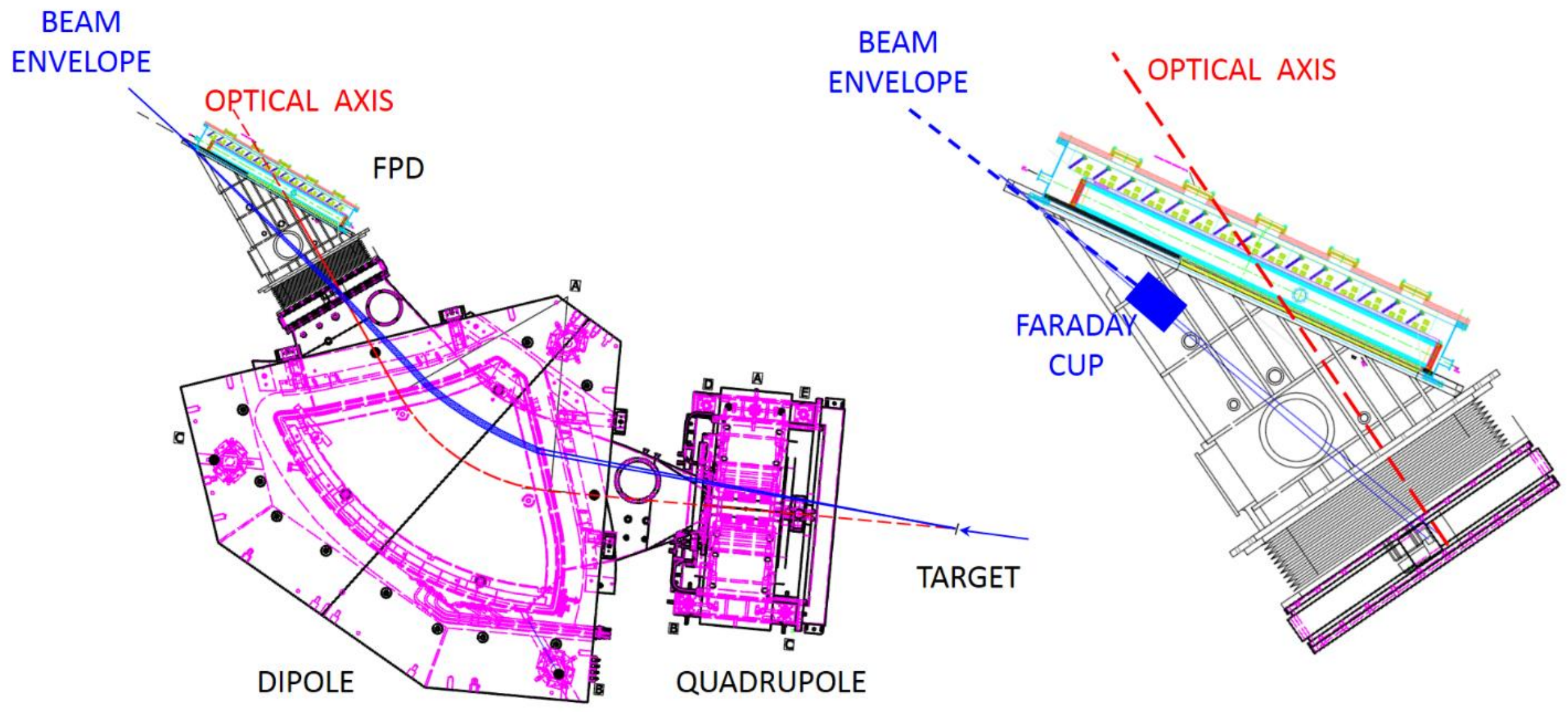

Figure 2.21(left) Plane view of the MAGNEX magnetic layout. The blue band corresponds to a typical beam envelope when the spectrometer is set at $\theta_{opt} = 4°$. The red line corresponds to the optical axis. A zoomed view of the region just upstream of the focal plane is shown in the right, where the beam envelope is drawn in blue. The Faraday cup, which intercepts the beam, is drawn as the blue rectangle. The envelope downstream of the Faraday cup is drawn just to guide the eye.

## 2.8 The MAGNEX-EDEN coupling

The investigation of the decay modes of nuclear states populated in direct reactions is a powerful tool for understanding their microscopic structure. When the reaction populates unbound or weakly bound neutron-rich systems, the neutron emission is the dominant decay mode of the excited states and sometimes of the ground state itself. In these cases the coincidence detection of the emitted neutrons and charged ejectiles and the high-resolution measurement of the neutron energy are crucial tasks for spectroscopic investigations of the residual nuclei.

Recently the neutron decay of the $^{15}C$ resonances up to 16 MeV excitation energy, populated via the $^{13}C(^{18}O,^{16}O\ n)$ reaction at 84 MeV incident energy was studied for the first time using a new method to determine the neutron kinetic energy by time-of-flight (TOF) in exclusive experiments [73]. It involves the use of MAGNEX coupled to the EDEN neutron detector array. A schematic view of the set-up is shown in Figure 2.22.

EDEN is an array of 36 cylindrical organic scintillators (NE213) by IPN-Orsay located around the MAGNEX scattering chamber. Usually they are positioned at a distance ranging from 1.8 to 2.4 m from the target, covering a total solid angle of 270 msr. A detailed description of the detectors is given in Ref. [74]. The neutron-gamma discrimination is provided by pulse-shape analysis of the fast and slow components of the scintillation signal, as described in ref. [75]. A timing signal is also an output of the module for each EDEN channel. A typical fast-slow distribution measured by EDEN is shown in Figure 2.23a.

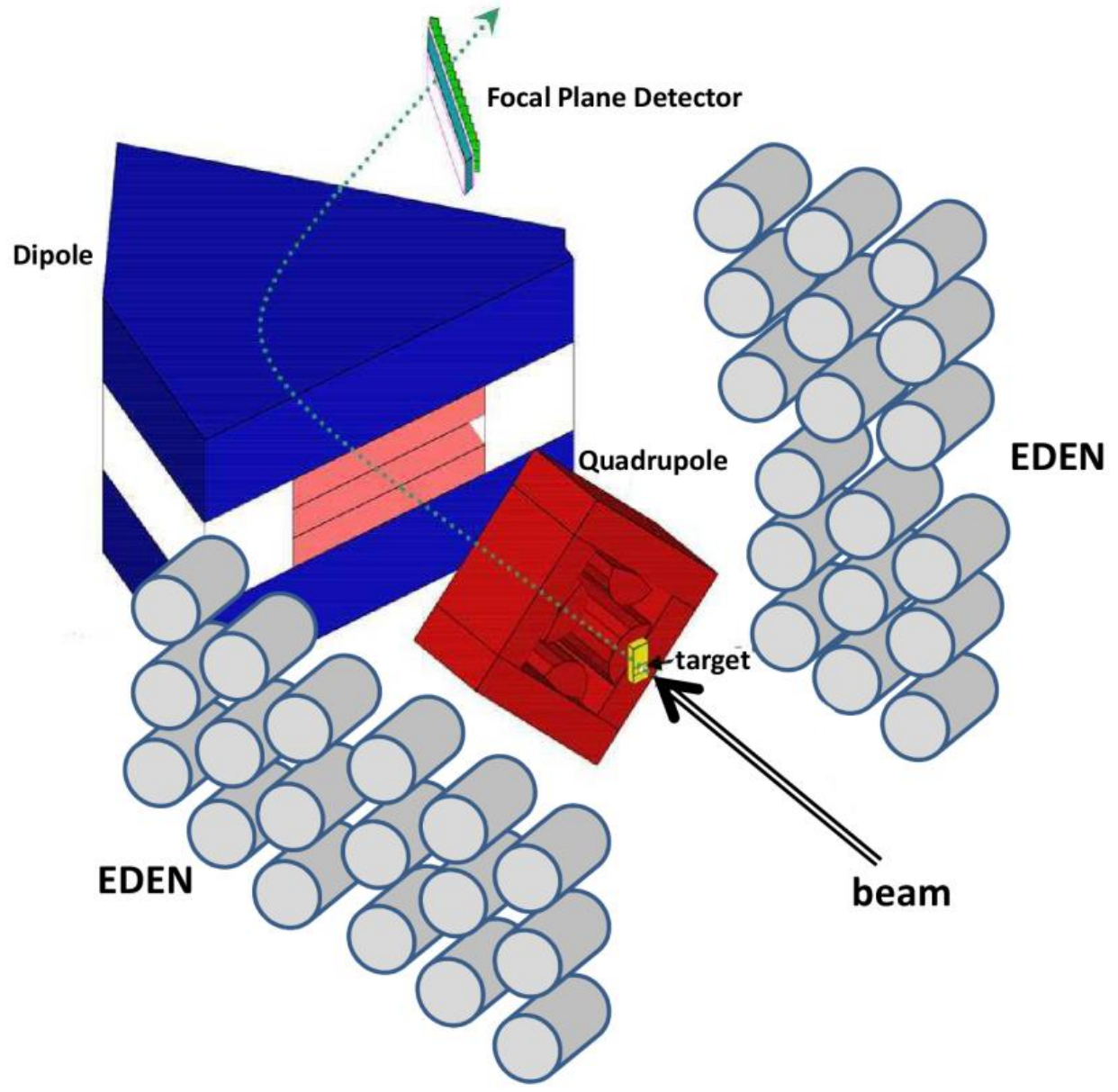

Figure 2.22 Schematic layout of the MAGNEX-EDEN facility.

In the experiment, the START signal is given by the detection of the ejectiles at the focal plane of MAGNEX. The STOP is provided by the EDEN time signal, delayed by a known quantity. The novelty is that the ion path length along the spectrometer is extracted, event by event, by solving the equation of motion of the ions detected at the focal plane and reconstructing the complete ion trajectory, according to eq. (2.4).

In order to measure the neutron TOF and hence the energy, the EDEN timing signal is sent to a high-stability delay line which introduces a delay of $\Delta T_{delay}$ = 400 ns and then to the STOP input of a Time to Digital Converter (TDC). The common START to the TDC is given by the logic OR of the timing signals of the MAGNEX silicon detectors. Thus, once the time-of-flight of the charged ejectiles along the spectrometer is known ($TOF_{ion}$) by ray-reconstruction, the time-of-flight of the neutrons from the target point to the EDEN detector ($TOF_{EDEN}$) can be deduced by the relation:

$$TOF_{EDEN} = TOF_{ion} + T_{TDC} - \Delta T_{delay} \tag{2.14}$$

A typical $TOF_{EDEN}$ spectrum for the $^{13}C(^{18}O,^{16}O\ n)$ reaction at 84 MeV with conditions on the identification of the $^{16}O$ ejectiles in MAGNEX is shown in Figure 2.23(b) . The γ-ray peak at about 5 ns < $TOF_{EDEN}$ < 9 ns and a bump due to the presence of correlated neutrons are visible at $TOF_{EDEN}$ ~ 90 ns above a flat uncorrelated background. The time resolution achieved with the reconstructed variables is 2.4 ns FWHM. This value is compatible with the intrinsic resolution of an EDEN scintillator coupled to a MAGNEX silicon detector, thus making the aberration compensation done by trajectory reconstruction quite satisfactory.

Then, the knowledge of the distances and angles of the EDEN array with respect to the target allows to determine the kinetic energy of the neutrons by relativistic relations and energy $E_n$ in the reference frame of the decaying nucleus. Examples of typical neutron energy spectra determined as described here are shown and discussed in Section 3.4.2.

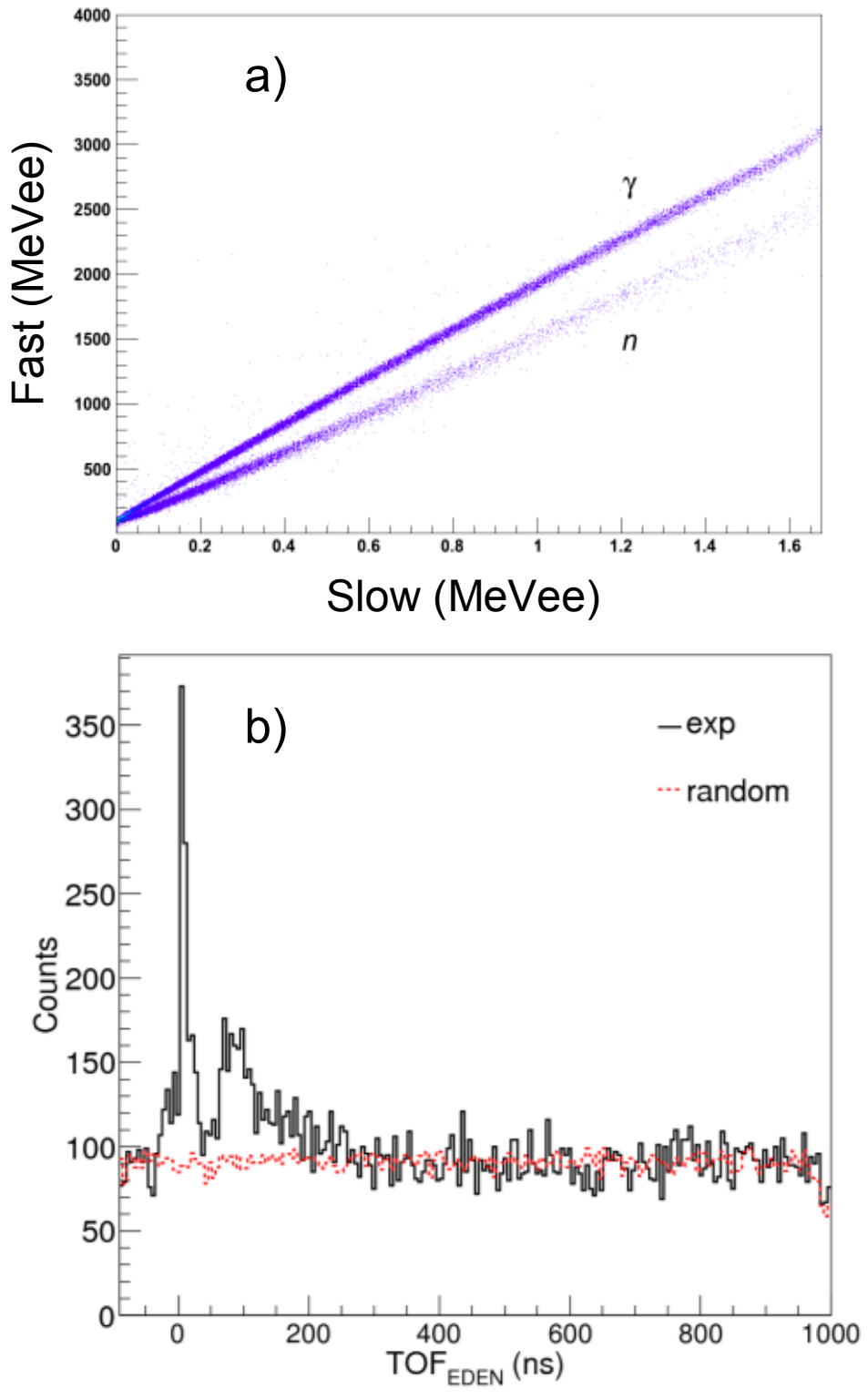

Figure 2.23 (upper) Example of fast versus slow distribution relative to a single EDEN detector. (lower) Neutron time-of-flight $TOF_{EDEN}$ gated on the $^{16}O$ ejectiles detected by MAGNEX (black spectrum) superimposed to a randomly distributed background (red dashed spectrum).

# 3 Scientific achievements

The appealing characteristics of MAGNEX, discussed in Section 2, has made it a unique tool for frontier research in the field of nuclear structure and direct reaction mechanisms. Several proposals have been submitted for experimental campaigns requiring the use of this facility. These were often based on ideas from foreigner scientists. As a result, many experiments were and will be performed. In the following Sections, a selection of some of the most relevant achieved results is presented.

## 3.1 Channel coupling in elastic scattering

Heavy-ion nuclear collisions are powerful tools to access effective nuclear potentials and coupling schemes. There, the structure of the colliding particles plays a major role on the scattering observables. Specific information on nuclear structure can be extracted by studying direct reaction channels, due to their selectivity. At energies around the Coulomb barrier, few channels are accessible and the coupled channel (CC) formalism is useful to take them into account.

### 3.1.1 Search for nuclear rainbow in heavy-ion elastic scattering

Above the Coulomb barrier, the elastic cross section can exhibit effects of important channels coupled to it with relevant deviations from simple optical model calculations. A striking feature is that state of the art CC calculations predict new kind of rainbow-like structures [76]. Experimentally these effects are stronger at large scattering angles, where elastic cross sections are sensibly smaller than inelastic and the channel coupling mechanism can be a dominant source of the elastic flux. However, in this angular region the elastic cross sections become extremely small ($d\sigma/d\omega$ ~ 100 nb/sr and less) and experiments are consequently quite challenging. Moreover, supplementary difficulties arise for heavy systems scattering, resulting in stringent requirements as listed below:

- particle identification for unambiguous selection of the scattering channels;
- high energy resolution, for separation of the elastic channel from the inelastic. This has a direct influence on the choice of the target thickness, due to the emphasized role of the energy straggling in heavy systems;
- the discrimination between the scattering events of interest and those from the target contaminants;
- high angular resolution, to characterize the measured cross section angular distributions;
- high angular accuracy, necessary for heavier systems, since the exact location of the minima in the angular distributions is fundamental to test different theoretical approaches against the experimental data;
- a large angular acceptance, needed to allow for statistically significant data especially at large angles.

All the above listed experimental requirements are fulfilled by MAGNEX, which has recently proven to be well suitable for these challenging experiments.

In particular, based on the proposal of D. Pereira from the University of São Paulo, three experimental campaigns have been set-up to study $^{16}$O + $^{27}$Al scattering at 100 MeV and 280 MeV and $^{16}$O + $^{60}$Ni at 280 MeV.

In a first experiment the $^{16}O^{8+}$ beam, delivered by the Tandem Van der Graaff accelerator at 100 MeV total incident energy, was focused on a $^{27}$Al self-supporting target [77]. The elastically and inelastically scattered $^{16}$O ions were momentum analyzed by the spectrometer and detected by the focal plane detector. Two different aluminum target foils were mounted for the runs at different scattering angles, a 100 μg/cm$^2$ target for the runs at forward angles (13° < $\theta_{lab}$ < 36°) and a 137 μg/cm$^2$ one for the runs at backward angles (31° < $\theta_{lab}$ < 52°).

Examples of energy spectra are shown in Figure 3.1 at 18.0° < $\theta_{lab}$ < 19.0° and 42.7° < $\theta_{lab}$ < 46.4°. The average angular and energy resolutions are 0.3° and 250 keV, respectively. The differential cross section in the forward angle range is about 37 mb/sr, while it decreases to about 780 nb/sr in average in the shown backward angles range. Smaller values down to few hundreds of nb/sr are measured at further backward angles. Despite the many orders of magnitude differences in the elastic cross sections, the obtained spectra present essentially the same features with similar resolution and acceptable signal to noise ratio. In particular, the first inelastic peaks (0.844 + 1.015 MeV) are still well distinguishable from the elastic one.

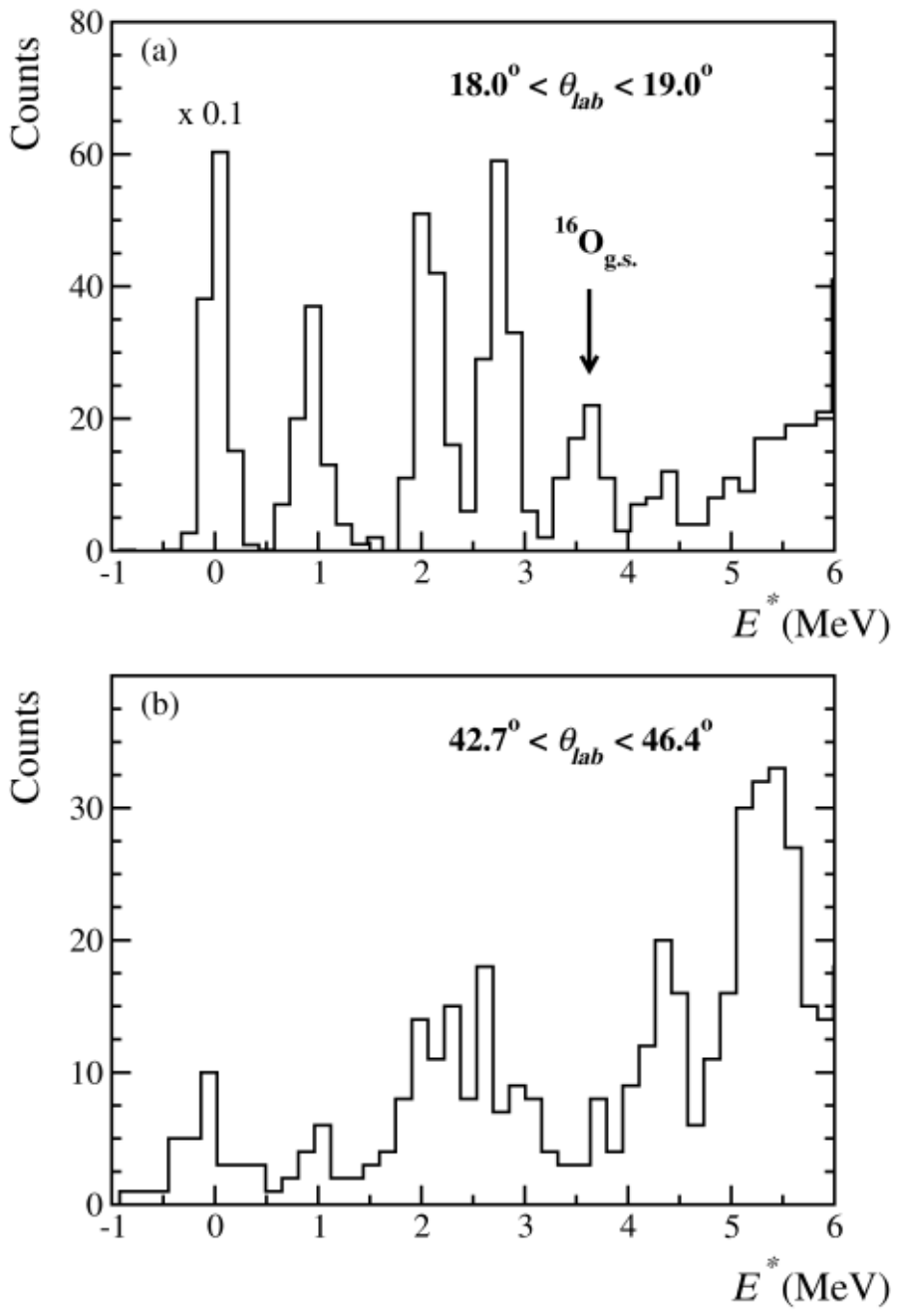

Figure 3.1 $^{27}$Al excitation energy spectra for two different scattering angles: (a) 18.0° < $\theta_{lab}$ < 19.0°, (b) 42.7° < $\theta_{lab}$ < 46.4°. The peak relative to the $^{27}Al_{g.s.}$ in the plot (a) has been scaled by a factor 0.1. The peak marked with an arrow refers to the $^{16}$O + $^{16}$O elastic scattering. From [77].

The cross section data were then analysed in the framework of Coupled Channel theory in refs. [76] [78]. The Sao Paulo double folding potential was used as the real part of the nuclear components of the nucleus-nucleus potential. The imaginary part was extracted from Glauber theory. An approximate rotational coupling for the $^{27}$Al low-lying states was assumed. This analysis indicates a crucial role of the coupling of the ground to the low-lying collective states of $^{27}$Al [76]. The observed rainbow-like pattern at large scattering angles is generated by the couplings only [78]. The role of couplings on the nuclear rainbow-like structure observed in this system shed a new light on the interpretation of the rainbow phenomenon. Even though, there is little understanding of the contribution of high-lying and collective states to the elastic channel.

In a second experiment the elastic and inelastic scattering of $^{16}$O + $^{27}$Al at 280 MeV was explored [79], [80]. This beam energy should correspond to a more pronounced rainbow-like structure, due to the coupling of the elastic channel to inelastic channel [81]. The achieved energy resolution (~ 550 keV FWHM) allowed a clear identification of the elastic peak. Transitions to known low-lying states of $^{27}$Al were also observed. Moreover, the $^{27}$Al excitation energy spectra shows two broad structures at about 20 and 24 MeV. The first one is identified as the GQR of the $^{27}$Al, also observed in previous (α,α’) experiments [82], while the second could be another collective mode, like the isoscalar GMR, not observed before. This is a qualitative interpretation that requires a careful analysis and proper theoretical explanation.

Experimental elastic and inelastic angular distributions are presented in Figure 3.2. The inelastic data shown here correspond to the differential cross section summed over the five low-lying states of $^{27}$Al, namely $1/2^+$ at 0.84 MeV, $3/2^+$ at 1.01 MeV, $7/2^+$ at 2.21 MeV, $5/2^+$ at 2.73 MeV and $9/2^+$ at 3.00 MeV. The high-precision angular distributions allow one to draw some important model independent insights directly from the data. Firstly, angular distributions for elastic and inelastic $L$-even transfer oscillate in opposition of phases at forward angles (see insert in Figure 3.2). This is an indication of the Fraunhofer diffraction and the consequent Blair phase rule occurring in these strongly absorbing systems [83]. The second argument concerns the optical model. Considering the Sommerfeld parameter for $E_{lab}$ = 280 MeV ($\eta$ = 3.9), a Fraunhofer-like behaviour is expected for the elastic distribution in the explored angular range. Scattering cross-sections of this type are characterized by a Bessel function of the first order [84], which exhibits oscillations that roughly fall off exponentially. The experimental elastic distributions show oscillations with successive maxima (minima) separated of about 3° (see insert in Figure 3.2).

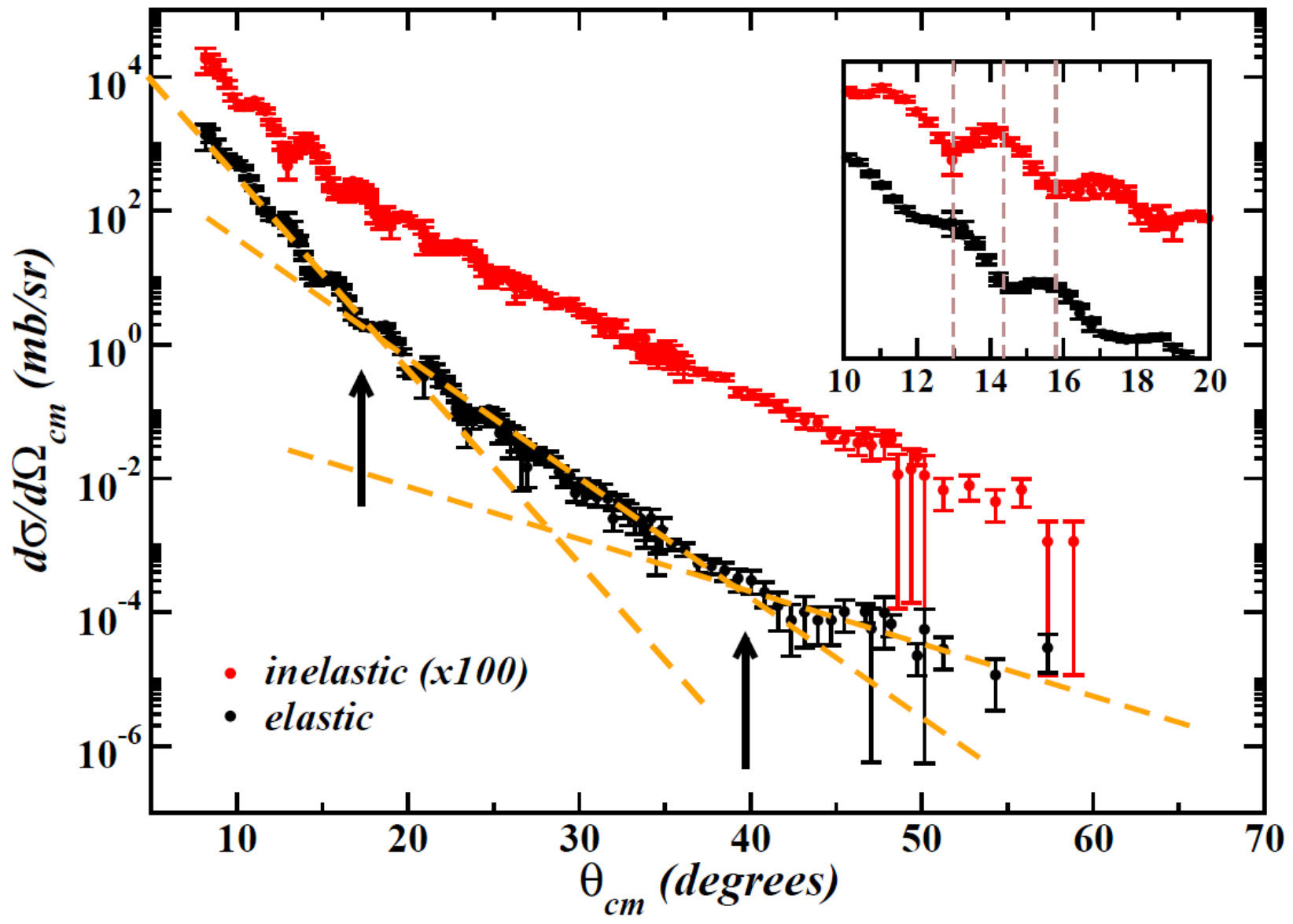

Figure 3.2: (Color online) Angular distribution in the centre-of-mass frame for elastic (black) and inelastic (red) scattering of $^{16}O$ + $^{27}Al$ system at 280 MeV. The arrows indicate angles of deviation from exponentials fall off (orange dashed lines). The insert shows in detail the angular distributions at forwards angles (between 10° and 20°). From [80].

Surprisingly two changes of slopes of the cross sections are observed. Three orange dashed lines are superimposed to the angular distribution in Figure 3.2 to guide the eye in order to highlight this feature. Changes of slope at ~20° and ~40° (indicated by vertical arrows) are observed, that may be related to the competition between the near/farside components. The first change of slope is located where a strong interference pattern is present. This is the typical behaviour due to near/farside coherent scattering in the Fraunhofer regime. The second change gives rise to a bow-like structure in the elastic scattering, which is a genuine effect beyond the interpretation of the optical model in such strongly absorptive systems. In ref. [80] this was further discussed in the light of coupled channel calculations. In particular, the calculation model, that successfully describes both elastic and inelastic dataset at 100 MeV, fails to reproduce precisely the dataset at 280 MeV. At such high energy, a strength normalization factor of 0.6 in the real part of the potential is necessary to account effectively for the polarization due to the couplings to all missed channels. It must be emphasized that the calculation simultaneously describes experimental data for elastic and inelastic angular distributions in more than seven orders of magnitude. However, the Airy-like minimum observed in the experimental data at about 40° is not well reproduced by this calculation. The topic is still ambiguous and demands efforts on the microscopic description of the coupling due to high lying excited states like the GQR. We see that the scattering of heavy ions far above the Coulomb barrier still retains information regarding specific details of internal degrees of freedom. Therefore, the description of the scattering observables requires a high degree of accuracy of the nuclear structure. To overcome some of the difficulties connected to the structure of the odd deformed $^{27}Al$ nucleus a promising option is to study the elastic scattering of an even and spherical heavy nucleus.

A third experiment exploring the $^{16}O$ + $^{60}Ni$ scattering at 280 MeV, using the same set-up was performed. Even in this case a good sensitivity was achieved and conclusive results are at reach. The analysis is still under way and the results will be published soon.

### 3.1.2 Elastic scattering in inverse kinematic

The MAGNEX capability to investigate inverse kinematics reactions at very forward angles with an excellent accuracy in energy and angles was probed in the campaign submitted by A. Pakou, from the University of Ioannina, for elastic scattering and break-up of weakly bound nuclei $^{6,7}Li$+p with the aim of probing the nuclear potential.

The final goal was the $^{6}Li + p \rightarrow {}^{4}He + d + p$ break-up measurement around zero degree, which was studied in inverse kinematics due to the confinement of the ejectiles at forward angles. A good energy and angular resolution was achieved measuring the $\alpha$ particles with MAGNEX in coincidence with the detection of the protons by a silicon detector. The analysis is still on the way and the results will be soon published. The $^{6}Li + p \rightarrow {}^{3}He + {}^{4}He$ channel was also investigated and the results are reported in ref. [85].

The elastic scattering measurement was then an unavoidable study for probing the potential in the same conditions. Results from the elastic scattering of protons by $^{6}Li$ in inverse kinematics are reported in ref. [86]. In the experiment, the $^{6}Li^{3+}$ beam was delivered by the Tandem accelerator at the energies of 16, 20, 25, and 29 MeV and impinged on a 240 $\mu$g/cm$^2$ $CH_2$ target. The elastically scattered lithium ions were momentum analysed and detected by MAGNEX, whose optical axis was set at $\theta_{opt}$ = +4°. The full kinematics of the reaction was explored by the use of three different magnetic fields. A typical reconstructed ($\theta_{lab}$,$E$) correlation plot at a projectile energy of 20 MeV is displayed in Figure 3.3. Overlapping regions among the different magnetic settings confirm the accuracy of the measurement and of the reconstruction procedure and the consistency between the different runs.

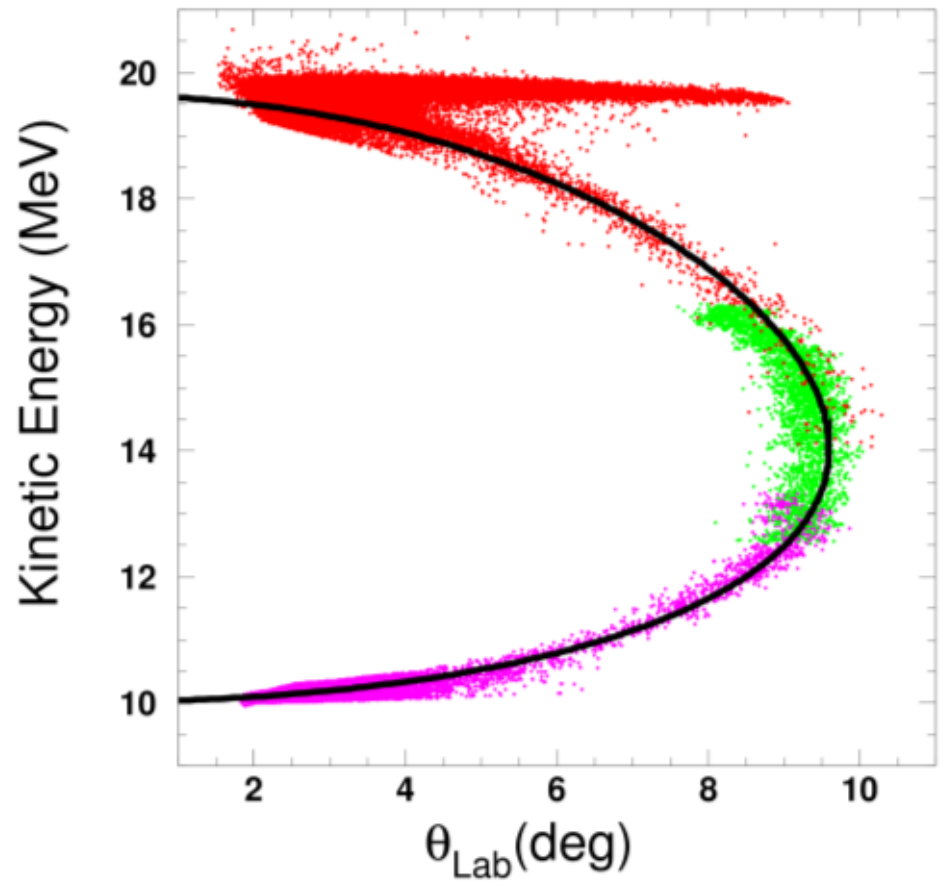

Figure 3.3 A reconstructed ($\theta_{lab}$,$E$) correlation plot for $^6$Li+$p$ at a projectile energy of 20 MeV. The measured kinematical plot was obtained with the superposition of the data of three runs at different magnetic sets, which are designated with different colours. The black solid line represents the kinematic curve. The horizontal locus at ~20 MeV corresponds to events coming from $^6$Li + $^{12}$C elastic scattering.

## 3.2 Reaction mechanisms in heavy-ion transfer reactions

Direct transfer reactions have the fundamental property to select specific degrees of freedom in the complex many body nuclear system. For this reason, they have always played a crucial role in the exploration of the nuclear structure. Examples are the use of one-neutron transfer reactions to study the single particle configurations in nuclei or the α-transfer reactions in the exploration of α-clustering phenomena. In particular, the relation between transfer probabilities in two-neutron transfer reactions and pairing correlations in nuclei is a subject of great interest [87]. Indeed, the strength of the pairing force is directly connected to microscopically derived transition densities for pair addition and removal modes [88]. Compared to other methods, in which the pairing force is studied by the exploration of ground state properties, transfer reactions are particularly interesting because a large variety of excited states is accessed. In this way, the effects of the pairing as a function of the occupation of different single-particle orbitals and linear and angular momentum transfer, including the rise of collective modes, can be determined.

A campaign of experiments of heavy-ion induced one- and two- neutron transfer reactions was performed by MAGNEX with the aim, among the others, of studying the reaction mechanism. In one of these experiments, a beam of $^{18}$O$^{6+}$ ions was accelerated at 84 MeV incident energy on 50 $\mu$g/cm$^2$ pure $^{12}$C and 99% enriched $^{13}$C targets. The experiment was performed at three angular settings, with the MAGNEX optical axis centered at $\theta_{opt}$ =8°, 12° and 18°.

The energy spectra measured in the $^{13}$C($^{18}$O,$^{17}$O)$^{14}$C and $^{12}$C($^{18}$O,$^{16}$O)$^{14}$C reactions are shown in Figure 3.4. An overall resolution of 160 keV in energy and 0.3° in angle was obtained. A first evidence is that the $^{14}$C states are populated with rather different cross sections by the two processes. Only states with a well-known structure based on a two-neutron cluster coupled to the $^{12}$C 0$^+$ core are efficiently populated in the ($^{18}$O,$^{16}$O) reaction. The suppression of single-particle states, which would require an uncorrelated transfer of two neutrons and the breaking of the initial neutron pair in the $^{18}$O ground state, reveals the minor role of the two-step dynamics. This strengthens the conclusions of Ref. [89], also discussed in Section 3.4.1 for a similar system, where calculations based on the removal of two uncorrelated neutrons from the $^{18}$O projectile describe only the continuum of the energy spectrum but not the narrow resonances with two-particle–three-hole configurations, because of the lack of neutron-neutron correlations.

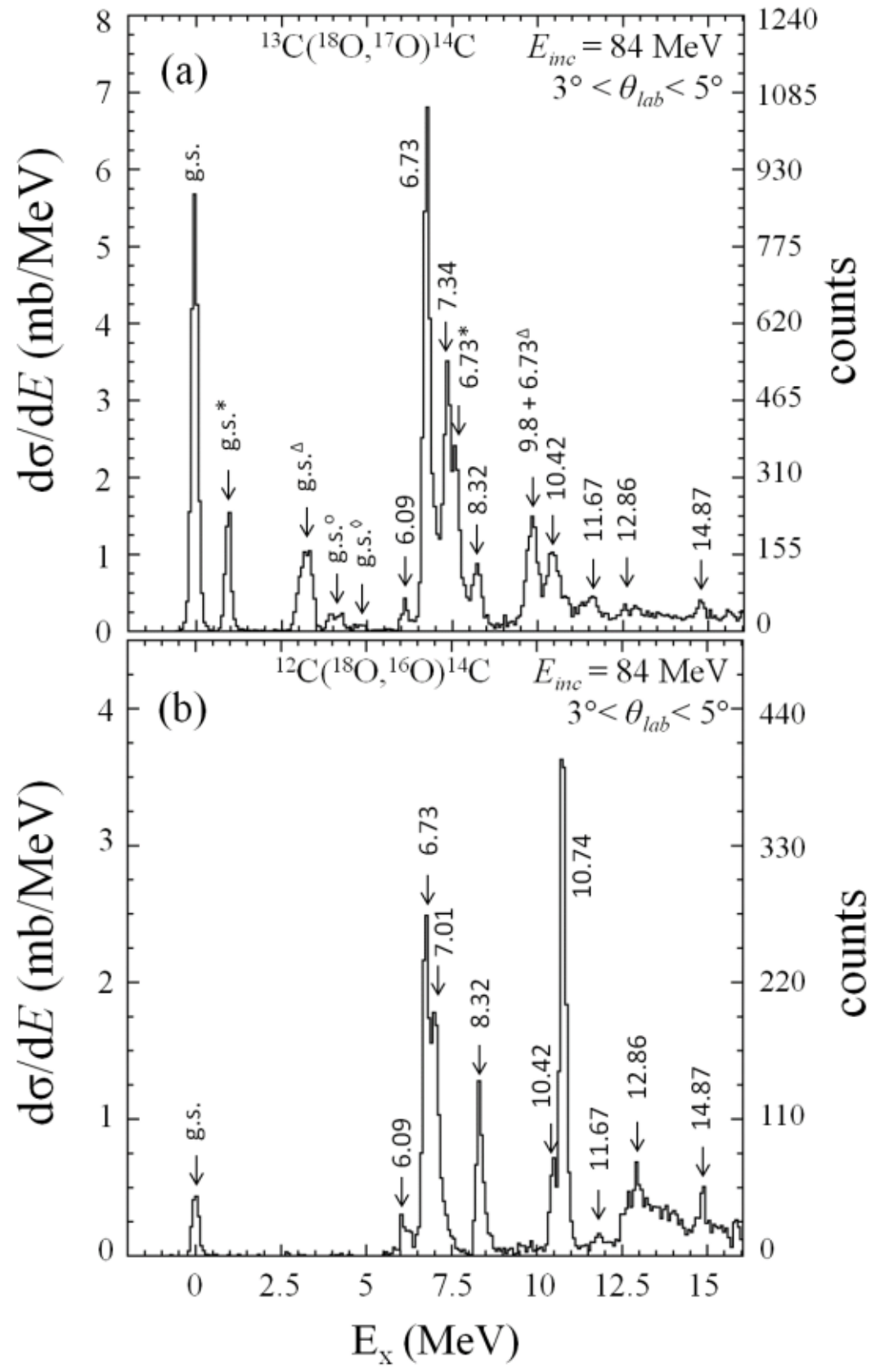

Figure 3.4 $^{14}$C energy spectra for the (a) one-neutron and (b) two-neutron transfer reactions. In the panel (a) the peaks marked with an asterisk, a triangle, a circle, and a diamond represent the transition to the excited states of the $^{17}$O ejectiles at 0.87, 3.05, 3.84, and 4.55 MeV, respectively. From [90].

The absolute cross-section angular distributions for some of the transitions induced by the ($^{18}$O,$^{17}$O) and ($^{18}$O,$^{16}$O) reactions are shown in Figure 3.5. In the $^{12}$C($^{18}$O,$^{16}$O)$^{14}$C reaction, the transition to the $^{14}$C ground state exhibits a pronounced oscillating pattern, characteristic of the expected $L$ = 0 angular momentum transfer. The other transitions, characterized by $L \neq 0$, do not show such oscillations, still preserving a certain degree of sensitivity to the transferred angular momentum in the slope. In Ref. [91] the damping of oscillations in heavy-ion reactions for $L \neq 0$ transitions was attributed to the different phases of the transfer amplitude components exhibited by the different angular momentum projections $M_L$. The $L$ = 0 mode are not affected by this phenomenon since they contain only one $M_L$ projection, giving a distinctive signature for them.

Exact finite range coupled reaction channel (CRC) cross-section calculations using the FRESCO code [92] were performed. The spectroscopic amplitudes were determined by performing a shell-model calculation with $^{12}$C treated as closed core and valence protons and neutrons in the orbits $1p_{1/2}$, $1d_{5/2}$ and $2s_{1/2}$. The details of the calculations and the used coupling schemes are reported in Ref. [90].

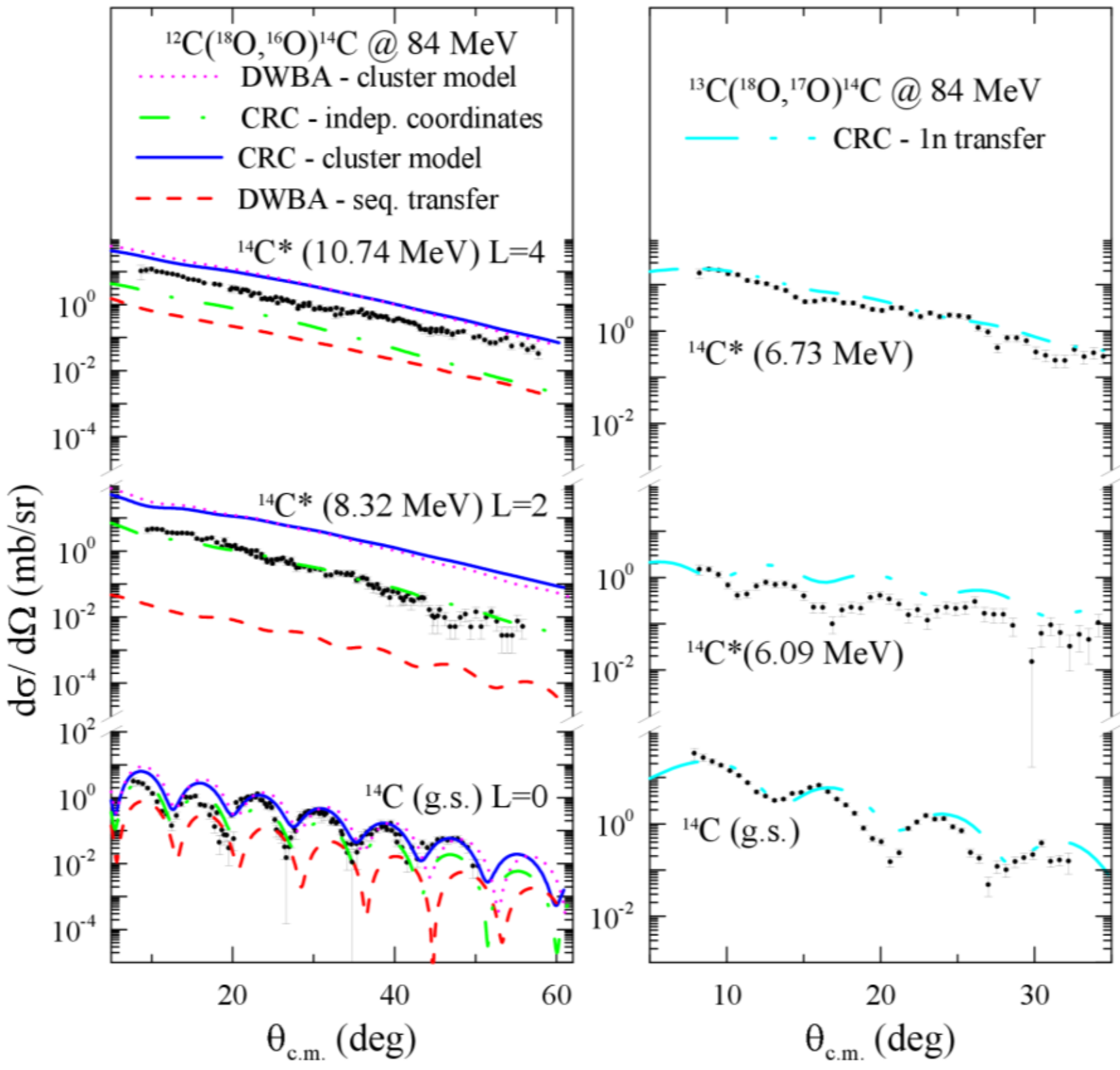

Figure 3.5 Comparison of the experimental angular distributions with theoretical calculations for the (left) $^{12}C(^{18}O,^{16}O)^{14}C$ and (right) $^{13}C(^{18}O,^{17}O)^{14}C$ reactions at 84 MeV. No scaling factors are used. From [90].

For the one-neutron transfer reaction, the CRC cross sections give an accurate estimate of the measured absolute values, together with a remarkable description of the period and amplitude of the observed oscillations.

For the transitions induced by the ($^{18}O$,$^{16}O$) reaction, both the extreme cluster model and the independent coordinate scheme were used to calculate the two-particle state wave functions in CRC. In the extreme cluster model the relative motion of the two-neutron system is frozen and separated from the center of mass and only the component with the two neutrons coupled antiparallel to a zero intrinsic angular momentum ($S$ = 0) participates in the transfer. In the independent coordinate scheme, the transfer of two neutrons is described taking into account single-particle information obtained by shell-model calculations in the available model space. Even if a two-nucleon wave function can be ideally described both in a cluster and in an independent particle basis representation, differences arise when limited model spaces are used, as in the present case.

In the ($^{18}O$,$^{16}O$) calculations the transition to the $^{14}C_{g.s.}$ is described, especially at the largest angles, by the cluster model (see Figure 3.5). Both the shape and the absolute value of the angular distribution are satisfactorily reproduced, even setting to 1 the spectroscopic amplitude for this configuration. An amplitude of 0.89 ± 0.05 is obtained by scaling the CRC calculated cross section to the experimental data. This value is in good agreement with the 0.913 predicted by shell model calculations for the $(p_{1/2})^2$ configuration.

A smaller cross section is obtained in the independent coordinate scheme, remarking a relevant contribution of components of the true wave function beyond the adopted model space. The difference in the two results is a consequence of the larger number of pairs of single-particle wave functions that would be necessary to describe the cluster structure. This finding is a strong evidence of the presence of two-neutron pairing correlations in the $^{14}C$ ground state.

The two-step sequential transfer calculations account only for a minor contribution to the absolute cross section, justifying *a posteriori* the approximated approach to separate the sequential transfer from the CRC calculations. Interference effects between direct and sequential processes were found to be important to describe the deep minima of the ground state angular distribution [93].

The calculated angular distribution of the transition to the $2^+$ state at 8.32 MeV does confirm the expected $L = 2$ transfer. In the cluster model a spectroscopic amplitude of 0.30 is extracted by scaling the CRC calculations to the experimental data. This indicates the need for a larger model space for the cluster wave function, which is limited to the $S = 0$ component in our approach. Instead, in the independent coordinate calculations, the cross section is accurately reproduced. These results confirm the weak nature of the coupling of the two neutrons in the $s_{1/2}\, d_{5/2}$ model space for this state.

For the transition to the $4^+$ state at 10.74 MeV, a spectroscopic amplitude equal to 0.55 is extracted for the cluster configuration. In this case the independent coordinate calculation underestimates the cross section, indicating that a larger space is required in the shell-model calculations. This is not surprising since one expects to find relevant ($d_{5/2}\, d_{3/2}$) contributions in the $4^+$ wave function, which are excluded in our model space.

Other systems different from the carbon isotopes and other incident energies were explored via ($^{18}O$,$^{16}O$) reactions to study the dependence on the target mass and on the energy. Preliminary results are published in refs. [94], [95], [96].

A comparison between our ($^{18}O$,$^{16}O$) measured cross-sections and the corresponding values from (t,p) reaction at 18 MeV incident energy, known from literature [97] is done in Ref. [98]. Besides the different kinematical conditions that can generate small differences, the important finding is the striking similarity between the cross-sections of the two reactions. This is a not trivial result since rather different dynamical conditions are in principle present in the two many body problems. Only assuming the negligible role of the $^{16}O$ core degree of freedom in the two-neutron stripping reaction mechanism, one can expect such a behaviour. The cross-section is thus determined to a large extent by the target polarizability to a two-neutron transfer from an external probe.

These results indicate that ($^{18}O$,$^{17}O$) and ($^{18}O$,$^{16}O$) reactions are valuable probes for accurate spectroscopic studies. In the next section, we present examples of such applications.

## 3.3 Continuum single-particle spectroscopy by one-neutron transfer reactions

In order to perform a systematic study aiming at the investigation of one-neutron excitations, the ($^{18}O$,$^{17}O$) one-neutron transfer reactions were explored using MAGNEX. Among the others, an interesting case is that of $^9Be(^{18}O,^{17}O)^{10}Be$, since single-particle configurations in $^{10}Be$ are essential to understand the details of the $^{11}Be$ halo nucleus.

In the experiment, an $^{18}O^{6+}$ beam at 84 MeV impinged on a 130 ± 6 μg/cm$^2$ self-supporting $^9Be$ target. Supplementary runs with a 59 ± 3 μg/cm$^2$ self-supporting $^{12}C$ target and a 212 ± 10 μg/cm$^2$ $WO_3$ target mounted on a 193 ± 10 μg/cm$^2$ Au backing were recorded for estimating the background in the $^{10}Be$ energy spectra coming from $^{12}C$ and $^{16}O$ impurities in the $^9Be$ target. MAGNEX was set for covering an angular range between 4° and 14° in the laboratory reference frame.

An example of the energy spectra measured at forward angles is shown in Figure 3.6. The contribution arising from $^{12}C$ and $^{16}O$ impurities in the target was subtracted from the final spectrum after normalization on known peaks. An overall energy and angular resolution of about 180 keV and 0.3°, respectively, were obtained. Some narrow states of $^{10}Be$ are recognized below the one-neutron emission threshold ($S_n$ = 6.812 MeV), as labelled in Figure 3.6. Each $^{10}Be$ state shows up as a doublet corresponding to the $^{17}O$ ejectile emitted in its ground and first excited state at $E_x$ = 0.87 (1/2$^+$) MeV. The contribution of the higher $^{17}O$ excited states is less relevant and undistinguishable among the other peaks. Above the one-neutron emission threshold, there are two narrow peaks at ~ 7.4 and ~ 9.4 MeV. Both structures are identified as the superposition of two resonances, as indicated in Figure 3.6. A large background is present above ~ 10 MeV excitation energy and no resonances are identified in this region. In order to understand what contributions are present in the region of the spectra above the one-neutron emission threshold the one-neutron continuum was studied by the transfer to the continuum model developed in Ref. [99].

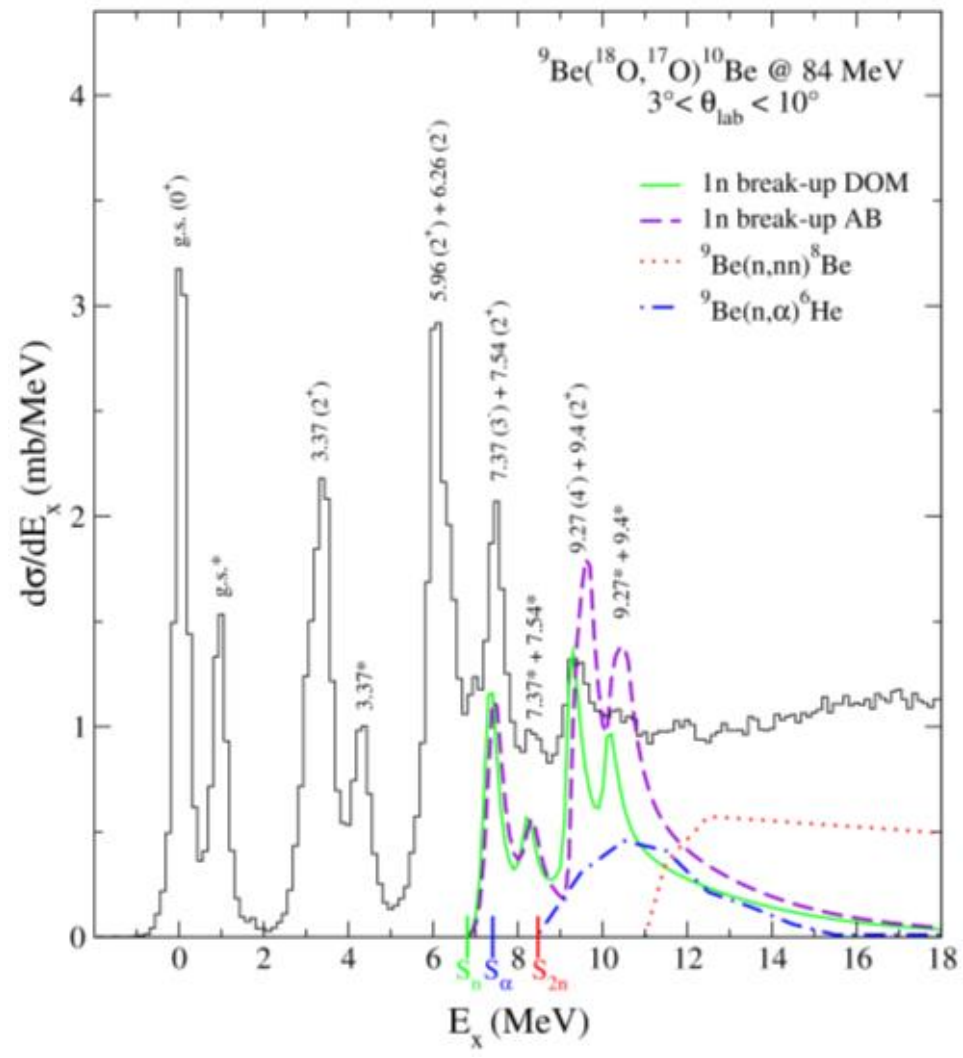

Figure 3.6 Inclusive excitation energy spectrum of the $^9Be(^{18}O,^{17}O)^{10}Be$ reaction at 84 MeV incident energy and 3°< $\theta_{lab}$ < 10°. The background coming from $^{12}C$ and $^{16}O$ impurities was subtracted. Peaks marked with an asterisk refer to the $^{17}O$ ejectile emitted in its first excited state at 0.87 MeV. Total 1-n break-up calculations resulting from the use of the DOM and the AB potentials (see text) [100] are shown as the green continuous and the violet dashed lines, respectively. The experimental data [101] of the $^9Be(n,nn)^8Be$ [102], [103], [104], [105] and $^9Be(n,\alpha)^6He$ [106], [107], [108] reactions are drawn as red dotted and blue dotted-dashed line, respectively. The 1n- ($S_{2n}$), 2n- ($S_{2n}$) and α- ($S_\alpha$) separation energies are also indicated. From [109].

Semi-classical approaches have proven to be accurate enough to explain integral properties such as the selectivity of the reaction, allowing also to treat the transfer to bound and unbound states in a coherent way. Among these, the formalism of the transfer to bound states [110] extended to the case of the unbound ones [111] [112] was applied to analyse the $^{10}Be$ continuum populated via the ($^{18}O$,$^{17}O$) reaction [109]. The same approach was also used to study two-neutron transfer reactions [89] [113] [114].

The calculations require both initial and final single-particle states of the transferred neutron. The initial state is bound in the projectile and the final states are unbound with respect to the target. In order to obtain the radial wave functions of the initial (projectile) bound states and the corresponding asymptotic normalization constant $C_i$, the Schrödinger equation was solved numerically by fitting the depths of the Wood-Saxon potentials $V_0$ to the experimental separation energies ε. It is known that the $^{18}O_{g.s.}$ wave function contains an admixtures of $(1d_{5/2}^2)_0$ (~75.9%) and $(2s_{1/2}^2)_0$ (~13.5%) [90]. As a consequence, the presence of peaks corresponding to the excitation of the $^{17}O$ first excited state were taken into account in the calculation considering the removal of the neutron from the $2s_{1/2}$ orbital.

The potentials used to calculate the energy dependent $S$-matrix were the $n$-$^9Be$ optical-model potentials developed by two methods (AB and DOM) in Ref. [100]. Both potentials reproduced the experimental cross section at all energies, in particular the $p_{3/2}$ resonance at $E_{lab}$ = 0.7 MeV and the $d_{5/2}$ at $E_{lab}$ = 3.1 MeV. Both potentials were used to extract the phase shift and $S$-matrix needed by the formalism to perform the one-neutron transfer to the continuum in the system $^{18}O$ + $^9Be$ and a comparison between the two results was done.

The results obtained are superimposed to the experimental continuum spectrum of $^{10}$Be in Figure 3.6. The displayed curves are the sum of two calculations. One corresponds to the elastic break-up and the absorption relative to the emission of a neutron that leaves the $^{17}$O in its ground state. The calculations included the spectroscopic factor, coming from the estimates of the configuration mixing in the $^{18}O_{g.s.}$ wave function [90]. In order to describe the peaks, in the experimental spectra, coming from the transition that leaves the $^{17}$O core in its first excited state at 0.87 MeV, supplementary calculations in which the $^{18}$O neutron is emitted from the $2s_{1/2}$ orbital were performed. The resulting energy spectrum of the scattered neutron was shifted by 0.87 MeV and included the spectroscopic factor according to the shell-model configuration admixtures in the $^{18}$O ground-state wave function [90].

Both calculations reproduce the $^{10}$Be continuum spectrum, without needing any scaling factor. In particular, two $^{10}$Be single-particle resonances are reproduced, at $E_x$ = 7.54 MeV, mainly built as $|\ ^9Be_{g.s.}(3/2^-) \otimes (1p_{1/2})_\upsilon >$ and at $E_x$ = 9.27 MeV, which shows a dominant $|\ ^9Be_{g.s.}(3/2^-) \otimes (1d_{5/2})_\upsilon >$ configuration. These peaks have been identified by looking at the contribution of each single partial wave $j_f$ to the total sum. It was found that the spectrum is dominated by the two resonances at 7.54 and 9.27 MeV, which correspond to the $p_{1/2}$ (red curve) and $d_{5/2}$ (blue curve) orbitals, respectively [109]. The contribution of other partial wave to the total cross section was found to be negligible.

The resonances at $E_x$ = 7.37 ($3^-$), 9.4 ($2^+$) MeV are not reproduced within the present approach because they are built on more complicated configurations which are not included in the adopted transfer to the continuum model.

At higher excitation energies, above 11 MeV, both calculations give a very small contribution to the inclusive experimental spectrum, which shows an almost flat behaviour. Other contributions are expected there, since the two-neutron ($S_{2n}$ = 8.476 MeV) and alpha ($S_\alpha$ = 7.409 MeV) emission thresholds are open. In order to estimate these contributions, at least in their shapes, the experimental data [101] of the $^9$Be($n$,$nn$)$^8$Be [102], [103] [104], [105] and $^9$Be($n$,$\alpha$)$^6$He [106] [107] [108] are superimposed on the experimental spectrum in Figure 3.6. The data were scaled by a factor $10^{-3}$ resulting from the ratio between the free $n$-$^9$Be cross-section [100] and the present $^{18}$O-$^9$Be reaction. As expected, these break-up channels manifest a shape compatible with the flat background above 11 MeV in the present experimental $^{10}$Be spectrum.

## 3.4 Neutron pairing correlations by two-neutron transfer reactions

A study of the neutron pairing correlations was pursued by the ($^{18}$O,$^{16}$O) two-neutron transfer reactions on different targets ($^{12}$C,$^{13}$C,$^9$Be,$^{11}$B,$^{16}$O) using MAGNEX. Thanks to its high resolution and large acceptance, high quality inclusive spectra were obtained, even in a largely unexplored region above the two-neutron emission threshold in the residual nucleus [90] [115]. New phenomena appeared, such as the dominance of the direct one-step transfer of the two neutrons [116] and the presence of broad resonances at high excitation energy in the $^{14}$C and $^{15}$C spectra. These structures were recently identified as the first experimental signature of the Giant Pairing Vibration [117], [118], predicted long time ago [119].

In this context, the neutron decay modes were also explored by detecting the neutrons in coincidence using the EDEN array (see Section 2.8) and an exploratory investigation was extended to the $^{120}$Sn(p,t)$^{118}$Sn reaction.

### 3.4.1 The Giant Pairing Vibration in $^{14}$C and $^{15}$C via the ($^{18}$O,$^{16}$O) reaction

In 1977, Broglia and Bes [119] predicted the existence of a giant collective mode in the atomic nuclei, named the Giant Pairing Vibration (GPV). The microscopic origin of such a mode is the coherence among elementary particle–particle excitations generated by the addition or removal to/from an atomic nucleus of two nucleons in a relative S-wave (motion component characterized by an orbital angular momentum $L$ = 0).

Two different experimental campaigns were devoted to this study. In the first a beam of $^{18}O^{6+}$ ions at 84 MeV incident energy impinged on a 49 ± 3 $\mu$g/cm$^2$ self-supporting $^{12}$C and a 50 ± 3 $\mu$g/cm$^2$ self-supporting 99% enriched $^{13}$C targets. The runs with $^{12}$C were used also for estimating the background coming from $^{12}$C impurities in the $^{13}$C target.

Examples of the reconstructed energy spectra for the two experiments are shown in Figure 3.7. For the $^{15}$C case, it is necessary to correctly evaluate and subtract the background contribution, which comes mainly from impurities of $^{12}$C in the $^{13}$C target. The data runs on $^{12}$C target were performed at the same experimental conditions and magnetic fields of the $^{13}$C one. The background spectrum superimposed on the $^{15}$C spectrum is shown in the bottom panel of Figure 3.7, after a careful normalization of the data runs on $^{12}$C. It appears that the background contribution is typically small. Several narrow peaks corresponding to well-known low-lying bound and resonant states are observed in both cases [120], [97]. An overall energy resolution of ~160 keV FWHM is obtained and the laboratory angle resolution is better than 0.5°.

A sudden increase of the yield is found just above the two-neutron separation energy ($S_{2n}$= 13.123 MeV for $^{14}$C and $S_{2n}$= 9.394 MeV for $^{15}$C). A large bump, superimposed to the continuum background, is observed in both spectra. A best-fit procedure with Gaussian shapes and a locally adjusted linear background model (see inserts in Figure 3.7) gives an energy of $E_x$ = 16.9 ± 0.1 MeV (FWHM 1.2 ± 0.3 MeV) in $^{14}$C and $E_x$ = 13.7 ± 0.1 MeV (FWHM 1.9 ± 0.3 MeV) in $^{15}$C. In the $^{14}$C case, two known narrow resonances at 16.43 and 16.72 MeV are weakly populated in the lower energy region of the bump and accounted for in the fit. The centroids and widths are confirmed by the results of the second experimental campaign performed at 270 MeV bombarding energy studying the same reactions by the same experimental set-up [121].

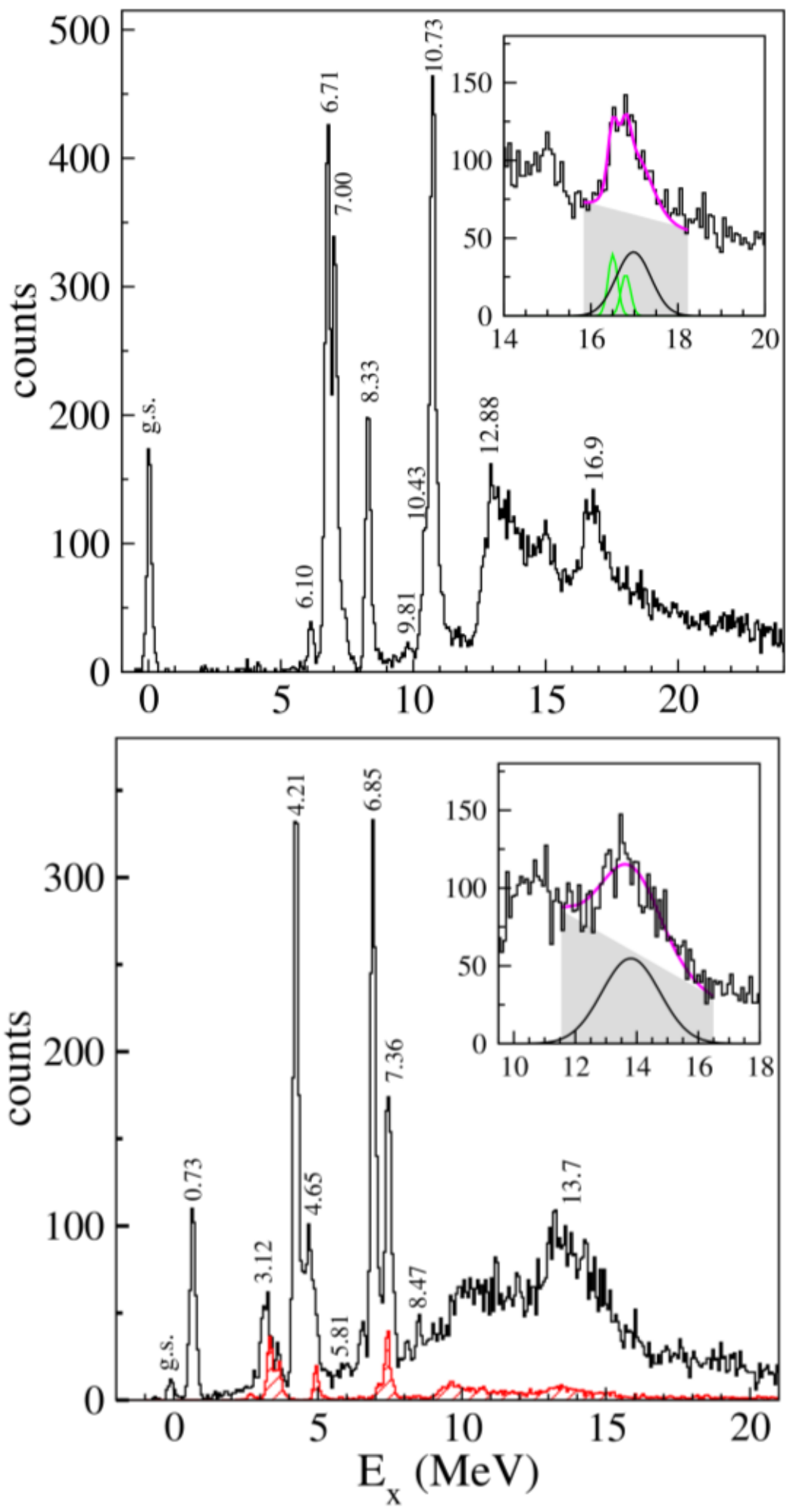

Figure 3.7 Upper panel: Excitation energy spectrum of the $^{12}C(^{18}O,^{16}O)^{14}C$ reaction at $10° < \theta_{lab} < 11°$. In the insert: a zoomed view of the spectrum with the modeled 3-body continuum (grey area), the fitted known resonances at 16.43 and 16.72 MeV (green Gaussians), the fitted GPV (black Gaussian) and the sum of them (magenta curve). Bottom panel: Excitation energy spectrum of the $^{13}C(^{18}O,^{16}O)^{15}C$ reaction at $9.5° < \theta_{lab} < 10.5°$. The red hatched histogram represents the normalized background of the $^{12}C(^{18}O,^{16}O)^{14}C$ reaction coming from $^{12}C$ impurities in the $^{13}C$ target. In the insert: a zoomed view of the spectrum with the modelled 3-body continuum (grey area), the fitted GPV (black Gaussian) and the sum of them (magenta curve). From [118].

The absolute cross section angular distributions were extracted, according to the procedure described in Section 2.6. Examples of the cross section angular distributions for the bound and narrow resonant states populated by the $^{12}C(^{18}O,^{16}O)^{14}C$ reaction are reported in Section 3.2, in which they are discussed and analyzed within a Coupled Reaction Channels (CRC) framework [90].

The resulting angular distributions for the bumps at 16.9 ± 0.1 MeV in $^{14}C$ and 13.7 ± 0.1 MeV in $^{15}C$ are shown in Figure 3.8. The contribution to the angular distribution of the bumps due to the continuous background in the spectra was estimated at each angle by a least-squared approach. A clear indication of an oscillating pattern is present in the $^{14}C$ $0^+$ ground state (see Figure 3.5) and in the two broad bumps. This behaviour reveals the dominance of a resonant state in that energy region, characterized by a well-defined angular momentum. The properties of such resonances have been analysed and compared to those expected for the GPV in refs. [117] [118] [122].

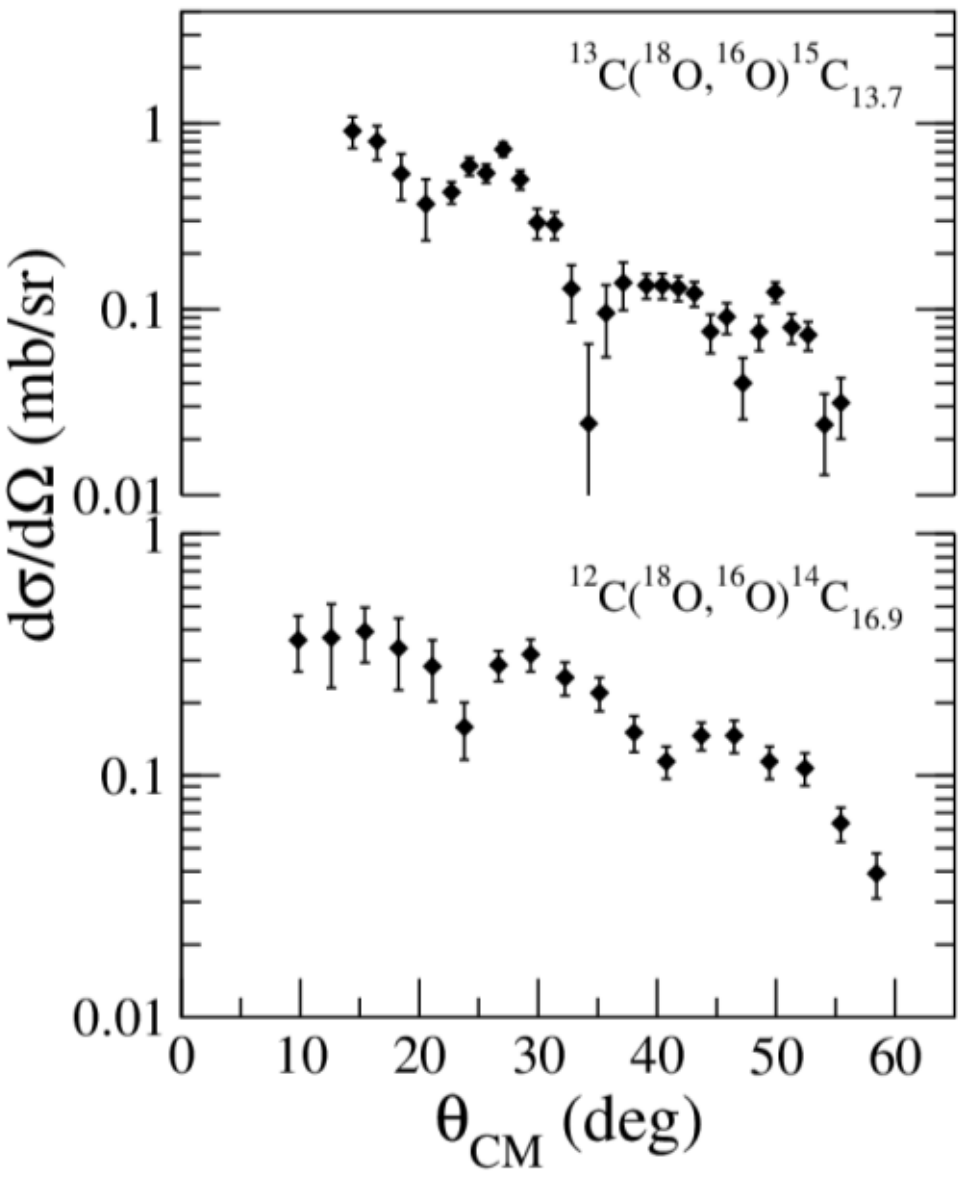

Figure 3.8 Angular distributions for $^{15}C$ and $^{14}C$ GPVs. From [118].

In order to isolate the spectral characteristics of the resonant-like excitations in the $^{15}C$ and $^{14}C$ nuclei, it is important to identify the various components of the projectile break-up present in the inclusive excitation energy spectra. The extreme semi-classical model applied in Section 3.3 for the description of the one-neutron removal in the $^{9}Be(^{18}O,^{17}O)^{10}Be$ reaction was used also to describe the removal of two neutrons from the projectile and to analyse the measured energy spectra of $^{14}C$ and $^{15}C$ in the region above $S_{2n}$ [89], [113]. The specific treatment of nucleon-nucleon correlations beyond the residual nucleus mean field, as, for example, those due to the neutron-neutron pairing in the *sd*-shells were not included.

The results are shown in Figure 3.9. In both cases, the calculations give a good account for the continuum background in the excitation energy spectra. Both bumps at 16.9 ± 0.1 MeV in $^{14}C$ and 13.7 ± 0.1 MeV in $^{15}C$ are not explained within this approach. Similarly, approaches based on the towing mode of the $2n$ cluster [123], [124] fail in account for the observed bumps [117]. These results indicate that a more complete description of the $^{12,13}C + n + n$ systems, including the $n$-$n$ correlations, is required to describe these structures.

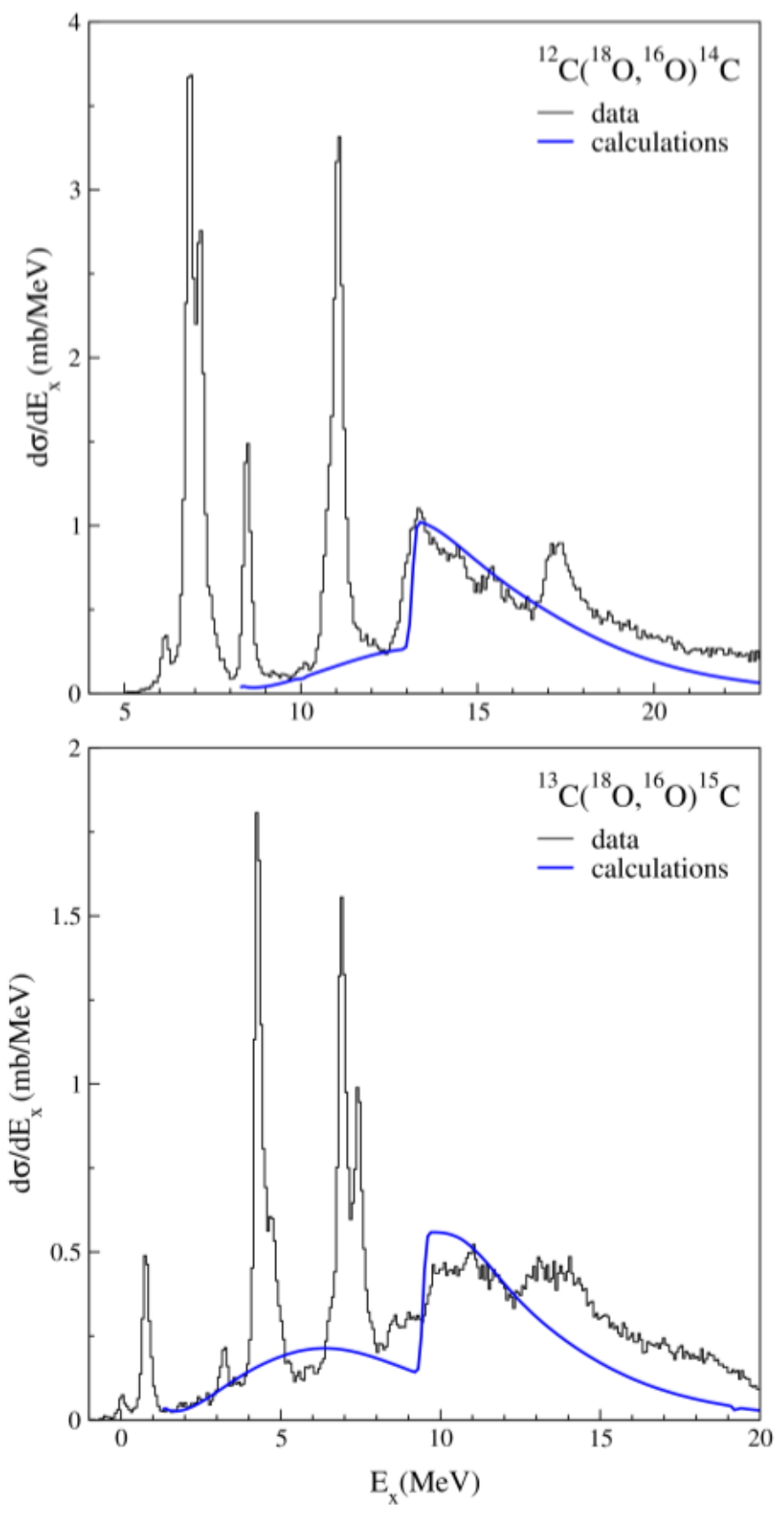

Figure 3.9 Experimental inclusive energy spectra for the $^{12}C(^{18}O,^{16}O)^{14}C$ (upper panel) and $^{13}C(^{18}O,^{16}O)^{15}C$ (lower panel) reactions at 84 MeV and 7° < $\theta_{lab}$ < 17° and independent removal of the two neutrons calculations (blue line).

As a first investigation of a possible common origin of the two observed structures in $^{15}C$ and $^{14}C$, the excitation energy compared to the target ground state was computed as $E^t_x = E_x + M_r - M_t$, where $M_t$ and $M_r$ represent the mass of the target and of the residue, respectively. The $E^t_x$ is a more suitable parameter when comparing the data with theoretically derived *p-p* excitations built on the target mean field. When applying such a scaling of the excitation energies, $E^t_x = 16.9 + 3.0 = 19.9$ MeV is obtained for the $^{14}C$ case and $E^t_x = 13.7 + 6.7 = 20.4$ MeV for the $^{15}C$ one. The energy of the two resonances is found at about the same value in both nuclei. This is consistent with the results of ref. [125], where Quasi-particle Random Phase Approximation (QRPA) calculations predicted a GPV mode at ~ 20 MeV in oxygen isotopes. The same kind of calculations were performed for the $^{12}C$ response to the monopole *p-p* operator to predict the $0^+$ two neutrons addition strength [118]. A broad distribution connected to the superposition of the $2p_{3/2}$, $2p_{1/2}$, $1d_{3/2}$, $1f_{7/2}$ and orbitals above is identified in the response function at $E_x^{theor} \sim 17$ MeV with respect to the $^{14}C_{g.s.}$, a value consistent with the experimental centroid of the bump ($E_x = 16.9 \pm 0.1$ MeV). A certain degree of collectivity is observed in such bump when the residual interaction is taken into account.

The measured widths of the two resonances (FWHM = 1.9 ± 0.3 MeV for the 13.7 ± 0.1 MeV bump in $^{15}C$ and FWHM = 1.2 ± 0.3 MeV for the 16.9 ± 0.1 MeV bump in $^{14}C$) are compatible with the GPV, for which the width is expected to be in the range of 1-2 MeV for heavy nuclei [126] [127]. The different width in the two systems indicates a different half-life, which is shorter in the case of $^{15}C$. This is mainly due the fact that the $^{15}C$ resonance is about 4.3 MeV above $S_{2n}$ and 12.5 MeV above the one-neutron emission threshold ($S_n$), whereas $^{14}C$ resonance is approximately 3.7 MeV above $S_{2n}$ and 8.7 MeV above $S_n$. The larger energy available in $^{15}C$ results in a substantially faster emission of both one and two neutrons.

In order to investigate the multipolarity of the observed resonances, two approaches were used: the analysis of the oscillating behaviour of the experimental angular distributions and the comparison of the measured cross-sections with state-of-the-art theoretical calculations.

A well-known phenomenon in heavy-ion induced transfer reactions above the Coulomb barrier is the presence of oscillations in the angular distributions only for the $L = 0$ transitions [91]. The distinctive feature of the $L = 0$ angular distributions provides a model independent identification of $L = 0$ among the other multipolarities. This phenomenon was experimentally observed for the same $^{12}C(^{18}O,^{16}O)^{14}C$ reaction at 84 MeV in ref. [90] for transitions below $S_{2n}$, as reported in Section 3.2. Therefore, the oscillating pattern of the broad resonances angular distributions supports their dominant $L = 0$ nature (see Figure 3.8).

From a theoretical point of view, the comparison of the experimental angular distributions of such resonances with state-of-the-art model calculations is not trivial. On one hand, the proper approach to analyse a resonance above the two-neutron emission threshold would be that proposed in ref. [128], which adopts a discretized scheme to describe the transfer in the continuum. However, the limitation of such a model is the need of a three-body assumption, since a complete four-body approach is not available. Consequently, these kind of calculations are not very accurate in describing the details of the cross section angular distributions. An alternative approach are the CRC calculations with an extreme cluster approximation, as those developed in ref. [90], but, in this case, the convergence is not reachable for transitions above $S_{2n}$.

Both approaches were attempted to study the multipolarity of the $^{14}C$ resonance in ref. [117]. Taking into account the limitations of the discretized continuum scheme, only the cross section absolute value was considered. The $L = 0$ transition cross section was found consistent with the experimental one, without the need of any scaling factor. Regarding the CRC calculations, the same ingredients of the calculations in ref. [90] were used. Since it is not possible to perform such calculations at the excitation energy of the observed resonance (16.9 MeV), an artificial value of 12 MeV, which is below $S_{2n}$, was chosen and separate calculations were performed assuming $L$ values from 0 to 5. In this case, only the shape of the resulting calculations was looked at, finding again that only the $L = 0$ transition is consistent with the bump experimental angular distribution.

Another observable defining the GPV is the cross section, which was predicted to be comparable with that of the $L = 0$ transition to the ground state pairing vibration [119]. In the present case, it was found $\sigma(^{14}C_{g.s.}) = 0.92$ mb and $\sigma(^{14}C_{GPV}) = 0.66$ mb integrated in the same angular range. This indicates the high-collectivity of the GPV. This property was also confirmed in terms of the strength, in order to remove trivial *Q*-value effects.

To summarize, the MAGNEX data showed the clearest signal compatible with the long searched GPV so far. The resonances at $E_x = 16.9 \pm 0.1$ MeV in $^{14}C$ and at $E_x = 13.7 \pm 0.1$ MeV in $^{15}C$ show properties consistent with those defining the GPV mode. These results were recently published in [117]. Further work is needed to confirm the results and provide clear evidence for the observation of the GPV. It is worth exploring other systems, where structural and reaction mechanism constraints are as favourable for the GPV excitation as the present case. We are presently engaged with this activity. Moreover, the study of the decaying features of these resonances, partially presented in Section 3.4.2, could reveal further information on the nature of the neutron-neutron correlation in the giant mode.

### 3.4.2 Neutron decay spectroscopy

An example of the reconstructed excitation energy spectrum ($E_x$) of the $^{15}$C system populated via the $^{13}$C($^{18}$O,$^{16}$O) reaction is shown in Figure 3.10. Above the one-neutron separation energy of $^{15}$C, $S_n$ = 1.218 MeV, the neutron decay of the observed resonances was studied using the EDEN array of neutron detectors [75] in coincidence with MAGNEX, as described in Section 2.8. The technique consists in measuring the ion energy spectra (see Figure 3.10) using MAGNEX, then gating on the different peaks of the $^{15}$C excitation energy spectrum ($E_x$) and plotting the corresponding neutron energy spectra ($E_n$) (see Figure 3.11) measured by EDEN in coincidence. An analysis of the neutron decay ratios was also performed in ref. [73]. The resulting neutron emission probability for the whole $^{15}$C spectrum above $S_n$ is 101% ± 8%, which is compatible to the values previously measured. Thanks to the high energy resolution in the $^{15}$C spectrum it was also possible to determine for the first time the neutron emission probability for each excited state, as described in detail in ref. [73].

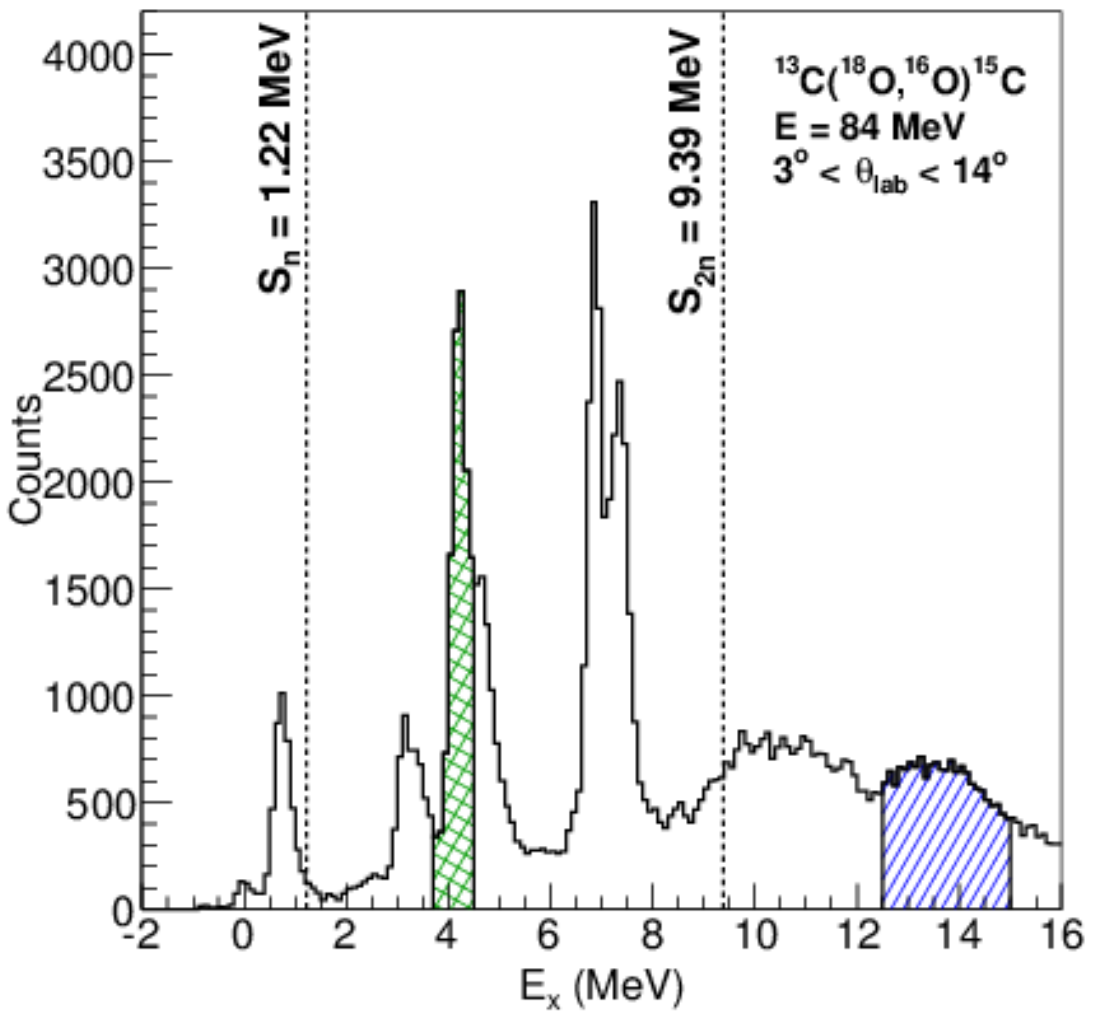

Figure 3.10 $^{15}$C excitation energy spectrum for the $^{13}$C($^{18}$O,$^{16}$O) reaction at 84 MeV incident energy and 3° < $\theta_{lab}$ < 14°. The color-filled areas are two of the different regions selected for the study of the neutron decay spectra.

Considering for example the $^{15}$C excited state at $E_x$ = 4.220 MeV, the neutron energies obtained by gating the excitation energy spectrum in the region 3.8 MeV < $E_x$ < 4.4 MeV, filled in Figure 3.10, is shown in Figure 3.11a. A structure in the neutron spectrum at around 3 MeV appears well visible. Such an energy corresponds, within the uncertainty, to the energy of the neutrons decaying from the $^{15}$C excited state at $E_x$ = 4.220 MeV to the $^{14}$C ground state, which is $E_n = E_x - S_n$ = 3 MeV. This means that such $^{15}$C resonance decays to the $^{14}$C ground state. A branching ratio of 88 ± 16% was measured. The analysis of the other states of $^{15}$C below $S_n$ is reported in ref. [73].

The neutron decay of the $^{15}$C resonance at $E_x$ = 13.7 MeV, associated to the Giant Pairing Vibration (see Section 3.4.1), to the $^{14}$C ground state is shown in Figure 3.11b. The measured neutron energy spectrum rules out the emission of single neutrons which are expected to have energy distributed around $E_n = E_x - S_n$ = 13.7 - 1.22 = 12.5 MeV. The high energy part of the neutron spectrum can be described as the decay to the group of $^{14}$C excited states between $E'_x(^{14}C)$ = 6.094 MeV and $E''_x(^{14}C)$ = 7.341 MeV, which would produce neutrons with energies distributed between 5 and 6.4 MeV. However, the most intense neutron distribution in coincidence with the GPV peak of $^{15}$C can be explained by the decay to the $^{13}$C ground state via a two-neutron emission. In this case, neutron energies ranging from zero to $E_n = E_x - S_{2n}$ = 4.3 MeV are expected and, indeed, observed. Due to the low yield, two-neutron coincidences were not observed in the present experiment. A dedicated run is foreseen as the next step of this research program.

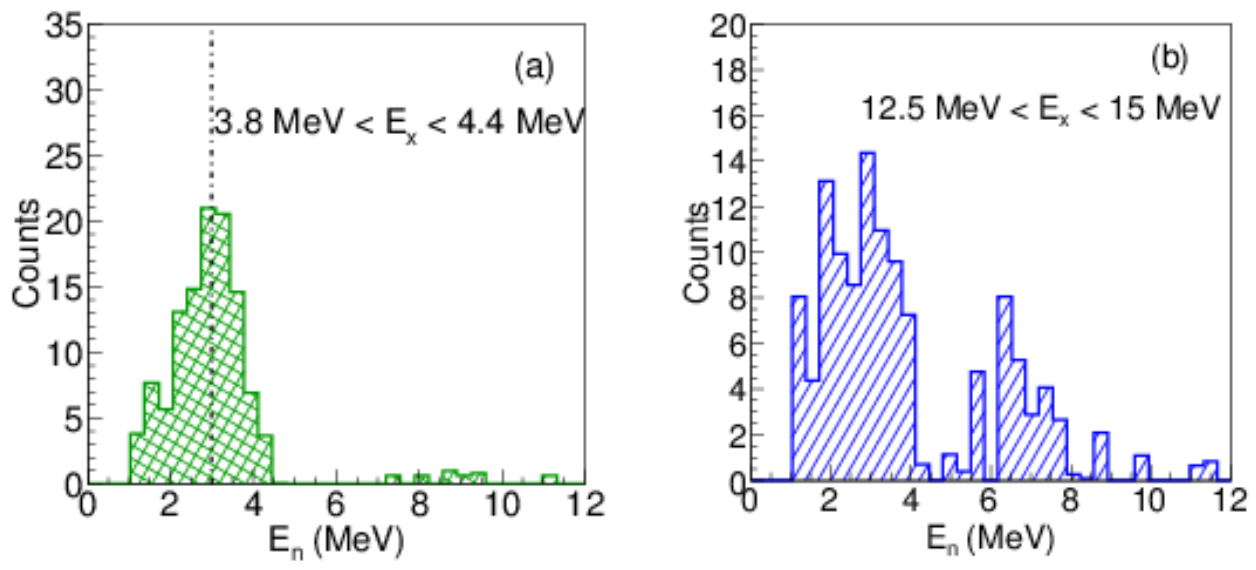

Figure 3.11(a) Neutron energy spectrum gated on the $^{15}$C excitation energy 3.8 MeV < $E_x$ < 4.4 MeV. (b) Neutron energy spectrum gated on the $^{15}$C excitation energy 12.5 MeV < $E_x$ < 15 MeV. In both plots the background neutron spectra are subtracted as described in ref. [73].

### 3.4.3 The $^{120}$Sn(p,t)$^{118}$Sn reaction

A study concerning the search for the GPV in the $^{120}$Sn(p,t)$^{118}$Sn reaction at 35 MeV was carried on with MAGNEX [129]. The experiment was performed using a CS proton beam at 35 MeV impinging on a 2.8 mg/cm$^2$ thick $^{120}$Sn target. The tritons produced in the reaction were detected by MAGNEX. In particular, six different magnetic field settings were used to explore the $^{118}$Sn excitation energy spectrum up to ~16 MeV, and four different angular settings of the spectrometer to measure the angular distributions between 8° and 24° in the laboratory reference frame. The kinetic energy and the scattering angle were obtained by the ray-reconstruction technique and the $^{118}$Sn excitation energy spectrum shown in Figure 3.12 was extracted.

The ground state and several known $^{118}$Sn excited states are populated. A residual component of deuterons, not completely eliminated from the selected tritons, generates the two narrow peaks between 9 and 10 MeV. At higher energies, the presence of a bump over the background is clearly visible in the insert of Figure 3.12.

This structure is located in the energy region where the GPV is expected, and where B. Mouginot et al. [126] found a possible signal of a bump populated by the same $^{120}$Sn(p,t)$^{118}$Sn reaction at 50 MeV bombarding energy. In order to study the observed structure, the spectrum was fitted in the 10 MeV< $E_x$ < 16 MeV region using a Lorentzian function plus a linear background. The result of the best-fit procedure is shown in the insert of Figure 3.12. The parameters extracted from the fit are $E_0 = 13.6 \pm 0.1$ MeV for the centroid and $\Gamma = 1.5 \pm 0.4$ MeV for the width. These values are compatible with those expected for the GPV (70 $A^{-1/3}$) [119]. Further analyses are in progress to extract the angular distributions to better characterize the angular momentum transfer and thus get more insight on the nature of this resonance.

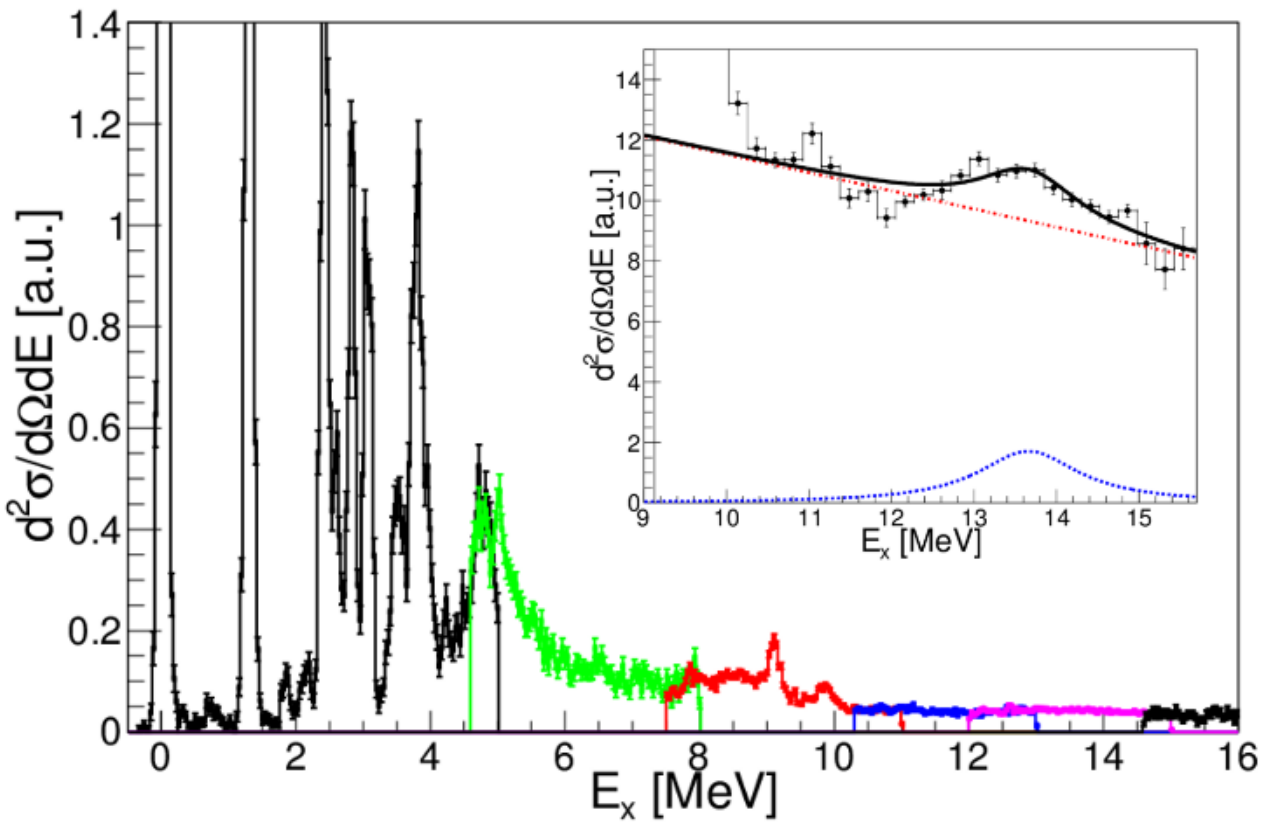

Figure 3.12 Excitation energy spectrum of $^{118}$Sn integrated over the angular range 8° ≤ $\theta_{lab}$ ≤ 12°. The different colors represent different magnetic field settings of the spectrometer. The error bars indicate statistical uncertainties. In the insert, a zoom of the high excitation energy region is shown. The black line is the fit with a Lorentzian (blue line) plus a linear (red line) function.

## 3.5 Gamow-Teller transitions in heavy-ion charge exchange reactions

The Charge-EXchange (CEX) reactions represent a major source of information on isovector excitations in stable and weakly bound nuclei. Beside the interest as spectroscopic tools, studies on CEX reactions like (*p, n*) or (*n, p*) give crucial information on the isovector component of the effective nucleon-nucleon interaction. In particular, two CEX probes were studied by MAGNEX: the ($^7$Li,$^7$Be) and the ($^{18}$O,$^{18}$F) reaction. Some results about these investigations are discussed in the following sections.

### 3.5.1 The ($^7$Li,$^7$Be) reaction

A systematic exploration of the ($^7$Li, $^7$Be) charge-exchange reaction at about 8 MeV/*u* was performed in light neutron rich nuclei, showing that at such incident energy this process proceeds with a considerable predominance of the direct one-step mechanism, thus being a useful probe for spectroscopic studies [130], [131], [89], [132], [133]. In particular, the $^{19}$F($^7$Li, $^7$Be)$^{19}$O reaction at 52 MeV incident energy was studied with MAGNEX. The target was an 80 $\mu$g/cm$^2$ $AlF_3$ foil evaporated on a 250 $\mu$g/cm$^2$ gold backing produced at the chemical laboratory of the LNS. Self-supporting $^{27}$Al, $WO_3$ and carbon target were also used in order to estimate the presence of contaminants in the target compound [134] [52].

A reconstructed spectrum of the $^{19}$O excitation energy measured in the angular range 9.5° < $\theta_{lab}$ < 13.5° is shown in Figure 3.13. An energy resolution of ~ 80 keV FWHM is obtained, mainly limited by the target thickness. Several excited states are identified in the low excitation energy region. Among them, the state at 1.47 MeV has a known structure dominated by a single-particle configuration with one neutron in the $2s_{1/2}$ orbital on a $^{18}$O(0$^+$) core. The transition from the $^{19}$F (g.s.,1/2$^+$) to this state can proceed via 0$^+$ or 1$^+$ spin transfer. The angular distribution for the $^{19}$F($^7$Li, $^7$Be)$^{19}$O$_{1.47}$ transition was also measured as shown in Figure 3.14. A theoretical analysis of this transition was done by DWBA calculations with QRPA-derived transition densities [134] [52]. The results show a good agreement with the experimental data, both in the shape of the angular distribution and in the magnitude of the cross section, without introducing any renormalization factor as shown in Figure 3.14. This result resembles those found for the $^{11}$B($^7$Li, $^7$Be)$^{11}$Be and $^{15}$N($^7$Li, $^7$Be)$^{15}$C reactions in similar conditions, thus supporting the one-step hypothesis for the reaction mechanism [130], [133].

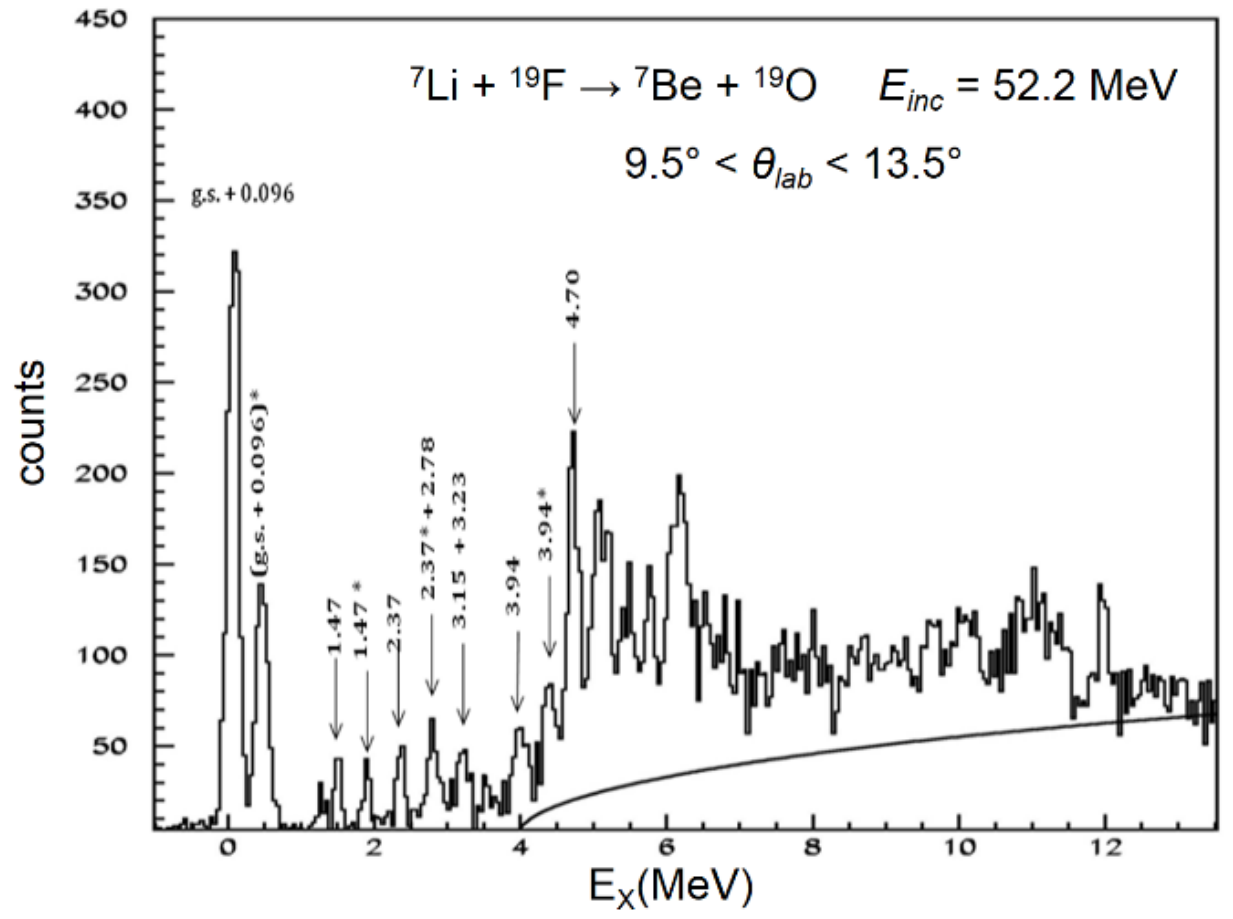

Figure 3.13 $^{19}$O excitation energy spectrum from the $^{19}$F($^7$Li, $^7$Be)$^{19}$O reaction at 52 MeV. The solid line is the estimated non-resonant three-body phase space $^7$Li + $^{19}$F → $^7$Be + $^{18}$O + n. The known $^{19}$O states are indicated with their energy. Peaks marked with an asterisk refer to the transitions in which the $^7$Be ejectiles are in the first excited state at 0.43 MeV.

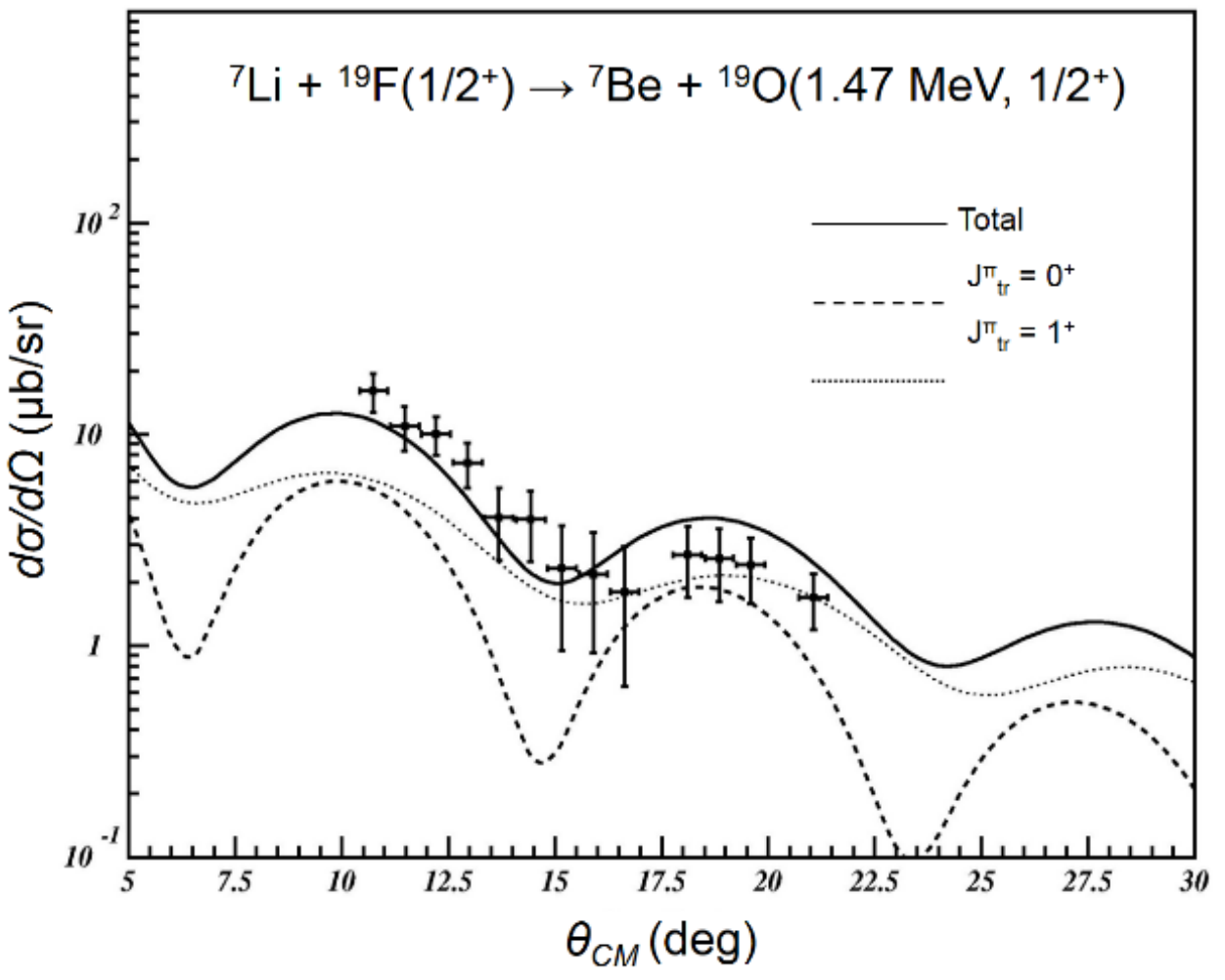

Figure 3.14 Angular distribution for the $^{19}$F($^7$Li, $^7$Be)$^{19}$O$_{1.47}$ transition. The DWBA calculations (the total and the two contributions due to the 0$^+$ and 1$^+$ target transitions) are shown separately. The calculated cross-sections are not scaled. From [134].

## 3.5.2 The ($^{18}O$,$^{18}F$) reaction

The $^{40}Ca(^{18}O,^{18}F)^{40}K$ CEX reaction was recently measured with MAGNEX by using a beam of $^{18}O^{4+}$ ions at 270 MeV incident energy on a 279 μg/cm$^2$ Ca target. An angular range of $+2.5° < \theta_{lab} < +10°$ in the laboratory frame was explored. Figure 3.15 shows an example of the measured energy spectrum.

Some structures are observed below 5 MeV excitation energy; however, the limited resolution (~ 500 keV) and the high level density do not allow to isolate single transitions. The strongest group is between 500 keV and 1.2 MeV where the transitions to the known $2^-$ and $5^-$ states of $^{40}K$ at 0.800 and 0.892 MeV and those to the excited states of the $^{18}F$ ejectiles at 0.937, 1.041, 1.080 and 1.121 MeV, if populated, are not resolved [135], [136].

In particular, the dominance of the excited states of $^{18}F$ at 1.041, 1.080 and 1.121 MeV is ruled out by a least square analysis, considering that they will generate Doppler broadened peaks with an extra width of about 300 keV. A number of $^{40}K$ states are known in the region between 1.8 and 2.8 MeV. Calculations based on the QRPA – Distorted Wave Born Approximation of ref. [130] indicate that the cross-section is mainly distributed among the $4^-$, $2^-$, $1^+$ and $3^-$ transitions. In particular the $1^+$ accounts for about 40 μb/sr, consistent with the 38 μb/sr extracted from literature.

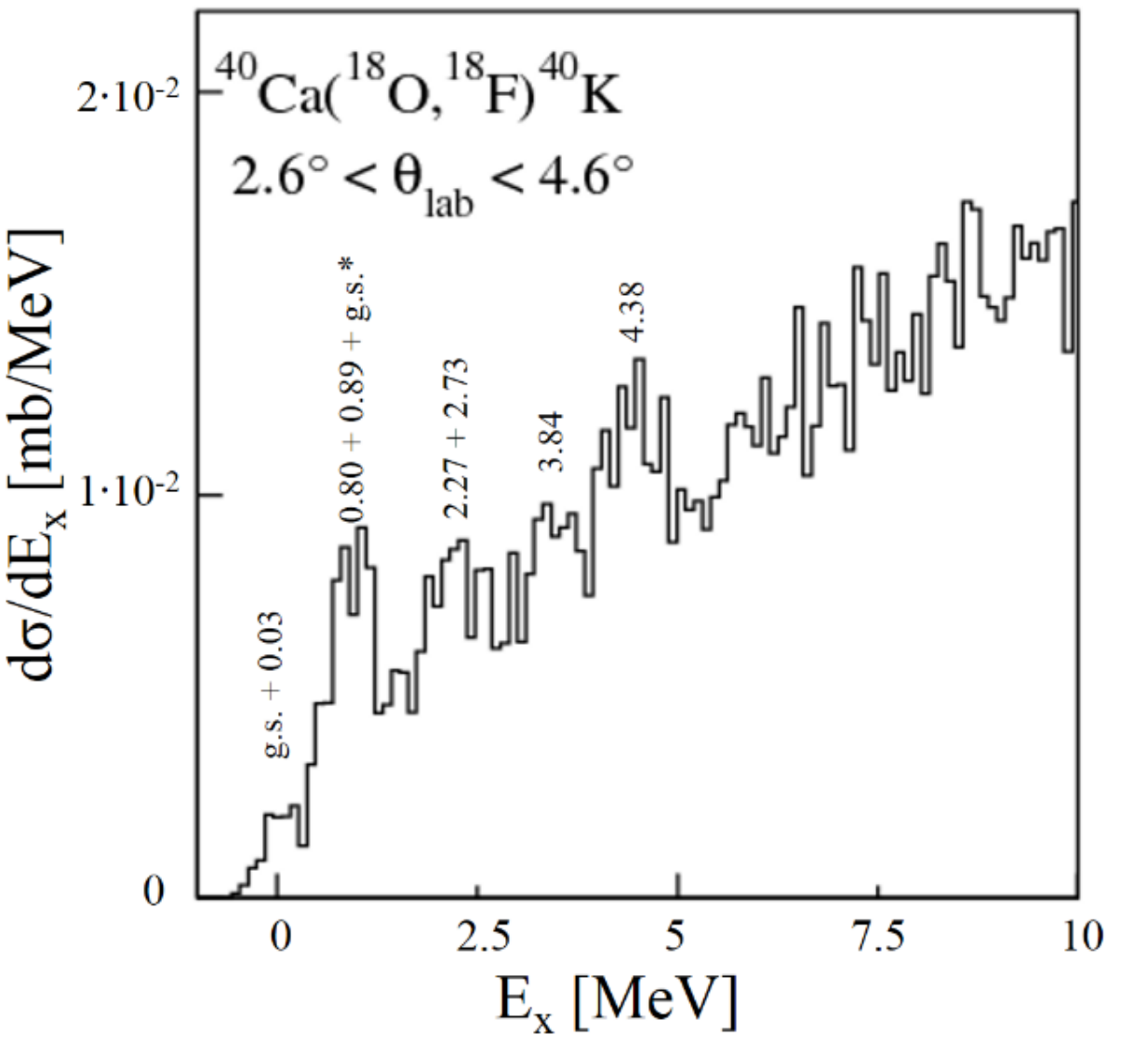

Figure 3.15 Energy spectrum measured by $^{40}Ca(^{18}O,^{18}F)^{40}K$ CEX reaction at 270 MeV. The symbol g.s.* indicates the $^{40}Ca(^{18}O,^{18}F_{0.937})^{40}K_{g.s.}$ transition.

A proportionality relation between β-decay strengths and charge exchange reaction cross sections is known to be valid for linear momentum transfer $q$ ~ 0 and under specific conditions [137], [138], [139], [140] for the reaction. This proportionality can be described by the following relation:

$$\frac{d\sigma}{d\Omega}(q,E_x)=\hat{\sigma}_\alpha\left(E_p,A\right)F_\alpha(q,E_x)B_T(\alpha)B_P(\alpha) \qquad (3.1)$$

where $B_T(\alpha)$ and $B_P(\alpha)$ are the target and projectile β-decay transition strengths (related to the nuclear matrix elements $M(\alpha)$) for the $\alpha$ = Fermi ($F$) or Gamow-Teller ($GT$) operators. The factor $F_\alpha(q,E_x)$ describes the shape of the cross section distribution as a function of the linear momentum transfer $q$ and the excitation energy $E_x$. For $L = 0$ transitions, it depends on the square of the $j_0(qr)$ spherical Bessel function [138], [137]. The quantity $\hat{\sigma}_\alpha$, named "unit cross section", is of primary interest since it almost behaves as a universal property of the nuclear response to $F$ and $GT$ probes. In a rigorous Distorted Wave approach as that proposed by Taddeucci et al. [138], the unit cross section for a CEX process is factorized as:

$$\hat{\sigma}\left(E_p,A\right)=K(E_p,0)|J_\alpha|^2N_\alpha \qquad (3.2)$$

where $K(E_p,E_x)$ is a kinematic factor, $J_\alpha$ is the volume integral of the effective isovector nucleon-nucleon interaction and $N_\alpha$ expresses the distortion of the incoming and outcoming waves in the scattering [140].

Eqs. (3.1)-(3.2) are routinely used for accurate (within few percent) determination of the strengths $B$ in light-ion induced reactions such as (n,p), (p,n), ($^3$He,t), (t,$^3$He), (d,$^2$He) [141]. For heavy-ion induced reactions, the data analyses are typically more involved, due to the projectile degrees of freedom and the sizeable amount of momentum transfer. In addition, the contribution of sequential nucleon exchange to CEX cross section should also be considered. A relevant simplification comes from the strong absorption of the scattering waves in the inner part of the colliding systems and the resulting surface localization of such reactions. As a consequence in these cases, the use of fully consistent microscopic approaches with double folded potentials for the reaction form factors still allows the determination of $B(\alpha)$ within 10-20% [130].

Under the hypothesis of the cross section factorization of eq. (3.1), in the case of the $^{40}Ca(^{18}O,^{18}F)^{40}K(1^+)$ transition, $B_{GT}$ = 0.087 is extracted applying eqs. (3.1)-(3.2). This number should be compared to the value obtained by ($^3$He,t) reactions $B_{GT}$ = 0.083. The closeness of these numbers is a hint of the validity of the proportionality relation also for ($^{18}O$,$^{18}F$) CEX reactions.

## 3.6 Double-charge exchange reactions in connection with 0*νββ* decay

An exploratory measurement of a double charge-exchange (DCE) reaction was performed using MAGNEX. In particular the $^{40}Ca(^{18}O,^{18}Ne)^{40}Ar$ DCE reaction was investigated using a beam of $^{18}O^{4+}$ ions at 270 MeV and a 279 μg/cm$^2$ Ca target. The scientific interest in the study of such reactions is related to their possible connection to the neutrinoless double beta decay ($0\nu\beta\beta$) Nuclear Matrix Element (NME), as discussed in Section 4.1.1.

Thanks to the high resolution achieved in the DCE energy spectrum (500 keV FWHM), the $^{40}Ar$ ground state is clearly separated from the not resolved doublet of states $^{40}Ar$ $2^+$ at 1.460 MeV and $^{18}Ne$ $2^+$ at 1.887 MeV (see Figure 3.16). At higher excitation energy, the measured yield is spread over many overlapping states.

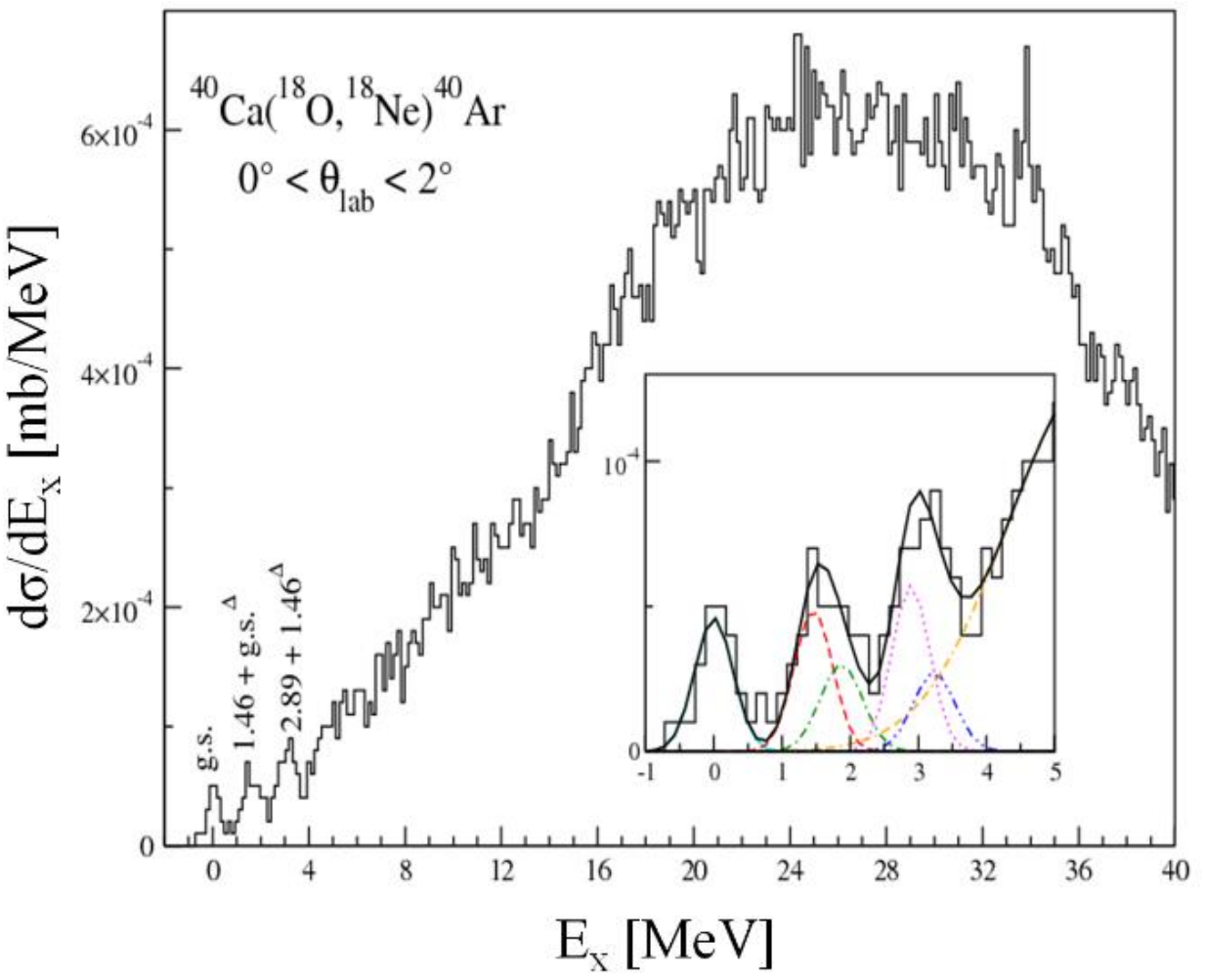

Figure 3.16 Energy spectrum from $^{40}$Ca($^{18}$O,$^{18}$Ne)$^{40}$Ar DCE. The symbols g.s.$^{\Delta}$ and 1.46$^{\Delta}$ indicate the $^{40}$Ca($^{18}$O,$^{18}$Ne$_{1.87}$)$^{40}$Ar$_{g.s.}$ and $^{40}$Ca($^{18}$O,$^{18}$Ne$_{1.87}$)$^{40}$Ar$_{1.46}$ transitions, respectively. In the insert, a zoomed view of the low-lying states and, superimposed (black solid line), a fit with 6 Gaussian functions are shown.

The angular distribution for the transition to the $^{40}$Ar $0^+$ ground state is shown in Figure 3.17. A clear oscillating pattern is observed. The position of the minima is described by a $|j_0(qR)|^2$ Bessel function, where $R = 1.4\ (A_1^{1/3}+A_2^{1/3})$ and $A_{1,2}$ is the mass number of projectile and target. Such an oscillating pattern is not expected in complex multistep transfer reactions, due to the large number of angular momenta involved in the intermediate channels that would determine a structure-less cross section slowly decreasing at larger angles. The experimental slope is shallower than the Bessel function as expected since a plane-wave description is not appropriate [142]. Despite that, a very simple model of $L = 0$ direct process reasonably well describes the main features of the experimental angular distribution.

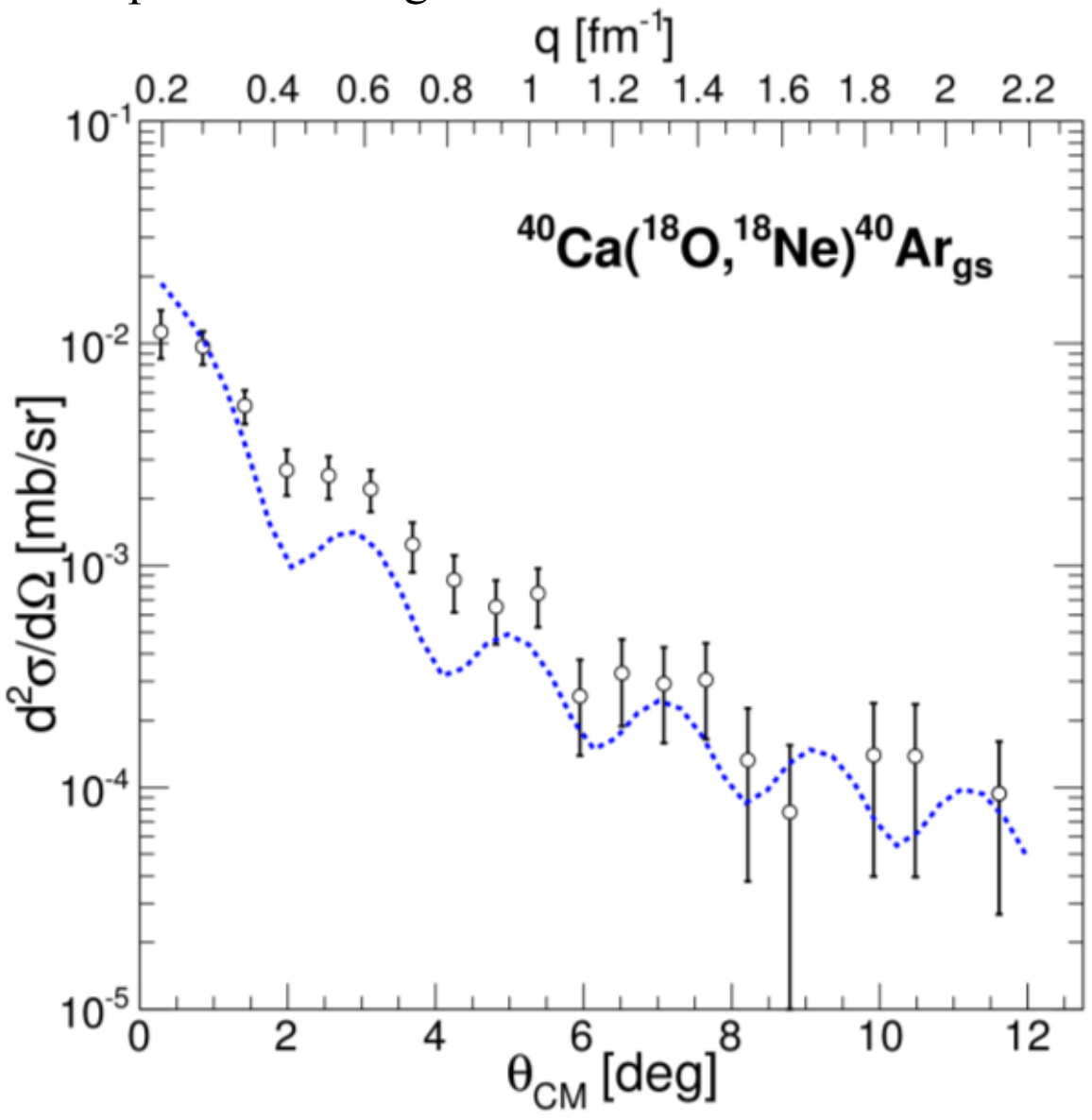

Figure 3.17 Differential cross section of the $^{40}$Ca($^{18}$O,$^{18}$Ne)$^{40}$Ar$_{g.s.}$ transition as a function of $\theta_{CM}$ and $q$. The error bars include a statistical contribution and a component due to the solid angle determination. The blue curve represents the $L = 0$ Bessel function folded with the experimental angular resolution (~ 0.6°) and scaled to reproduce the incoherent sum of the predicted double *F* and *GT* cross sections.

The $L = 0$ Bessel function shown in Figure 3.17 is scaled to give a cross section of 21 μb/sr at $\theta_{CM} = 0°$, which is the incoherent sum of the predicted Gamow-Teller (*GT*) and Fermi (*F*) cross sections extracted in ref. [143]. The comparison with the experimental data show a remarkable quantitative agreement. However, the effects of the interference should be studied in detail. The fact that both pure *GT*- and *F*-like extreme models give comparable contributions to the final cross section is a direct consequence of the similar volume integrals for both operators. The relation between these volume integrals resembles that for nucleon-nucleon interaction at 15 MeV. This indicates that the reaction mechanism is largely determined by the effective nucleon-nucleon interaction.

Assuming a proportionality relation between DCE cross-section and NME, in analogy to single $\beta$-decay, the NME for the transition $^{40}$Ca$_{gs}$ to $^{40}$Ar$_{gs}$ was extracted. More details on the approximations used are published in Ref. [143]. Under the hypothesis that the considered transition is pure *GT*, the extracted NME results $M^{DCE}(GT) = 0.42 \pm 0.21$. Analogously, in the case of pure Fermi process, we extract $M^{DCE}(F) = 0.28 \pm 0.14$. The systematic error is about ±50%. It mainly comes from the uncertainty on volume integrals of the effective two-body interaction. The contribution of the experimental error (±10% systematic, ±25% statistical) is less relevant in this case. We notice that the NME in the case of pure Fermi or pure GT transitions are very similar, so even the weighted average, representing a more realistic combination of both contribution, will be. Assuming the known GT and F strengths from literature, an estimate of the weights can be deduced and the matrix element for the $0\nu\beta\beta$ decay of $^{40}$Ca can be inferred: $M^{DCE}(^{40}\mathrm{Ca}) = 0.37 \pm 0.18$.

To speculate, a comparison between the present result for $^{40}$Ca and the NME of $0\nu\beta\beta$ decay of $^{48}$Ca can be done assuming pure F and GT and artificially removing the effect of the Pauli blocking, since the same single particle shells are involved but no Pauli blocking is active in the $^{48}$Ca case. In $^{40}$Ca indeed the transitions take place only through the small $1f_{7/2}$ $1f_{5/2}$ particle and $1d_{3/2}$ hole components of the $^{40}$Ca$_{gs}$ wave function, which account for about 14% of the total. So, it is possible to approximately deduce the $^{48}$Ca$_{gs}$ NME by just multiplying the $^{40}$Ca NME for a factor 7, obtaining $M^{DCE}(^{40}\mathrm{Ca}) = 2.6 \pm 1.3$. It is noteworthy that this number is compatible with literature for the state of the art calculations of the $^{48}$Ca $0\nu\beta\beta$ NMEs.

This pilot work showed for the first time high resolution and statistically significant experimental data on heavy-ion DCE reactions in a wide range of transferred momenta. Strengths factors and NMEs are extracted under the hypothesis of a two-step charge exchange process. Despite the approximations used in this model, which determine an uncertainty of ±50%, the result is compatible with the values known from literature, signaling that the main physics content has been kept. The DCE unit cross section is likely to be a predictable quantity, in analogy to the CEX processes. We believe that this finding is mainly due to the particularly simple transitions which take place in the $^{18}$O → $^{18}$F → $^{18}$Ne, characterized by a strong dominance of single $1^+$ and $0^+$ $^{18}$F states in both *GT* and *F* transitions, respectively. This makes the ($^{18}$O,$^{18}$Ne) reaction very interesting to investigate the DCE response of the nuclei involved in $0\nu\beta\beta$ research.

Experiments at different incident energies are envisaged in order to explore conditions characterized by different weights of *GT*-like and *F*-like contributions and disentangle the role of each operator. In all cases, the contextual measurements of the multi-nucleon transfer and single charge exchange channels are mandatory. A rigorous treatment of the process in a multistep direct reaction quantum-mechanical framework will be the next step toward a more accurate determination of 0νββ NMEs.

# 4 Future perspectives

As it was shown in the previous Sections, MAGNEX was designed to study different processes, mainly characterized by very small yields, in different fields of nuclear physics, ranging from nuclear structure to the characterization of reaction mechanisms in a broad range of energies and masses. The MAGNEX optical properties have permitted to achieve scientific results not addressed before. In addition, the use of the spectrometer can be extended even to research fields traditionally beyond heavy-ion physics with accelerators. A relevant case is that of neutrinoless double beta decay with its fundamental implications, in which an original contribution can be given by facilities like MAGNEX. However, in this scenario a much higher beam luminosity is requested, which represents a challenge for present accelerators and detection technologies for heavy ions. This is the new frontier for most of the future research programs with MAGNEX and thus a substantial change in the technologies used is necessary.

In the following an overview of the scientific activities that are in program with MAGNEX and the main items of the foreseen upgrade are summarized.

## 4.1 Scientific perspectives

### 4.1.1 Challenges in the investigation of double charge-exchange nuclear reactions towards neutrinoless double beta decay: the NUMEN project at LNS

Neutrinoless double beta decay (0*νββ*) is at the present time strongly pursued both experimentally and theoretically [144]. Its experimental observation would determine whether the neutrino is a Dirac or Majorana particle and would provide a measurement of the average neutrino mass, which nowadays is one of the most fundamental problems in physics. However, an accurate determination of nuclear matrix elements entering in the expression of the lifetime of the 0*νββ* is fundamental to this purpose. An innovative technique to access the NMEs by relevant cross sections of double charge exchange reactions has been proposed. The basic point is the coincidence of the initial and final state wave-functions in the two classes of processes and the similarity of the transition operators, which in both cases present a superposition of Fermi, Gamow-Teller and rank-two tensor components with a relevant implicit momentum transfer. First pioneering experimental results for the $^{40}$Ca($^{18}$O,$^{18}$Ne)$^{40}$Ar reaction, described in Section 3.6, give encouraging indication on the capability of the proposed technique to access relevant quantitative information towards the determination of the NMEs for 0*νββ* decay [143].

On the basis of the above mentioned ground-breaking achievement, the ambitious NUMEN (NUclear Matrix Elements of Neutrinoless double beta decays) project has been proposed with the aim to go deep insight in the heavy-ion double charge exchange studies on nuclei of interest in 0*νββ* decay, looking forward at the 0*νββ* NME determination [145], [146], [147], [148].

The availability of MAGNEX for high-resolution measurements of very suppressed reaction channels was essential for the first pilot experiment [143] (see Section 3.6). However, with the present set-up it is difficult to suitably extend this research to nuclei which are candidates for 0*νββ* decay, where studies are and will be concentrated ("hot" cases) [144]. The present limit of low beam current both for the CS and for the MAGNEX focal plane detector (see Table 2.3) must be sensibly overcome. For a systematic study of the many "hot" cases of *ββ* decays an upgraded set-up, able to work with two orders of magnitude more current than the present, is thus necessary. This goal can be achieved by a substantial change in the technologies used both in the beam extraction and in the detection of the ejectiles. For the accelerator the use of a stripper induced extraction is an adequate choice. The main foreseen upgrades for MAGNEX are:

1. The substitution of the present focal plane detector [53] gas tracker with a new one based on micropattern gas detectors;
2. The substitution of the wall of silicon pad stopping detectors with a wall of telescopes based on SiC detectors;
3. The enhancement of the maximum magnetic rigidity;
4. The introduction of an array of detectors for measuring the coincident γ-rays around the target.

In this framework, the NUMEN project is structured in four phases:

*Phase1: the experiment feasibility*

The pilot experiment $^{40}$Ca($^{18}$O,$^{18}$Ne)$^{40}$Ar reaction at 270 MeV, with the first experimental data on heavy-ion double charge-exchange reactions in a wide range of transferred momenta, which was already done. The results demonstrated the technique feasibility (see Section 3.6).

*Phase2: toward "hot" cases*

The necessary work for the upgrading of both the CS and MAGNEX, briefly described in Section 4.2, will be carried out still preserving the access to the present facility. Due to the relevant technological challenges connected, the Phase2 is foreseen to last 3-4 years. In the meanwhile, experiments with integrated charge of tens of mC (about one order of magnitude more than that collected in the pilot experiment) will be performed. These will require several weeks (4-8 depending on the case) of data taking for each reaction, since thin targets (a few $10^{18}$ atoms/cm$^2$) are mandatory in order to achieve enough energy and angular resolution in the energy spectra and cross section angular distributions. Only few favourable cases will be explored, with the goal to achieve conclusive results for them.

*Phase3: the facility upgrade*

Once all the building block for the upgrade of the accelerator and spectrometer facilities will be ready at the LNS, a Phase3, connected to the disassembling of the old set-up and re-assembling of the new will start. An estimate of about 24 months is considered.

*Phase4: the experimental campaign*

The Phase4 will consist of a series of experimental campaigns at high beam intensities (some pμA) in order to reach, in each experiment, integrated charge of hundreds of mC up to C, for the experiments in coincidences, spanning all the variety of candidate isotopes for $0\nu\beta\beta$ decay.

Once selected the optimal experimental conditions for the different cases in the Phase2, the Phase4 will be devoted to collect data using the upgraded facilities. The purpose is to give a rigorous determination of the absolute cross sections for all the system of interest. New theoretical analysis will allow to extract from these data accurate NMEs for the $0\nu\beta\beta$ that is the most ambitious goal of NUMEN project.

### 4.1.2 Exploring the Breathing Mode in Exotic Nuclei

Giant resonances have proven to be a valuable source of information on both nuclear structure and nuclear matter. In particular, the isoscalar Giant Monopole Resonance (GMR), also called the breathing mode of nuclei, is of major importance because its properties are related to the nuclear matter incompressibility $K_\infty$. The latter is a fundamental ingredient in defining the equation of state (EOS) for nuclear matter and is a basic parameter in the description of neutron stars and supernovae [149].

The question about the proper value of the nuclear incompressibility is still open. Part of the existing uncertainty in $K_\infty$ is due to the poor knowledge of the symmetry energy [150]. The asymmetry term in the expansion of the nuclear incompressibility $K_\tau$ is poorly determined, since its determination requires studies of the compression modes in an isotopic chain spanning a wide range of $(N - Z)/A$ values. This was recently done for the stable Sn isotopes and a value of $K_\tau$ = -550 ± 100 MeV was obtained [151]. In order to determine $K_\tau$ with more accuracy and to probe the role of the symmetry energy in the nuclear incompressibility, extending the GMR measurement to unstable nuclei is mandatory.

Although extremely appealing, the GMR measurement in unstable nuclei remains a major experimental challenge due to the low radioactive beam intensities and the unfavourable conditions in inverse kinematics. Indeed, the GMR cross-section peaks at 0° in the centre of mass reference frame, resulting in very low recoil velocities for the light isoscalar probes, i.e. deuterons, alpha particles and $^{6}$Li [152].

The small kinetic energy of the recoils implies the need to use thin targets. An interesting future technique that is proposed at FAIR-GSI consists in the use of a helium-gas-jet target in a storage ring [153]. The drawback of a thin target can thereby be compensated in a storage ring by the revolution frequency of the ions.

A 120 mg/cm$^2$ thick liquid helium target was developed at RIKEN and used to excite $^{14}$O in inverse kinematics at a beam energy of 60 AMeV and a beam intensity of 50 kHz [154]. The isoscalar monopole and dipole compressional strength were extracted but a clear resonance peak was not observed. The authors suggested, as a possible explanation, the fragmented nature of the compressional strength distribution typical of light stable nuclei.

Another approach was applied at GANIL using the active target MAYA to measure deuteron scattering of $^{56}$Ni [155]. However, it was not possible to clearly disentangle, in the excitation energy spectrum, between the monopole and the quadrupole resonances. Indeed, the impossibility for the MAYA active target to measure the recoil at 0° made the observation of the GMR less clear. More recently, with the same technique, the isoscalar monopole response was measured in the unstable nucleus $^{68}$Ni, using inelastic alpha scattering at 50 AMeV in inverse kinematics [156].

The In-Flight FRIBS facility at the LNS [157] produces radioactive ion beams at Fermi energies with enough intensity to successfully perform nuclear physics experiments [158]. Moreover, the capability MAGNEX to perform experiments with beams accelerated by the CS at 0° is established (see Section 2.7).

Based on such considerations, we propose to measure the isoscalar GMR in exotic nuclei via deuteron inelastic scattering, measuring the recoiling deuterons emitted at around 0° by MAGNEX. To explore the feasibility of such method, we propose to investigate in this first experiment the unstable nuclei in the $^{38}$S region, since they were already produced at LNS in the past [157].

This could be the first of a series of isoscalar GMR studies on unstable nuclei that certainly will benefit of the planned upgrade of the LNS CS to higher beam intensities, as mentioned in Section 4.1.1.

### 4.1.3 The isoscalar monopole resonance of the alpha particle

A recent ab-initio calculation of the monopole transition form factor of $^{4}$He [159] pointed to a strong dependence on the different realistic potentials used. Moreover, a large disagreement (almost a factor of two) with respect to existing electron scattering data was found when using modern three-body Hamiltonians from chiral perturbation theory. Experimental studies were mainly focused on $(e,e')$ scattering, which did not guaranteed enough accuracy in the extraction of the form factors to be conclusive.

In this framework, a new project has been proposed. The purpose is to measure the $^{4}$He($^{4}$He,$^{4}$He)$^{4}$He* inelastic scattering cross section with MAGNEX in the region of the first $0^+$ excited state at ~20.4 MeV. Compared to the $(e,e')$ probe, the ($^{4}$He,$^{4}$He) presents two basic advantages, the cross section is sensibly larger and pure isoscalar ($T$=0) spin scalar ($S$=0) modes are excited from the nuclear interaction. The Coulomb interaction, which is the only one determining the electron scattering, does not filter isoscalar modes from isovector ($T$=1) which generates a relevant background in the spectra underneath the $0^+$ resonance. The Coulomb as well as the isovector nuclear interaction are negligible in the ($^{4}$He,$^{4}$He) at the energies of about 10-20 MeV, which is thus well suited for the study of pure isoscalar transitions.

On the other hand, the ($^{4}$He,$^{4}$He) cross section gives an indirect access to the form factor, being it folded with distorted waves in the transition amplitude integrals. However, modern techniques of nuclear reaction mechanism, based on the distorted waves formalism, allow to reliably disentangle the form factors from the measured cross sections by the use of double folding optical potentials.

The plan is to produce data of high quality in terms of energy and angular resolution as well as in term of signal to noise ratio thanks to the use of MAGNEX for the detection of the $^{4}$He ejectiles. The cross sections will be measured in absolute value and used to extract the form factors from an optical model analysis. In particular, the distorted waves will be obtained by the use of the double folding San Paulo potential [160], [81]. Coupled Channel (CC) calculations will guarantee an adequate accuracy of the extracted form factor to allow a comparison with the ab-initio calculations. A 20% accuracy is at least expected. Conversely, the calculated ab-initio monopole form factor can be used as an external input for CC calculations of the differential cross section. In this way the analysis of the shape of the experimental angular distribution can also signal the presence of non- monopole contributions in the 20.4 MeV resonance region. The extraction of the position and width of the $0^+$ resonance with better accuracy with respect to the previous ($^4$He,$^4$He) measurements [161], [162] is also planned, to hopefully shed light on the disagreement with the extractions from (*e*,*e'*) data [163]. Precise data will be useful to the ab-initio theory community, in view of a better understanding and constraining of the nuclear Hamiltonians.

### 4.1.4 Challenges in nuclear physics cluster states studies

Recently, R. Bijker and F. Iachello [164] showed the evidence of tetrahedral symmetry in $^{16}$O. In their work, the authors show that the low-lying states of $^{16}$O can be described as the rotation-vibration of a 4α cluster with tetrahedral symmetry. In particular, the proposed model predicts very peculiar rotational bands as experimental signature of the underlying cluster symmetry. Moreover, studies connected with cluster degrees of freedom showed that the low-lying states of $^{12}$C could be described as the rotation-vibration of a 3α cluster with equilateral triangle symmetry [165], [166] [167]. These results emphasize the occurrence of α-cluster states in light nuclei and stimulate further experimental work on the structure of $^{16}$O.

From the experimental point of view, an accurate determination of energy, width, spin and parity of the populated states is a key requirement to test the theoretical predictions. Interesting probes for this kind of investigations are $\alpha$-transfer reactions, as for example ($^6$Li,*d*) or ($^7$Li,*t*), or ($\alpha$,$\alpha'$) inelastic scattering, due to their selectivity of $\alpha$-cluster configurations. For $\alpha$-transfer reactions, for example, angular correlation measurements of the ejectile with a spectrometer at zero degrees and $\alpha$ with a separate detector in coincidence allows to extract unambiguous information on spin and parity [168].

A research program is foreseen at LNS in order to investigate this topic of great interest, using MAGNEX that is an ideal tool for these studies.

## 4.2 Technological perspectives

A manifold upgrade of the MAGNEX spectrometer is mandatory in order to face with the challenging experimental conditions foreseen by the future projects, as discussed in Section 4.1.

### 4.2.1 A new gas tracker

The first direct consequence of the beam current increase, foreseen for the NUMEN project (see Section 4.1.1), is the need of a specially tailored tracker at the MAGNEX focal plane. The present FPD gas tracker, based on a series of drift chambers and on the use of long multiplication wires (see Section 2.3) is intrinsically limited to about 5 kHz (see Table 2.3). This limit can be overcome substituting the multiplication wires with a series of micropattern gas detectors, such as GEM foils, or equivalently THGEM or Micromegas [169] without changing the geometry of the drift sections of the detector. The GEM technology, for example, is extremely fast since the primary electrons are multiplied in a series of many independent holes and since a mesh can be mounted very close to the multiplication planes. In recent developments, detectors based on this technology were proved to work up to several Mhz/mm$^2$, i.e beyond the rates expected in NUMEN. In addition, analog signals, preserving the information of the charge generated by the particle track and the crossing time, can be processed by the read-out of segmented electrodes downstream of the multiplication foils. This allows to reconstruct with sub-millimetric resolution the tracks of the primary electrons and consequently that of the impinging particles. However, large part of the R&D studies on these technologies deal with the applications at atmospheric pressure and beyond, features not available for the NUMEN project, where the ideal working pressure for guaranteeing a high energy resolution in MAGNEX, is about 10 mbar. In addition, most of the experience is in using these detectors at energies where all particles behave as minimum ionizing particles. In the cases foreseen for NUMEN, ions with rather different ionizing power will be detected, thus requiring a broad dynamic range.

The development of suitable technologies for the construction of a GEM-based tracker, working at low pressure and wide dynamic range, will be a key issue of the R&D activity for MAGNEX in the next years.

### 4.2.2 Ion Identification

NUMEN Phase 2 will also investigate promising technologies for stopping detectors, which need also to be upgraded in view of the high detection rate. Standard technologies, based on silicon pad detectors or plastic scintillators, require a high degree of segmentation (and thus high costs) in order to avoid double-hit events. At the beam currents expected for NUMEN, the probability of a double hit at the focal plane is considerable starting for 5 cm$^2$ area detectors for ($^{18}$O,$^{18}$Ne) reaction at 0°. In addition, the radiation hardness of such devices is not enough to avoid a short lifetime of these detectors. For example, in the same ($^{18}$O,$^{18}$Ne) reaction the rate limit of about $10^8$ ions/cm$^2$, above which a silicon detector starts to deteriorate, is reached in a few days or even less. Interesting opportunities arise from the new technology of SiC crystals, which preserves many of the good properties of silicon detectors, but are much harder to radiation [170] [171]. Improvements in epitaxial SiC growth means that semi-insulating epitaxial SiC layers have recently become available, with thicknesses up to 100 μm. However, R&D is still necessary to explore the possibility to build a reliable number of detectors for heavy ions by these epitaxial SiC. In addition, SiC detectors up to 1 mm thickness can be in principle constructed on semi-insulating SiC technology.

Starting from these considerations, the exploration, characterization and building, after the new tracker, of a wall of telescopes based on thin epitaxial SiC for energy loss followed by thick SiC or CsI detectors for residual energy will be an important part of the R&D. This solution looks like to be promising because it decouples the tracker from particle identification (PID) function and it is based on existing SiC technology, even if not yet implemented in commercial large area detectors. A SiC-SiC or SiC-CsI telescope module will be assembled including the basic pieces of the read-out electronics and the PID sensitivity studied under heavy ion beams. The expertise will be then spent to design and construct the final PID-wall for MAGNEX.

### 4.2.3 Exclusive measurements

An array of scintillators will also be studied within NUMEN Phase2 (see Section 4.1.1). These detectors are intended for detecting γ-rays from the de-excitation of the residual nucleus (and ejectile) in coincidence with the spectrometer, thus improving the resolution in the energy spectra. For example, going to 50 MeV/u the energy resolution for heavy ions can be rather poor (1/1000 of 1 GeV $^{18}$Ne corresponds to about 1 MeV, or even worse if the beam finite resolution and energy straggling at the target are considered). In this sense, the use of an array of detectors for γ-rays is mandatory for DCE reactions. Similarly to the focal plane, the challenge here is to work in a very intense flux of γ-rays and neutrons (up to ~200 kHz) produced also by the interaction of the beam with the target. This implies a good energy resolution in order to optimize the signal-to-noise ratio and reduce the probability of spurious coincidences. Interesting options as the $LaBr_3$(Ce) [172] [173] or CsI [174] will be studied in detail.

The strategy is to build small prototypes for different detector materials and, after a characterization with radioactive sources, to use them under realistic experimental conditions (intense beams, coincidence with MAGNEX, study of DCE). The results of these tests and the consequent choice of the "best" high-flux technology for γ-rays and neutrons will be an important delivery of NUMEN Phase2, which will conduct to the design of the final detector assembly.

# Conclusions

Since the very first conceptual idea, the MAGNEX spectrometer was conceived as a multipurpose device for the detection of reaction products emitted in a broad range of energies, masses and angles. After the commissioning in 2007, it has been used in several experimental campaigns, both in stand-alone and coincidence configurations, dealing with a large variety of topics of modern research in nuclear physics.

As a distinctive feature, MAGNEX fully implements the powerful technique of trajectory reconstruction, based on the algorithms of differential algebra, a careful interpolation of measured fields and a focal plane detector with tracking and particle identification capabilities. The high energy, mass and angle resolution, achieved by the spectrometer, has allowed to explore, with unprecedented sensitivity, important aspects of the nuclear structure and reaction mechanisms, mainly at bombarding energies above the Coulomb barrier. The precise control of transport efficiency and the consequent capability to measure accurate absolute cross sections have also been key technical features of MAGNEX. More recently, the spectrometer has been used for successful experiments at very forward angles, including zero degree, further expanding its discovery potential.

Studies of elastic and inelastic scattering have demonstrated the crucial role of the coupling between different reaction channels for a proper description of the measured cross sections. In particular, deviations from a pure optical model picture were observed in the scattering of $^{16}$O projectiles on $^{27}$Al targets at 100 and 280 MeV [76]. For lighter systems, the analyses of $^{6}$Li + p scattering data in inverse kinematics, at energies around the Coulomb barrier, has proven the major role of the break up channel of the weakly bound projectile in the elastic cross section [86]. More studies are on the way in this research program with new experiments already performed and planned for the future.

Detailed studies of transfer reactions, induced by $^{18}$O, $^{6}$Li and proton beams at different incident energies, have allowed to extract conclusive information on single-particle, pairing and cluster degrees of freedom in the description of nuclear states. Among the other, a special mention is given to the discovery of robust signatures of the long searched Giant Pairing Vibration in $^{14}$C and $^{15}$C nuclei [117]. These studies turned out to be precious even for a quantitative and detailed understanding of one- and two-neutron transfer mechanisms in ($^{18}$O,$^{17}$O) and ($^{18}$O,$^{16}$O) reactions above the Coulomb barrier [90]. New important results are found in newly collected data, whose analyses are on the way, which will be published soon.

Another important research field, approached in experiments performed with MAGNEX, is the study of nuclear response to charge exchange reactions. Initial experiments have investigated the structure of light neutron rich nuclei by ($^{7}$Li,$^{7}$Be) reactions [134]. The most important finding, achieved also thanks to a fully consistent microscopic analysis of the measured energy spectra and absolute cross sections, was the dominance of a direct reaction mechanism over the two-step multi-nucleon transfer, at least at forward angles. Consequently, $\beta$-decay strengths can be accessed, within reasonable accuracy. This was also confirmed by new results of the never explored ($^{18}$O,$^{18}$F) single charge exchange reaction, which turns out to be an interesting spectroscopic probe for Gamow-Teller and Fermi transitions in nuclei.

The results of single charge exchange reactions have also suggested to explore the more ambitious double charge exchange processes and namely the ($^{18}$O,$^{18}$Ne) reaction [143] as a tool for nuclear structure. Measurements around zero degree, performed with MAGNEX, have shown that despite the vanishing cross sections, valuable information can be extracted from the data. This result is particularly interesting towards the experimental access to the nuclear matrix elements entering in the expression of the half-life of $0\nu\beta\beta$ decay. Actually, first results are indicating that the MAGNEX data can give a crucial contribution especially if state of art analyses, based on microscopic and consistent theories of nuclear structure and reaction will be developed to this purpose.

Also due to the intrinsic relevance of the $0\nu\beta\beta$ decay in modern physics, we consider the development of double charge exchange research line as the leading one for the future of the experimental activity of MAGNEX at the LNS laboratory. A major upgrade of both the superconducting cyclotron accelerator and the spectrometer facilities is indispensable, in the view of an increase of the beam current of almost two orders of magnitude up to 10 kW or so. The NUMEN project, recently funded by INFN, is proposing this scientific content and the solution to the relative technological challenges, with the aim of giving a relevant contribution to the puzzle of neutrino mass.

# Acknowledgments

The authors are particularly grateful to Prof. A. Cunsolo, who first proposed the MAGNEX spectrometer and lead the group for more than one decade up to its successful commissioning. Among the other members of the team, who have participated to this enterprise we should specifically mention Dr. A. Lazzaro, for his key contribution to the ion optics, Dr. J.S. Winfield for having initially set up the FPD, the electronic chain and the DAQ. Special thanks to Prof. A. Foti for his continuous activity on the MAGNEX project. Thanks are also due to the past and present members of the group and collaborators for their precious help in specific aspects of the project. We want to thank the INFN-LNS accelerator, technical and research divisions for the fundamental support in all the phases of the project, from the installation up to the scientific activity. For nuclear rainbow in $^{16}O+^{27}Al$ scattering, we wish to acknowledge Prof. D. Pereira (deceased in 2012) and the group of Universidad de São Paulo and Universidad Federal Fluminense, who proposed this research. For the $^{6}Li$ + p, we acknowledge Prof. A. Pakou and the group of University of Ioannina, for proposing this interesting subject. We also acknowledge Dr. J.A. Scarpaci from IN2P3 CSNSM for the proposal of (p,t) reaction and Dr. F. Azaiez from IN2P3 IPN-Orsay for proposing and finalizing the installation of the EDEN array in coincidence with MAGNEX.